\pdfoutput=1
\documentclass[11pt,twoside,a4paper,cmspaper,final,collab]{cms-tdr}
\def\svnVersion{6318740}\def\svnDate{2025/12/29}\def\cmsCernNoTag{CERN-EP-2025-154}\def\cmsCernDate{\today}\def\cmsMessage{Published in Physical Review D as \href{http://dx.doi.org/10.1103/b26z-zmpy}{\doi{10.1103/b26z-zmpy}.}}
\begin{document}\cmsNoteHeader{SUS-23-003}

\newlength\cmsTabSkip\setlength{\cmsTabSkip}{1ex}
\newlength\cmsTabSkipLarge\setlength{\cmsTabSkipLarge}{3ex}

\ifthenelse{\boolean{cms@external}}{\providecommand{\cmsTable}[1]{#1}}{\providecommand{\cmsTable}[1]{\resizebox{\textwidth}{!}{#1}}}

\ifthenelse{\boolean{cms@external}}{\providecommand{\cmsLeft}{upper\xspace}}{\providecommand{\cmsLeft}{left\xspace}}
\ifthenelse{\boolean{cms@external}}{\providecommand{\cmsRight}{lower\xspace}}{\providecommand{\cmsRight}{right\xspace}} 

\newcommand{\argmin}{\operatornamewithlimits{argmin}}
\newcommand{\argmax}{\operatornamewithlimits{argmax}}
\newcommand{\I}[1]{\ensuremath{{\mathbf{#1}}}}
\newcommand{\V}[1]{\ensuremath{{\mathbf{#1}}}}
\newcommand{\D}[1]{\ensuremath{{\mathbf{#1}}}}
\newcommand{\lab}{\ensuremath{\mathbf{lab}}}
\newcommand{\pfour}[2]{\ensuremath{\mathbf{p}_{\mkern1mu#1}^{#2}}}
\newcommand{\pthree}[2]{\ensuremath{\vec{p}_{\mkern1mu#1}^{\mkern4mu#2}}\xspace}
\newcommand{\pone}[2]{\ensuremath{p_{\mkern1mu#1}^{#2}}}
\newcommand{\phat}[2]{\ensuremath{\hat{p}_{\mkern1mu#1}^{\mkern2mu#2}}\xspace}
\newcommand{\vbeta}[2]{\ensuremath{\vec{\beta}_{\mkern1mu#1}^{\mkern4mu#2}}\xspace}
\newcommand{\sbeta}[2]{\ensuremath{\beta_{#1}^{#2}}}
\newcommand{\hbeta}[2]{\ensuremath{\hat{\beta}_{\mkern1mu#1}^{\mkern2mu#2}}\xspace}
\newcommand{\mass}[2]{\ensuremath{m_{#1}^{#2}}}
\newcommand{\Mass}[2]{\ensuremath{M_{#1}^{#2}}}
\newcommand{\risr}{\ensuremath{R_{\mathrm{ISR}}}\xspace}
\newcommand{\ptisr}{\ensuremath{p_{\mathrm{T}}^{\mkern3mu\mathrm{ISR}}}\xspace}
\newcommand{\mperp}{\ensuremath{M_{\perp}}\xspace}
\newcommand{\gperp}{\ensuremath{\gamma_{\perp}}\xspace}
\newcommand{\ptcm}{\ensuremath{p_{\mathrm{T}}^{\mkern3mu\mathrm{CM}}}\xspace}
\newcommand{\dphicmi}{\ensuremath{\Delta \phi_{\mathrm{CM,I}}}\xspace}
\newcommand{\NjetISR}{\ensuremath{N_{\text{jet}}^{\mathrm{ISR}}}\xspace}
\newcommand{\NjetS}{\ensuremath{N_{\text{jet}}^{\mathrm{S}}}\xspace}
\newcommand{\NbS}{\ensuremath{N_{\text{\PQb tag}}^{\mathrm{S}}}\xspace}
\newcommand{\NbISR}{\ensuremath{N_{\text{\PQb tag}}^{\mathrm{ISR}}}\xspace}
\newcommand{\NsvS}{\ensuremath{N_{\mathrm{SV}}^{\mathrm{S}}}\xspace}
\newcommand{\etaSV}{\ensuremath{\eta_{\mathrm{SV}}^{\mathrm{S}}}\xspace}
\newcommand{\elel}{\ensuremath{\Pe\Pe}\xspace}
\newcommand{\MUMU}{\ensuremath{\PGm\PGm}\xspace}
\newcommand{\emu}{\ensuremath{\Pe\PGm}\xspace}
\newcommand{\SUL}{\ensuremath{S_{\mathrm{UL}}^{95\%}}\xspace}
\newcommand{\ttjets}{\ensuremath{{\PQt\PAQt}+\text{jets}}\xspace}
\newcommand{\ttx}{\ensuremath{{\PQt\PAQt}\PX+\text{jets}}\xspace}
\newcommand{\PGgst}{\ensuremath{\HepParticle{\PGg}{}{\ast}}\xspace}
\newcommand{\zdy}{\ensuremath{\PZ/\PGgst+\text{jets}}\xspace}
\newcommand{\Wjets}{\ensuremath{\PW{}+\text{jets}}\xspace}
\newcommand{\Nbins}{\ensuremath{N_{\text{bins}}}\xspace}
\newcommand{\PSeLpm}{{\HepSusyParticle{\Pe}{L}{\pm}}\xspace}
\newcommand{\PSeRpm}{{\HepSusyParticle{\Pe}{R}{\pm}}\xspace}
\newcommand{\PSGmLpm}{{\HepSusyParticle{\PGm}{L}{\pm}}\xspace}
\newcommand{\PSGmRpm}{{\HepSusyParticle{\PGm}{R}{\pm}}\xspace}
\newcommand{\PSlLpm}{{\HepSusyParticle{\Pell}{L}{\pm}}\xspace}

\cmsNoteHeader{SUS-23-003}
\title{General search for supersymmetric particles in scenarios with compressed mass spectra using proton-proton collisions at \texorpdfstring{$\sqrt{s}=13\TeV$}{sqrt(s) = 13 TeV}}

\date{\today}

\abstract{
A general search is presented for supersymmetric particles (sparticles) in scenarios featuring compressed mass spectra using proton-proton collisions at a center-of-mass energy of 13\TeV, recorded with the CMS detector at the LHC. The analyzed data sample corresponds to an integrated luminosity of 138\fbinv. A wide range of potential sparticle signatures are targeted, including pair production of electroweakinos, sleptons, and top squarks. The search focuses on events with a high transverse momentum system from initial-state-radiation jets recoiling against a potential sparticle system with significant missing transverse momentum. Events are categorized based on their lepton multiplicity, jet multiplicity, number of \PQb-tagged jets, and kinematic variables sensitive to the sparticle masses and mass splittings. The sensitivity extends to higher parent sparticle masses than previously probed at the LHC for production of pairs of electroweakinos, sleptons, and top squarks with mass spectra featuring small mass splittings (compressed mass spectra). The observed results demonstrate agreement with the predictions of the background-only model. Lower mass limits are set at 95\% confidence level on production of pairs of electroweakinos, sleptons, and top squarks that extend to 325, 275, and 780\GeV, respectively, for the most favorable compressed mass regime cases.  
}

\hypersetup{%
pdfauthor={CMS Collaboration},%
pdftitle={General search for supersymmetric particles in scenarios with compressed mass spectra using proton-proton collisions at sqrt(s) = 13 TeV},%
pdfsubject={CMS},%
pdfkeywords={CMS, compressed spectra, leptons, missing energy, supersymmetry search}} %

\maketitle

\section{Introduction}\label{sec:intro}
The standard model (SM) is a tremendously successful theoretical framework that essentially 
describes all known phenomena in high-energy physics. With the
demonstration of the existence of the Higgs boson~\cite{Aad:2012tfa,Chatrchyan:2012ufa},
the field is at a crossroads. On the one hand, as of yet,
there is no direct experimental evidence from colliders for new
phenomena beyond the SM (BSM), 
such as the new fundamental particles envisaged in supersymmetry (SUSY)
models~\cite{PDG2024}. On the other hand, the very existence of
the Higgs boson and the presence of dark matter in the universe are
compelling motivations for a model such as SUSY to be
realized in nature; it can stabilize the Higgs boson mass and has the potential to
provide a particle physics explanation for dark matter~\cite{Jungman:1995df}.
It is therefore crucial to confront such possibilities with experiment.
Supersymmetry~\cite{Drees:2004jm,Baer:2006rs,Martin:1997ns} has attracted much
interest as a result of its perceived strong motivation,
its tractability as a weakly coupled theoretical
framework for perturbative calculations and thus predictions, and the
rich set of potential new experimental signatures.

With a wide variety of search results
from the LHC experiments based on the data sets collected in the years 2016--2018,
many supersymmetric particle (sparticle) production scenarios have been constrained 
by a number of searches at the LHC~\cite{Aaboud:2017leg,Sirunyan:2019mlu,Sirunyan:2019zfq,
Aad:2019qnd,Sirunyan:2020eab,Sirunyan:2020tyy,Aad:2019vvf,
Aad:2019vvi,Aad:2021zyy,CMS:2021edw, CMSPRD.104.052001, 
Atlas2004.14060, Atlas2012.03799, ATLASPRD.103.112006, ATLAS:2024lda, 
ATLAS:2019lff,ATLAS:2021moa,ATLAS:2021yqv,ATLAS:2022zwa,
ATLAS:2022hbt,ATLAS:2023act,ATLAS:2024fub,ATLAS:2024qxh,
CMS:2021cox,CMS:2021few,CMS:2022vpy,CMS:2022sfi,ATLAS:2024rcx,ATLAS:2024umc,CMS:2024gyw,ATLAS:2025evx,ATLAS:2025dns}.
These results are primarily in the context of simplified model interpretations 
with the experimentally most favorable realizations leading to lower mass limits 
at the TeV scale and beyond for specific scenarios.
Nevertheless, there is still very strong experimental and phenomenological 
motivation for a focus on \textit{compressed} sparticle mass spectra, where 
the mass differences ($\Delta m$) between the initially produced (parent) sparticles and the 
lightest sparticle (LSP) are small. 
Supersymmetry searches are often least sensitive in corridor regions with small
mass differences; if SUSY is to be tested comprehensively, further exploration of these regions is essential.
Phenomenologically, the lowest lying states in the
electroweakino sector ($\PSGczDo, \PSGcpmDo, \PSGczDt$) may
form a nearly mass-degenerate dominantly higgsino-like triplet~\cite{Han:2014kaa}.
This scenario is particularly challenging as a result of
the suppressed production cross sections in addition to
the compressed mass spectrum, and is attracting
much interest~\cite{Canepa:2020ntc,Baer:2020sgm}. Furthermore, probing slepton production models 
including those with the SUSY partners of the muon (smuons), 
could give insight into potential supersymmetric contributions 
to the muon 
g-2 measurements~\cite{ref:gmin1, ref:gmin3, ref:gmin4} 
as calculated in for example~\cite{Moroi:1995yh, Martin:2001st, Stockinger:2006zn}.

A general search for sparticles is performed in proton-proton ($\Pp\Pp$) collisions 
at a center-of-mass energy of 13\TeV by 
the CMS experiment at the CERN LHC. The data were collected
from 2016 to 2018, with a total integrated luminosity 
of 138\fbinv. 
The focus for this paper is on SUSY scenarios
featuring compressed mass spectra, with the LSP expected to be the weakly interacting lightest neutralino, $\PSGczDo$.
Only $R$-parity\footnote{$R$-parity is a multiplicative quantum number defined by $R = (-1)^{3B+L+2S}$ where $B$, $L$, $S$ are 
the baryon number, lepton number, and spin, respectively, with all SM particles having $R=+1$ and all sparticles having $R=-1$.} 
conserving SUSY scenarios~\cite{Farrar:1978xj}, 
where sparticles are produced in pairs and the LSP is stable, are considered.
We target SUSY scenarios that include the 
associated production of a chargino and neutralino ($\PSGcpmDo \PSGczDt$) and the pair production of charginos, 
top squarks, and charged sleptons (selectrons or smuons). 
Observed event yields for various event signatures 
are also reported in a model-independent manner that does not assume a particular BSM particle production model.

Searching for the production and decays of such sparticles appearing in compressed 
mass spectra is experimentally challenging, as small mass splittings between sparticles imply that the visible products of
those decays will be of low momentum, and can be difficult to reconstruct, or even detect.
Normally reliable signatures of SM particles, such as the reconstructed mass of heavy vector bosons,
can be significantly distorted when forced off-shell and produced in
such decays, thereby degrading our ability to detect them.
For decays resulting in weakly interacting massive particles,
this can also mean that these invisible decay products will receive very little momentum 
from the decays of their parents, such that the resulting missing transverse momentum may 
also be small and indistinguishable from that of backgrounds.
The approach taken is primarily kinematic,
and a wide range of object multiplicities are used to incorporate
the potential decay signatures of the targeted sparticle systems. 
Events are selected with significant initial-state
radiation (ISR), where the high transverse momentum recoil
from the ISR can often lead 
to measurable missing transverse momentum associated with sparticle
decays in compressed scenarios, despite each invisible LSP acquiring only a small momentum in the parent rest frame from 
the parent sparticle decay. 
The method adopted is more general than, and complementary to, previous searches by CMS for signatures of 
compressed sparticle mass spectra, such as Ref.~\cite{CMS:2021edw}, which focused on events with two or three soft leptons.

The paper is organized as follows. The CMS detector and event reconstruction are described in
Section~\ref{sec:det}. The SUSY signal and background process modeling
and simulation are described in Section~\ref{sec:samples}.  
The following section (Section~\ref{sec:reco}) describes  
the selection of 
physics objects including electrons, muons,
jets,  \PQb-tagged jets, and  \PQb-tagged secondary vertices for use in the analysis. 
Section~\ref{sec:RJR} describes the kinematic reconstruction of events for this search  
considering multilepton final states, corresponding to exactly 0, 1, 2, and 3 leptons (electrons or
muons) with jets. The event selection and categorization (Section~\ref{sec:analysis}) has two elements.
Firstly, preselection criteria and event clean-up requirements are applied that remove events that are not consistent with 
the compressed phase space of interest for the analysis. Secondly, events are categorized into mutually exclusive analysis 
regions that are defined according to a combination of object multiplicities in the supersymmetric or ISR systems. 
A fit based on control samples in data used to constrain the background contributions in tests of various signal
hypotheses is described in Section~\ref{sec:fit}, including discussion of the treatment of systematic uncertainties.
Results are given in Section~\ref{sec:results} and the paper is summarized in Section~\ref{sec:summary}.
Tabulated results are provided in the HEPData record for this analysis~\cite{hepdata}.

\section{The CMS detector and event reconstruction}\label{sec:det}
A central feature of the CMS detector is a superconducting solenoid of 6\unit{m} internal diameter, providing a magnetic field of 3.8\unit{T}. 
Within the volume of the solenoid are a silicon pixel and strip tracker, a lead tungstate crystal electromagnetic calorimeter (ECAL), and a brass 
and scintillator hadron calorimeter (HCAL), each with a barrel and two endcap sections. Muons are detected using the gas-ionization chambers embedded in 
the steel flux-return yoke outside the solenoid. A detailed description of the CMS detector, including the definition of the 
coordinate system used, can be found in Refs.~\cite{cmsdet,CMS:2023gfb}. 

For this analysis, physics objects, such as jets, electrons, muons, and missing transverse momentum, are considered. 
The reconstruction and identification 
of individual particles in an event is performed 
using the particle-flow algorithm \cite{ref:partflow} with an optimized combination of information 
from the various elements of the CMS detector. The energy of photons is directly obtained from 
the ECAL measurement. 
Reconstructed energies of electrons are determined from a combination of the electron momentum at the primary interaction vertex as determined by the tracker, 
the energy of the corresponding ECAL cluster, and the energy sum of all bremsstrahlung photons spatially compatible with the 
origin of the electron track. The momentum of muons is estimated from the curvature of the corresponding track. The energy of charged hadrons 
is determined from a combination of their momentum measured in the tracker and the matching ECAL and HCAL energy deposits, corrected 
for the response of the calorimeters to hadronic showers. Finally, the energy of neutral hadrons is obtained from the corresponding corrected ECAL and HCAL energy.

Hadronic jets are found from these reconstructed particles and clustered using the infrared- and collinear-safe 
anti-\kt algorithm~\cite{ref:antikt}, implemented with the \textsc{FastJet} package~\cite{ref:fastjet}. 
Jets in this analysis use the anti-\kt distance parameter of 0.4. Jet momentum is determined as the 
vector sum of all particle momenta clustered in the jet. 
Additional $\Pp\Pp$ interactions within the same or nearby
bunch crossings (pileup) can contribute additional tracks and calorimetric energy deposits to the
jet momentum. To mitigate this effect, tracks identified as originating from pileup vertices
are discarded, and an offset correction is applied to correct for remaining contributions. Jet
energy corrections are derived from simulation to equalize the average measured response of jets 
to that of particle level jets.
\textit{In situ} measurements of the momentum balance in dijet, photon+jet, $\PZ$+jet, 
and multijet events are used to account for any residual differences in jet energy scale between data and simulation~\cite{ref:JES}. 
Typical residual response corrections are less than 3\% in the barrel region and less than 10\% in the endcap region.
Additional selection criteria are applied to each jet to remove jets potentially dominated by anomalous contributions 
from various subdetector components or reconstruction failures. The ultimate jet energy resolution 
typically ranges from 15\% at 10\GeV, 8\% at 100\GeV, to 4\% at 1\TeV~\cite{ref:JES}.

The momentum resolution for electrons 
with \pt of 45\GeV from $\PZ\to \Pe\Pe$ decays ranges from 1.7\% for barrel electrons that do not generate showers in the tracker to 4.5\% for showering electrons 
in the endcaps~\cite{ref:eres,ref:eideff}.
Muons are measured in the pseudorapidity range $\abs{\eta} < 2.4$, with detection planes made using three technologies: drift
tubes, cathode strip chambers, and resistive-plate chambers.  Matching muons to tracks measured in the silicon tracker results in a relative transverse momentum resolution for muons
with $20 < \pt < 100\GeV$ of 1.3--2.0\% in the barrel and better than 6\% in the endcaps. The \pt resolution in the barrel is better than 10\% for 
muons with \pt up to 1\TeV~\cite{ref:mures}.

The missing transverse momentum (\ptvecmiss) 
is defined as the negative vector sum of the transverse momenta 
of all particle-flow candidates in the event and its magnitude 
is denoted by \ptmiss.  Anomalous high-\ptmiss events can occur as a result of a 
variety of reconstruction failures, detector malfunctions, or noncollision backgrounds. Such events are rejected by event filters that are designed to identify 
more than 85--90\% of the spurious high-\ptmiss events with a mistagging rate less than 0.1\%~\cite{ref:metperf}.  The \ptvecmiss is modified to account for 
corrections to the energy scale of the reconstructed jets in the event.

Vertices are reconstructed from tracks according to the deterministic annealing algorithm~\cite{ref:determin}.  
The primary vertex is taken to be the vertex corresponding to the hardest scattering in the event, evaluated using 
tracking information alone, as described in Section 9.4.1 of Ref.~\cite{CMS-TDR-15-02}.
Events are required to have at 
least one reconstructed vertex with longitudinal position $\abs{z} < 24$\unit{cm} and radial position $r<2$\unit{cm} relative to the nominal mean collision point.

Events of interest are selected using a two-tiered trigger system. 
The first level, composed of custom hardware processors, uses information from the calorimeters 
and muon detectors to select events at a rate of around 100\unit{kHz} within a time interval of less than 4\unit{ms}. 
The second level, known as the high-level trigger, consists of a farm of processors running 
a version of the full event reconstruction software 
optimized for fast processing, and reduces the event rate to a few kHz before data storage~\cite{ref:trigsys, CMS:2024aqx}. 
The data sample for this analysis was collected in 2016--2018 
using inclusive \ptmiss triggers with $\ptmiss > 120$\GeV, corresponding to a total integrated 
luminosity of 138\fbinv of $\Pp\Pp$ collisions.

\section{Signal and background simulation}\label{sec:samples}
The SUSY signal and SM background processes are simulated using Monte Carlo (MC) event generators. 
Several different SUSY simplified models are used to study electroweakino, slepton, and top squark production and decay~\cite{ref:sms1, ref:sms2, ref:sms3}. 

Simulated samples are generated either at leading order (LO) or next-to-leading order (NLO) and use parton distribution functions (PDF) from either the NNPDF3.0~\cite{ref:nnpdf} set for 2016 data or 
the NNPDF3.1~\cite{ref:nnpdf2} set for 2017 and 2018 data. 
Hadronization and showering of events in all generated samples are simulated using \PYTHIA 8.230~\cite{ref:pythia}; these use 
the CUETP8M1, CP2, and CP5 tunes of the underlying-event simulation~\cite{ref:tune}.
All simulations corresponding to 2016 data use the CUETP8M1 tune, 
the CP2 tune is used for signal simulations corresponding to the 2017 and 2018 data, while 
the CP5 tune is used for background simulations corresponding to the 2017 and 2018 data. 
The background events are passed through a full simulation of the CMS apparatus, with the response of the detector 
modeled using the \GEANTfour~\cite{Agostinelli:2002hh} simulation toolkit. The detector simulation of signal samples is performed 
with the CMS fast simulation package \textsc{FastSim}~\cite{ref:fastsim1, ref:fastsim2}.  
Several sets of simulations are processed 
so that the version of the CMS event reconstruction software used matches the run conditions of the collected data sets.
Additional $\Pp\Pp$ collisions from pileup interactions are simulated and overlaid on the main interaction in the MC samples, 
with vertex distributions that reproduce conditions observed year-to-year in data. There was a trigger inefficiency during 2016 and 2017 
caused by a gradual shift in the timing of the inputs of the ECAL first-level trigger 
in the region $2.5 < \abs{\eta} <3.0$. The resulting efficiency loss is 10--20\% for events triggered by 
an electron (a jet) with \pt larger than ${\approx}50$ (${\approx}100$)\GeV in the specified $\abs{\eta}$ region, and 
is a function of \pt, $\eta$, and time. Correction factors are estimated 
from the data to model this effect in simulation.

For the simplified model based signal models, all SUSY particles other
than the electroweakinos, sleptons, or top squarks under study are assumed to be too massive to
affect the analysis observables. 
These simulated samples have sparticle decays 
with 100\% branching fraction 
to a particular final state and always include the $\PSGczDo$ as the LSP.
The signals are all generated using the \MGvATNLO (v2.2.2 for 2016 and v2.4.2 for 2017--2018) generator 
with LO precision and up to two additional partons at the matrix element level~\cite{ref:madgraph} and 
interfaced to \PYTHIA for sparticle decay.
The production cross sections are computed at NLO plus next-to-leading logarithmic precision
with all the other sparticles assumed to be heavy and decoupled using a number 
of calculations and computational tools~\cite{ref:sigcross1, ref:sigcross2, ref:sigcross3, ref:sigcross4, ref:sigcross5, Beenakker:1997ut, Beenakker:2010nq, Beenakker:2016gmf}.

\begin{figure}[ht]
	\centering
	\includegraphics[width=0.40\textwidth]{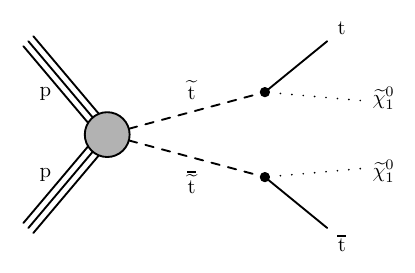}
	\includegraphics[width=0.40\textwidth]{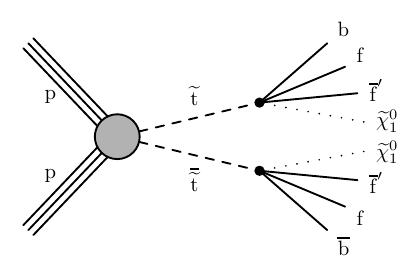}	
	\caption{Diagrams for top squark pair production. The \cmsLeft panel shows the T2tt model 
	with decay via top quarks and the \cmsRight panel illustrates the four-body phase space used in modeling the most compressed region.}
	\label{fig:grid_T2tt}
\end{figure}

\begin{figure}[ht]
	\centering
	\includegraphics[width=0.40\textwidth]{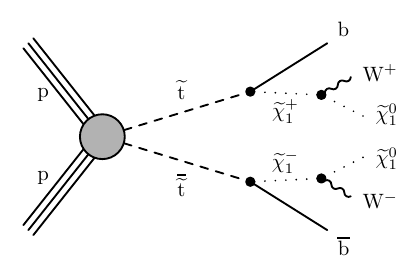}
	\includegraphics[width=0.40\textwidth]{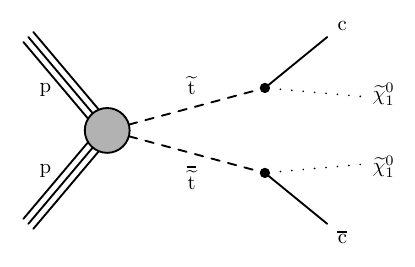}	
	\caption{Diagrams for top squark pair production. The \cmsLeft panel shows the T2bW model with decay via 
	an intermediate mass chargino and the \cmsRight panel shows the T2cc model with decay via charm quarks.}
	\label{fig:T2bW}
\end{figure}

For the direct top squark pair production and decay three simplified models are explored: T2tt (with top squark decay 
via a top quark), T2bW (with the top squark decaying through an intermediate chargino that 
subsequently decays 
to a \PW boson and the lightest neutralino), and T2cc (with top squark decay 
via a charm quark). 
A range of top squark and LSP masses is considered with mass differences
ranging from 6 to 200\GeV. For the T2tt model when $\Delta m \le 80$\GeV, where the top quark 
and the \PW boson from the top quark decay would both be off-shell, the decay phase space is modeled 
as a four-body decay ($\PSQt\to \PQb \mathrm{f} \bar{\mathrm{f'}} \PSGczDo$) whereas for 
intermediate mass differences, $80 < \Delta m \le 175$\GeV, where 
the top quark must be off-shell but the \PW boson can be resonant, the stop quark decay 
phase space is modeled as a three-body decay ($\PSQt \to \PQb \PW \PSGczDo$).
For the T2bW model, 
the mass of the intermediate chargino is set to $\tfrac{1}{2} [m(\PSQt) + m(\PSGczDo)]$. 
Example diagrams are shown in Figs.~\ref{fig:grid_T2tt} and ~\ref{fig:T2bW}.

\begin{figure}
	\centering
	\includegraphics[width=0.40\textwidth]{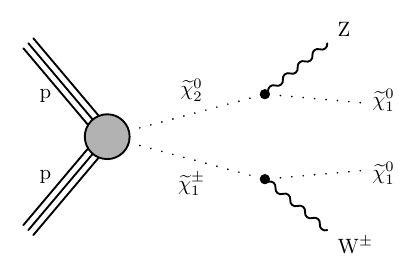}
    \includegraphics[width=0.40\textwidth]{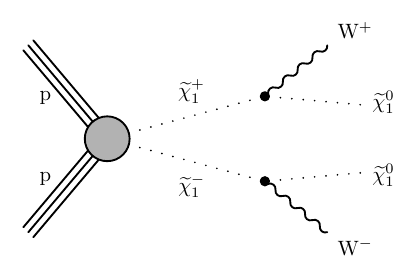}
	\caption{Diagrams for electroweakino production. 
	The \cmsLeft panel shows associated production of the lightest chargino and second-lightest neutralino ($\PSGcpmDo\PSGczDt$) 
	in the TChiWZ model and the \cmsRight panel shows pair production of the lightest chargino 
	($\PSGcpDo\PSGcmDo$) in the TChiWW model.}
	\label{fig:grid_TChiWZ}
\end{figure}

The primary simplified models for electroweakino production and decay explored assume that a 
chargino-neutralino pair $\PSGcpmDo\PSGczDt$ (TChiWZ) 
or a chargino pair $\PSGcpDo\PSGcmDo$ (TChiWW) 
is produced. 
Each chargino decays to a \PW boson and the \PSGczDo, while the 
second-lightest neutralino, \PSGczDt, 
decays to a \PZ boson 
and the \PSGczDo, where the \PSGczDo is the LSP.  
The diagrams for these production and decay processes are shown in Fig.~\ref{fig:grid_TChiWZ}. 
The TChiWZ model has the \PSGcpmDo and \PSGczDt (the initially-produced parent sparticles) 
with the same mass. 
For the TChiWW model, the \PSGcpmDo's are pair produced.  
Interpretations with purely wino- and higgsino-like \PSGcpmDo and \PSGczDt are included for the TChiWZ model,
while the pure wino-like interpretation is used for TChiWW.  
A range of \PSGcpmDo, \PSGczDt, 
and LSP masses is considered with mass 
differences ranging from 3 to 200\GeV; consequently the \PW and \PZ bosons 
are off-shell for much of the (mass, $\Delta m$) plane considered.

\begin{figure}[ht]
	\centering
	\includegraphics[width=0.40\textwidth]{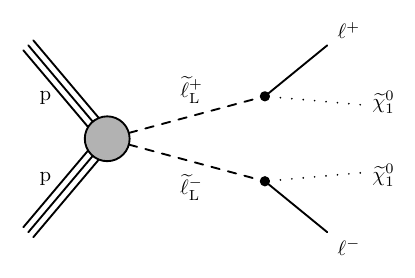}
	\includegraphics[width=0.40\textwidth]{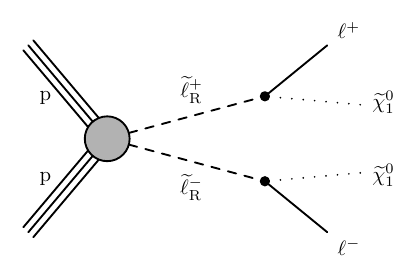}	
	\caption{Diagrams for pair production of charged sleptons with subsequent decay to $\Pellpm\PSGczDo$ where $\Pell = \Pe,\PGm$.}
	\label{fig:grid_TSlepSlep}
\end{figure}

For slepton pair production, the four charged sleptons of the first and second 
generation (\ie, selectrons and smuons), namely, the superpartners of both lepton 
chiralities ($\PSeLpm$, $\PSeRpm$, $\PSGmLpm$, $\PSGmRpm$),  
are pair produced and decay with a 100\% branching fraction 
to $\Pellpm\PSGczDo$. 
These possibilities are illustrated in Fig.~\ref{fig:grid_TSlepSlep}. 
The simplified model where all four states have the same mass 
is referred to as the TSlepSlep model.
Mass differences ranging from 3 to 100\GeV are considered and the generated event samples based on the 
TSlepSlep model permit exploration of appropriate combinations of the four states. 
 
\begin{figure}[ht]
	\centering
	\includegraphics[width=0.40\textwidth]{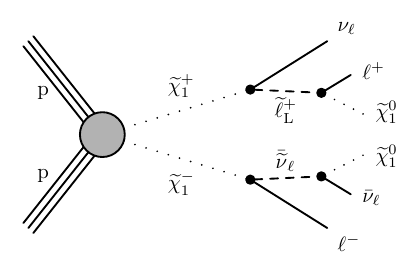}
	\includegraphics[width=0.40\textwidth]{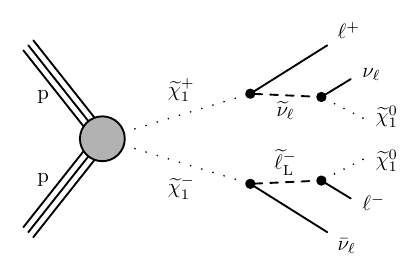}	
	\caption{Diagrams for pair production of the lightest chargino with subsequent leptonic decays via an intermediate 
	mass charged slepton or sneutrino, where $\Pell = \Pe,\PGm,\Pgt$. In addition to the illustrated diagrams, the 
	other two combinations where either both charginos decay to an intermediate charged slepton or both charginos decay to 
	an intermediate sneutrino are also included in this TChiSlepSnu model.}
	\label{fig:grid_TChiSlepSnu}
\end{figure}

An additional model for chargino pair production, denoted the TChiSlepSnu model, is studied. In this scenario, 
the chargino decays via 
an intermediate charged slepton ($\PSlLpm$) or sneutrino, as illustrated in Fig.~\ref{fig:grid_TChiSlepSnu}. 
In this case, the mass of the intermediate state is set halfway between the masses of the chargino and the LSP,  
and it is assumed that the chargino decays with equal probability to each slepton 
and sneutrino flavor (branching fraction of $1/6$). 
Mass differences between the chargino and the LSP exceeding 50\GeV are considered for this specific model.

The background samples are partitioned in seven groups:
\begin{enumerate}
	\item \zdy: composed of the $\PZ$+jets and Drell--Yan backgrounds and 
	generated at LO using \MGvATNLO.
	\item \Wjets: generated at LO using \MGvATNLO.
	\item \ttx: \ttjets is generated at LO using \MGvATNLO with the final states separated into dilepton and single lepton. 
	             Also included are \ttbar + boson processes generated with \MGvATNLO.
	\item ST: this group covers the three single top quark production processes corresponding to the $s$-channel, $t$-channel, 
	and $\PW$-associated production ($tW$). The $s$-, $t$-channel and leptonic $tW$ processes are simulated using \MGvATNLO, while the inclusive $tW$ samples 
	are simulated at NLO with the \POWHEG v1 generator~\cite{ref:pow1, ref:pow2, ref:pow5, ref:pow6}.
	\item VV: includes diboson processes. The $\PW\PW$, $\PZ\PZ$, $\PW\Pgg$ production and non-\bbbar Higgs boson decays for $\PW\PH$ and $\PZ\PH$ production 
	are generated at NLO precision with \MGvATNLO using the FxFx merging scheme~\cite{ref:fxfx}. The $\PW\PZ$ process, the \bbbar Higgs boson decays for $\PW\PH$ and $\PZ\PH$ production, and Higgs bosons from gluon-gluon fusion are 
	all generated with \POWHEG v2~\cite{ref:pow3, ref:pow4}. 
	\item VVV: includes triboson processes generated using \MGvATNLO at NLO.
	\item QCD: quantum chromodynamics (QCD) multijet background 
	using samples generated with \MGvATNLO. 
	Due to shortcomings in its simulation, this background is accounted for by using an approach almost 
	fully relying on control samples in data.
\end{enumerate}

\section{Physics object reconstruction}\label{sec:reco}
After the basic event reconstruction described in Section~\ref{sec:det}, physics objects such as electrons, muons,  \PQb-tagged jets, 
and tagged secondary vertices are reconstructed. Because the analysis targets compressed SUSY signatures, we prioritize the identification of objects 
with as low a \pt as can be efficiently analyzed.

Electrons with $\pt > 5$\GeV and $\abs{\eta} < 2.4$ are identified using a multivariate discriminant based on track quality 
variables and the energy distribution in both the ECAL and HCAL, with the very loose selection initially applied~\cite{ref:eideff}. 
Candidate electrons must have tracks that have a hit in every pixel detector layer and which are not associated with 
a reconstructed photon conversion vertex.
Candidate muons with $\pt > 3$\GeV and $\abs{\eta} < 2.4$ are selected based on the quality of the tracks both in 
the tracker and in the muon system, with the condition that they are matched to each other and satisfy the 
loose and soft identification criteria from Ref.~\cite{ref:muonid}. 
Initially, loose requirements are applied on the track quality of the leptons to qualify them for the analysis. 
These track quality criteria include requirements on the three-dimensional (3D) impact parameter significance ($\text{IP}_{\mathrm{3D}}/\sigma_{\text{IP}_{\mathrm{3D}}}<8$), 
the two-dimensional (2D) transverse distance of closest approach to the primary vertex ($\abs{d_{xy}}<0.05$\unit{cm}), and 
the longitudinal distance of closest approach to the primary vertex ($\abs{d_z}<0.1$\unit{cm}). 

Electrons and muons that satisfy these preselection requirements are separated into three mutually exclusive categories: gold, silver, and bronze. 
The gold category represents the most signal-like prompt and isolated leptons, while the silver category is used to recover efficiency from 
isolated secondary leptons from sources such as semileptonic decays of \PQb hadrons. 
The remaining leptons that do not qualify as gold or silver, but which satisfy the 
loose quality criteria described previously, are classified as bronze. 
Gold and silver electrons with $\pt > 10$\GeV are additionally required to pass the 
tight identification criteria, while muons are required to pass the medium 
identification selection~\cite{ref:muonid}. 
Gold and silver leptons must also satisfy further isolation criteria. Absolute isolation requirements 
separate the leptons from jets using the \pt sum deposited by the particle-flow candidates in a 
cone of radius $\DR=0.3$ around the lepton, where $\DR \equiv \sqrt{\smash[b]{(\Delta\eta)^2+(\Delta\phi)^2}}$, and $\phi$ 
is the azimuthal angle measured in radians. Mini-isolation is defined as the 
\pt sum of charged hadron, neutral hadron, and photon particle-flow candidates within a 
cone in $\eta$-$\phi$ space around the lepton, correcting for pileup using an effective area method~\cite{miniiso}. 
The cone size depends on the lepton \pt and has radius $\cal{R}$, defined as 
\begin{equation}
\mathcal{R}=\frac{10\GeV}{\text{min}(\text{max}(\pt,\,50\GeV),\,200\GeV)}. 
\end{equation}
Both absolute and mini-isolation are required to be less than 4\GeV for
gold and silver leptons.

Gold and silver leptons are then differentiated by their consistency with originating directly 
from the primary vertex, with gold categorization requiring a tighter 3D impact parameter 
significance, $\text{IP}_{\mathrm{3D}}/\sigma_{\text{IP}_{\mathrm{3D}}} < 2$, with the criterion reversed for silver leptons.

\begin{figure}[tbp]
	\centering
    \includegraphics[width=0.49\textwidth]{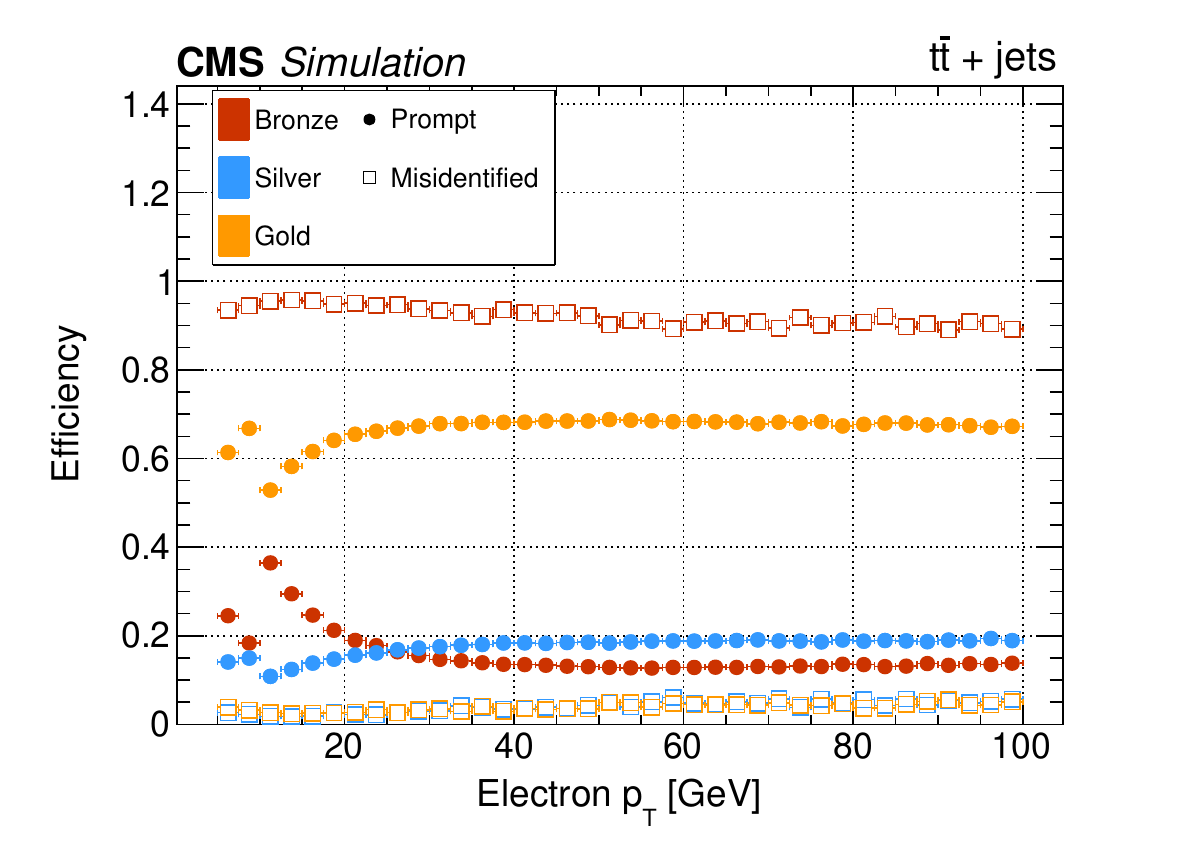}
	\includegraphics[width=0.49\textwidth]{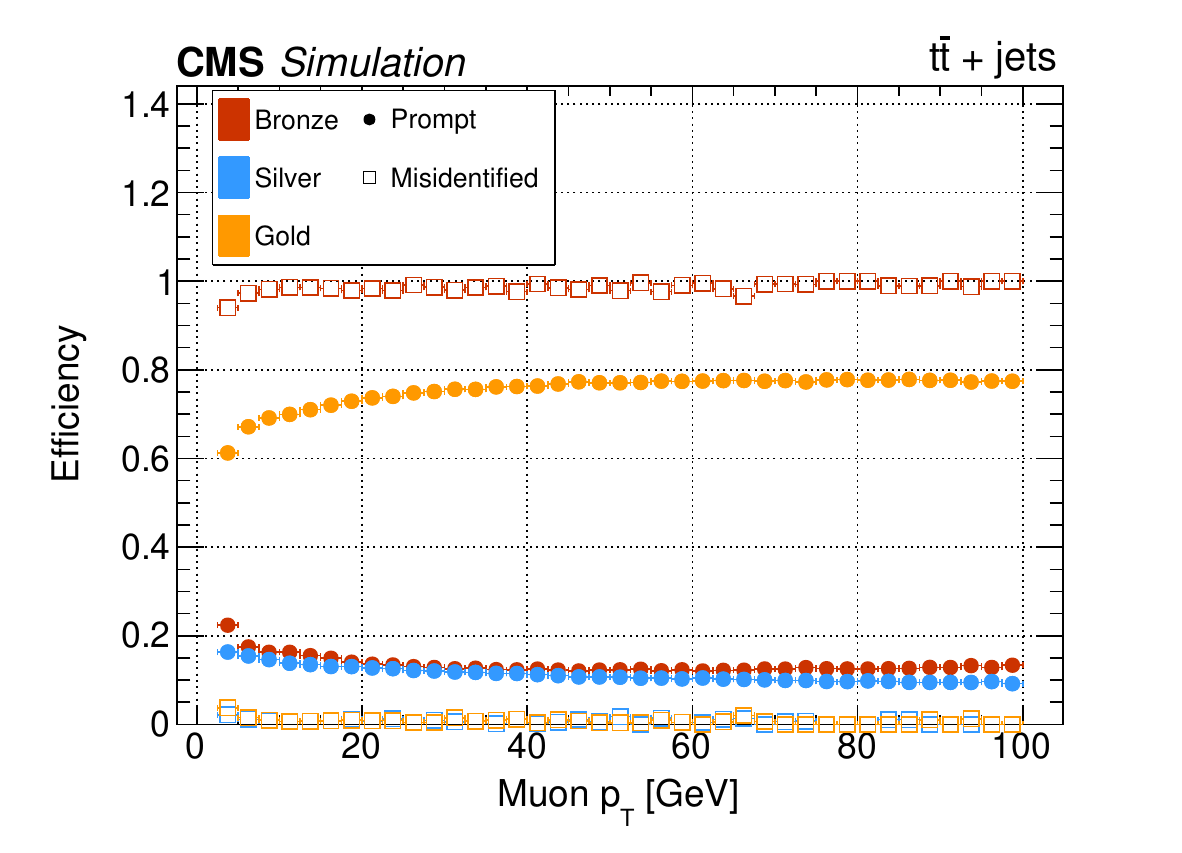}
	\caption{Efficiencies of lepton candidates satisfying baseline requirements 
          to be identified in the gold, silver, and bronze categories for prompt leptons (solid circles) 
	and misidentified leptons (open squares), evaluated in
        simulated \ttjets events. 
	Electrons (muons) are shown in the \cmsLeft (\cmsRight) panel. 
	As the three categories are mutually exclusive and exhaustive for
        baseline leptons, these efficiencies sum to one for each source
      in each lepton \pt bin.}
	\label{fig:effeleMC}
\end{figure}

Electron identification efficiencies range from above 98\% for the very loose criteria to 70\% for the tight criteria. 
The corresponding misidentification rates range from 2 to 3\% for tight identification and from 5 to 15\% for very loose identification, depending on the 
electron's $\eta$~\cite{ref:eideff}.  
Loose (medium) muons have identification efficiencies above 99\% (95--99\%) with hadron 
misidentification rates below 0.5\%~\cite{ref:muonid}. The
identification and misidentification efficiencies of these selection
criteria are studied using simulated \ttjets and \zdy samples. 
In these samples, lepton candidates are labeled as prompt if there is a generator lepton 
within 0.01 in $\DR$ of the lepton candidate and the lepton originates from the primary vertex.
Lepton candidates are considered misidentified 
if there is no generator-level lepton within 0.01 in $\DR$ of the lepton candidate. 
Figure~\ref{fig:effeleMC} shows these prompt 
and misidentified efficiencies as a function of lepton \pt. 
Further exploration of the sources of the leptons shows that the gold category is the most efficient at 
keeping genuine leptons originating from prompt sources, while rejecting most misidentified leptons as well as 
leptons from nonprompt sources. The silver category is also very good at rejecting misidentified leptons and accepting 
genuine leptons that were produced from 
secondary sources (primarily semileptonic \PQb hadron and $\Pgt$ lepton decays).  Finally, bronze leptons consist of 
prompt leptons that failed both the gold and silver requirements, followed by genuine leptons from nonprompt sources 
as well as particles misidentified as leptons. 

To account for observed small differences in reconstruction, identification, and isolation efficiencies between data and simulation, the simulation is corrected by factors estimated from data using the ``tag-and-probe" method~\cite{ref:muonid}, with both \PZ boson and \JPsi meson decays. 
These factors are derived as a function of \pt, $\eta$, and data-taking period, and take into account extrapolations of vertexing and isolation parameters.  Further scale factors 
account for differences found between the \textsc{FastSim} signal sample simulations and the background samples generated using the full 
detector simulation. 

Jets found with $\pt > 20$\GeV and $\abs{\eta} <2.4$ are selected if they pass criteria designed to remove jets dominated by instrumental 
effects or reconstruction failures~\cite{ref:jets}. Additionally,
jets are required to be a distance of at least $\DR=0.2$ from any
identified leptons. 
Jets that pass the medium working point of the \textsc{DeepJet} tagger~\cite{ref:deepjet} are classified as \PQb jets. 
The identification efficiency for \PQb quark jets ranges from 60 to 85\%, depending on the jet \pt, with a misidentification 
rate of about 15 to 25\% for charm quark jets and 1 to 7\% for light-quark or gluon jets. Differences between these efficiencies in data and simulation 
are corrected for as functions of jet \pt~\cite{ref:bjetcor}.

For compressed SUSY signal events, especially those originating from top squark decays, low-\pt \PQb hadron decays are 
an important signature. A soft secondary vertex (SV)  \PQb-quark-finding deep neural network (DNN) algorithm was developed 
for this analysis to identify these decay products. The SV candidates were reconstructed 
using the inclusive secondary vertex finder~\cite{ref:ivf}. 
The $\pt$ and $\eta$ of the SV are evaluated from the vector sum of the momenta of the tracks belonging to the SV. 
It is required that $2 < \pt <20~$\GeV, $\abs{\eta} <2.4$ and that 
the 3D displacement significance with respect to the primary vertex must exceed 3.
The SV must not be matched to any jet with a \pt above 20\GeV within a cone size of $\DR<0.4$, 
or to any lepton within a cone size of $\DR<0.2$ where the cone is centered on the SV momentum.

The SV  \PQb-tagger is built using the \textsc{DeepJet} framework \cite{mehta_2313218, kieseler_jan_2020_3670882} and 
a machine-learning algorithm based on the DeepCSV tagger~\cite{CMS:2017wtu}. 
Eight variables are used as input for the SV  \PQb-tagger DNN: the \pt, $\eta$, mass, number of degrees of freedom,
displacements from the primary vertex in both 2D and 3D, 3D displacement significance,  
and the cosine of the pointing angle between the primary and secondary vertices.
The discriminant was trained and tested using simulated \ttjets and \Wjets samples.

\begin{figure}[ht]
	\centering
	\includegraphics[width=0.49\textwidth]{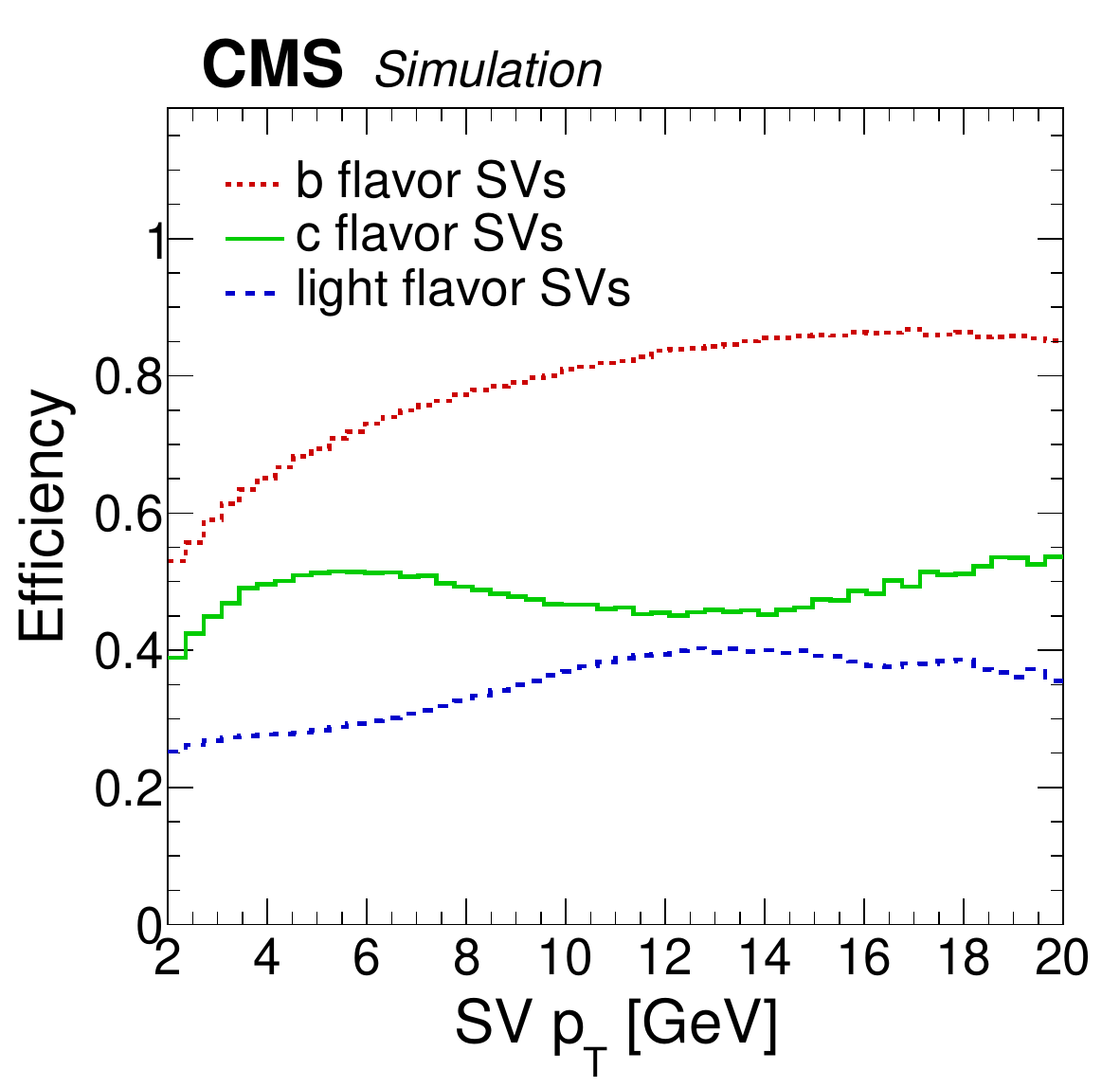}
	\caption{
    Distributions of the b, c, and light-quark SV tagging efficiencies, as 
	functions of the SV candidate \pt, for the chosen working point. 
	The SV flavor identities are determined from the generator-level flavor information and $\DR$ matching to SV candidates.}
	\label{fig:roc_final}
\end{figure}

A working point 
for the training sample was chosen for the analysis, yielding 
a good balance between rejecting light-flavor SVs and 
retaining events corresponding to compressed SUSY signals. 
Figure~\ref{fig:roc_final} shows the 
b, c, and light-quark SV tagging efficiencies as functions of \pt. 
Averaged over the \pt range, the \PQb quark SV tagging efficiency is 
approximately 80\%. The light-quark misidentification efficiency 
is about 25\% for the \pt distribution found in \ttjets events. Scale factors are derived to take into account differences 
between the \textsc{FastSim} signal sample simulations and the full detector simulation, with values within 5\% of unity. Scale factors which 
take into account differences between the data and simulation are found using the fit to control samples in data described in Section~\ref{sec:fit}.

\section{Kinematic event reconstruction}\label{sec:RJR}
\begin{figure}[htbp]
\centering
\includegraphics[width=0.49\textwidth]{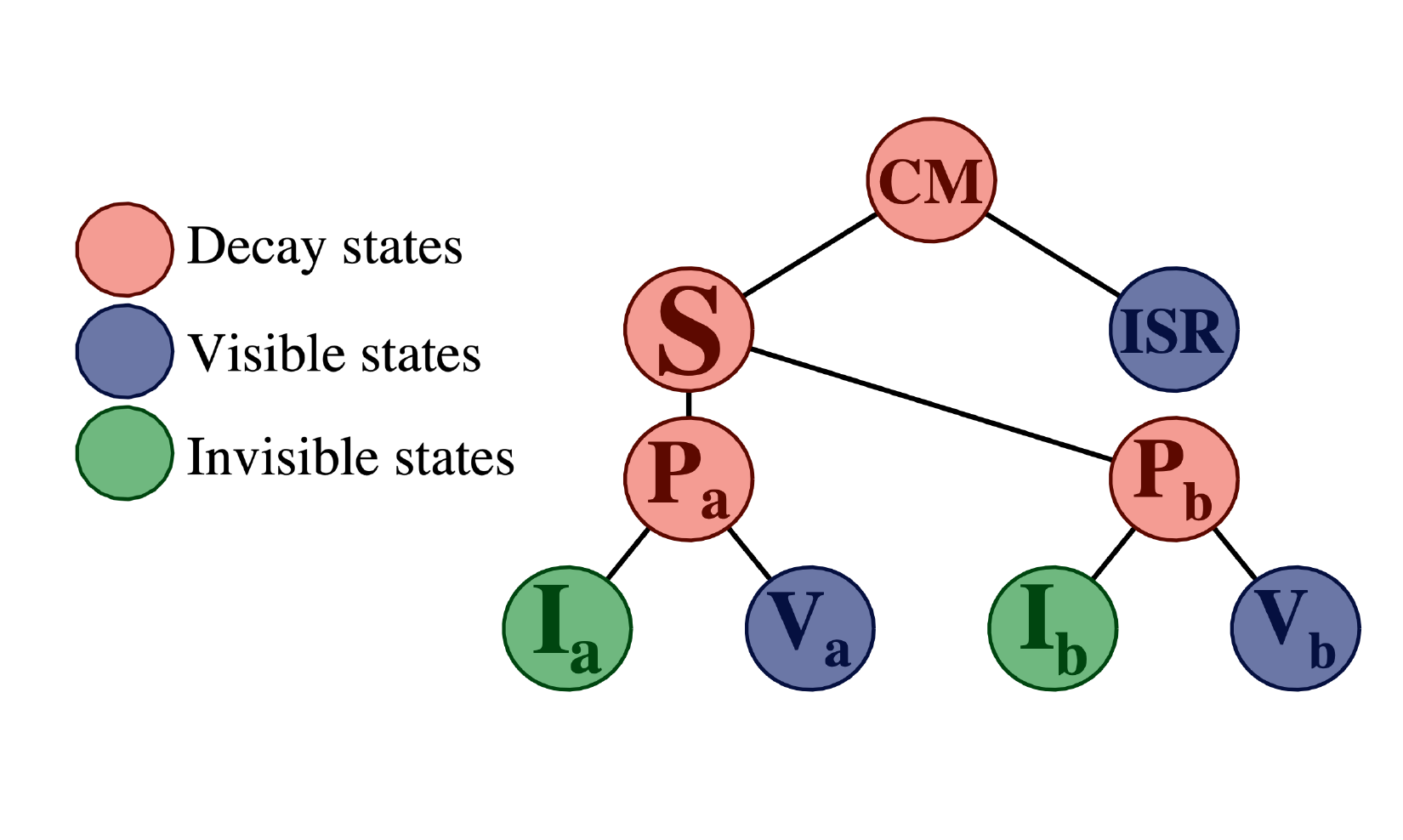}
\caption{ Decay tree diagram used to analyze events. Here \D{S} represents
  the total system of candidate sparticles, with \D{P_{a/b}}
  representing pair-produced SUSY parent particles; \I{I_{a/b}} and
  \V{V_{a/b}} represent the systems of invisible and visible sparticle
  decay products, respectively. The \D{S} system, along with the
  recoiling \V{ISR} system, are viewed as decay products of 
  the entire center-of-mass (\D{CM})
  system of the colliding partons with constituent center-of-mass energy, $\sqrt{\hat{s}}$.}
\label{fig:tree_all}
\end{figure}

In order to address the challenges associated with compressed sparticle mass spectra, 
events are selected based on significant ISR activity that causes the sparticle system to recoil resulting 
in observable \ptmiss from the 
momentum received by the invisible sparticles.
This event topology is illustrated in Fig.~\ref{fig:tree_all}, with \I{I_{a/b}} and \V{V_{a/b}} representing 
the systems of invisible and visible sparticle decay products, respectively, associated with the decay chains 
resulting from the parent particles \D{P_{a}} and \D{P_{b}}.
Analyzing events according to this generic decay tree, with a variable identity and number of particles corresponding 
to \I{I_{a/b}} and \V{V_{a/b}}, allows one to specifically tailor observables to exploit the features of this scenario 
using the recursive jigsaw reconstruction (RJR)~\cite{SUPERRAZOR,JIGSAW1,JIGSAW2} algorithm. 
Ideally, this leads to assignment of the \I{I_{a/b}} and \V{V_{a/b}} systems to individual candidate 
parent sparticle systems, \D{P_{a/b}}, collectively 
referred to as the \D{S} system, and an accompanying recoiling \V{ISR} system. Correspondingly, the \D{S} and \V{ISR} systems 
are treated as decay products of a singular center-of-mass (\D{CM}) system.

In the SUSY events targeted by this analysis, there are two types of unknowns: \textit{kinematic} unknowns resulting 
from undetected invisible particles and \textit{combinatorial} unknowns associated with the correct assignment and interpretation of visible particles. 
Within the decay tree framework, the kinematic unknowns correspond to the four-vectors 
of the two invisible systems \I{I_{a}} and \I{I_{b}}. The combinatorial unknowns involve determining how the reconstructed 
particles (leptons, jets, SVs) are assigned to  
the \V{V_{a}}, \V{V_{b}}, and \V{ISR} groups.  Assuming that the combinatorial assignments have been made, 
such that the four-vectors \pfour{\V{V_{a/b}}}{\lab} and \pfour{\V{ISR}}{\lab} are measured, 
the kinematic unknowns associated with the invisible particles are determined by applying a combination of assumed and measured constraints, along with 
algorithmic \textit{jigsaw rules} (JRs)~\cite{JIGSAW2}, which match the structure of the decay tree shown in Fig.~\ref{fig:tree_all}. 
In this case, the masses\footnote{Lower case $m$ denotes the correct (or true) mass, and upper case, $M$, denotes the calculated (or measured) 
mass that approximates the correct mass for the system that is indicated by the subscript.} 
of the individual invisible particles ($\Mass{\I{I_{a}}}{}, \Mass{\I{I_{b}}}{}$) are assumed to be zero, while the measured \ptvecmiss is interpreted as the vector sum of their transverse momenta, with
\begin{equation}
\Mass{\I{I_{a}}}{} = 0,\quad\Mass{\I{I_{b}}}{} = 0,\quad\pthree{\I{I_{a}},\mathrm{T}}{\lab} + \pthree{\I{I_{b}},\mathrm{T}}{\lab} = \ptvecmiss.
\end{equation}

With these constraints, the remaining kinematic unknowns are the total mass of the system of all invisible particles 
in the event (\mass{\I{I}}{}), the longitudinal momentum of 
that system in the lab (\pone{\I{I},z}{\lab}), and the direction of the ``decay axis'' corresponding to how the momentum of \I{I} is shared 
between \I{I_a} and \I{I_b}. The RJR algorithm proceeds by parametrizing these unknowns as components of the velocities relating the reference frames appearing in Fig.~\ref{fig:tree_all}. As all of the measured visible four-vectors correspond to the lab frame, the first of the velocities to consider is \vbeta{\D{CM}}{\lab} relating the \D{CM} frame to the \lab~frame. 
Among the kinematic unknowns, \vbeta{\D{CM}}{\lab} depends only on \mass{\I{I}}{} and \pone{\I{I},z}{\lab}, meaning these quantities can be determined independently of others.  
 
The first JR applied is the \textit{invisible mass rule}~\cite{JIGSAW2}, which 
assigns to \mass{\I{I}}{} the smallest Lorentz-invariant quantity (as a function of measured visible particles' four-vectors) that will ensure consistency of the approximate event reconstruction (no tachyonic particles), with
\begin{equation}
\Mass{\I{I}}{2} = \Mass{\V{V}}{2} - 4\Mass{\V{V_a}}{}\Mass{\V{V_b}}{}.
\end{equation}
This can be qualitatively understood as giving mass to the invisible particle system 
resulting  
from sparticle decays, based on the mass of the corresponding visible decay products.
This essentially exploits the fact that the orientation of the invisible
particles relative to each other is correlated with that of the visible ones, as
both arise from the same decays depicted in Fig.~\ref{fig:tree_all}.
As the individual invisible particles have masses constrained 
to zero, \Mass{\I{I}}{} is adjusted for the individual visible system masses, \Mass{\V{V_{a/b}}}{}.

The unknown \pone{\I{I},z}{\lab}, or equivalently \sbeta{\D{CM},z}{\lab}, is assigned 
via the \textit{invisible rapidity rule}~\cite{JIGSAW2}, 
by assigning the value of \sbeta{\D{CM},z}{\lab} which minimizes $\Mass{\D{CM}}{}$:
\begin{equation}
\sbeta{\D{CM},z}{\lab} = \argmin_{\sbeta{\D{CM},z}{\lab}} \Mass{\D{CM}}{}.
\end{equation}
This choice has two important consequences. Firstly, it avoids assigning erroneously high values of \Mass{\D{CM}}{} with an incorrect choice, avoiding promoting background events to appear more interesting 
than they actually are. Secondly, it ensures that the analytic 
expression for \Mass{\D{CM}}{} (and all observables following from this choice) is independent 
of the true value of \sbeta{\D{CM},z}{\lab} and so \textit{longitudinally boost invariant}. 
In fact these quantities are invariant to transverse boosts up to order $(\sbeta{\D{CM},\mathrm{T}}{\lab})^2$. The estimator for \Mass{\D{CM}}{} becomes the well-known \textit{transverse mass} of the \I{I} and \V{V} systems.

With \vbeta{\D{CM}}{\lab} assigned, the four-vectors of all the visible particles can be evaluated in the \D{CM} frame, and subsequently the \D{S} frame. Determining the remaining kinematic unknowns can now be viewed as assigning \vbeta{\D{P_{a/b}}}{\D{S}}, the velocities relating the \D{S} frame to the rest frames of its children, conditioned on the previous assignments made earlier in the decay tree. 
Using the \textit{invisible MinMasses$^2$ rule}~\cite{JIGSAW2}, these velocities are determined according to the equation,
\begin{equation}
\vbeta{\D{P_{a}}}{\D{S}},\vbeta{\D{P_{b}}}{\D{S}} = \argmin_{\vbeta{\D{P_{a}}}{\D{S}},\vbeta{\D{P_{b}}}{\D{S}} } \:(\Mass{\D{P_{a}}}{2}+\Mass{\D{P_{b}}}{2}),
\end{equation}
and subject to the constraints implied by previous choices and measurements. The practical effect of this choice is similar to that of the previous longitudinal boost, in that the mass estimators (or more accurately, this sum) become independent of the true, unknown, \vbeta{\D{P_{a/b}}}{\D{S}}. With the application of this last JR, all of the under-constrained kinematic quantities associated with invisible particles have been assigned.

The combinatorial unknowns associated with how visible objects are assigned to groups in the decay tree interpretation are determined in a manner similar to their kinematic analogs, where choices are factorized recursively according to when they appear in the decay. In this case, the first partitioning choice is how to split the total collection of visible, reconstructed objects in the event, \V{VIS}, between the \V{V} and \V{ISR} systems.
As for kinematic quantities, these assignments are determined by effectively picking the grouping that minimizes \Mass{\D{S}}{} and \Mass{\V{ISR}}{} simultaneously. Explicitly, the \textit{combinatorial MinMasses rule}~\cite{JIGSAW2} prescribes
\begin{equation}
\left\{ \V{V}, \V{ISR} \right\} = \argmax_{ \V{V}, \V{ISR} } \pone{\D{S}}{\D{CM}},
\end{equation}
where $\pone{\D{S}}{\D{CM}} = \abs{\pthree{\D{S}}{\D{CM}}} = \abs{\pthree{\V{ISR}}{\D{CM}}}$ is the momentum of the \D{S} system
in the \D{CM}, which, with \Mass{\D{CM}}{} fixed, will increase as \Mass{\D{S}}{} and \Mass{\V{ISR}}{} decrease. The maximization is over all the different ways jets can be exclusively partitioned into V and ISR groups.

The next partitioning choice is where the group \V{V} is 
split into \V{V_a} and \V{V_b} according 
to the \textit{combinatorial MinMasses$^2$ rule}~\cite{JIGSAW2}, with
\begin{equation}
\left\{ \V{V_a}, \V{V_b} \right\} = \argmin_{ \V{V_a}, \V{V_b} } 
\:(\Mass{\D{P_{a}}}{2}+\Mass{\D{P_{b}}}{2}).
\end{equation}
The practical effect of choosing partitions of objects based on mass minimization
is similar to exclusive jet clustering, where the invariant mass is used as a distance metric to group objects that are traveling in similar directions.

A hierarchy of JRs is defined to remove the combinatorial dependencies on the kinematics of the invisible particles, with combinatorial decisions proceeding down the decay tree and followed by the kinematical ones. With this  prescription applied, an event is fully reconstructed, as all of the four-vectors of the states shown in Fig.~\ref{fig:tree_all} are either 
measured or assigned. 

Observables are constructed to be sensitive to the mass of invisible particles in the event, characteristic of the compressed signals being sought. LSPs will receive little momentum from the decays where they are produced, and the resulting \ptmiss will be typically negligible. 
The massive invisible particles (which are nearly at rest in the \D{S} frame) will receive an out-sized fraction of the momentum from the ISR kick among the \D{S} decay products, as their rest energy is largest, leading to potentially large \ptmiss. This mechanism introduces a correlation between the ISR system and \ptvecmiss, such that
\begin{equation}
\ptvecmiss \approx - \frac{m_{\I{I}}}{m_{\D{P}}} 
\pthree{ \V{ISR},\, \mathrm{T}}{\D{CM}}.
\end{equation}
This relation can be further refined using the RJR reconstructed quantities and defining the variable \risr as
\begin{equation}
\risr =  \frac{\abs{\pthree{ \I{I}}{\D{CM}} \cdot \phat{ \V{ISR}}{\D{CM}}}}{ \abs{\pthree{ \V{ISR}}{\D{CM}}}} \approx \frac{m_{\I{I}}}{m_{\D{P}}}.
\end{equation}

\begin{figure}[!ht]
\centering
\includegraphics[width=0.48\textwidth]{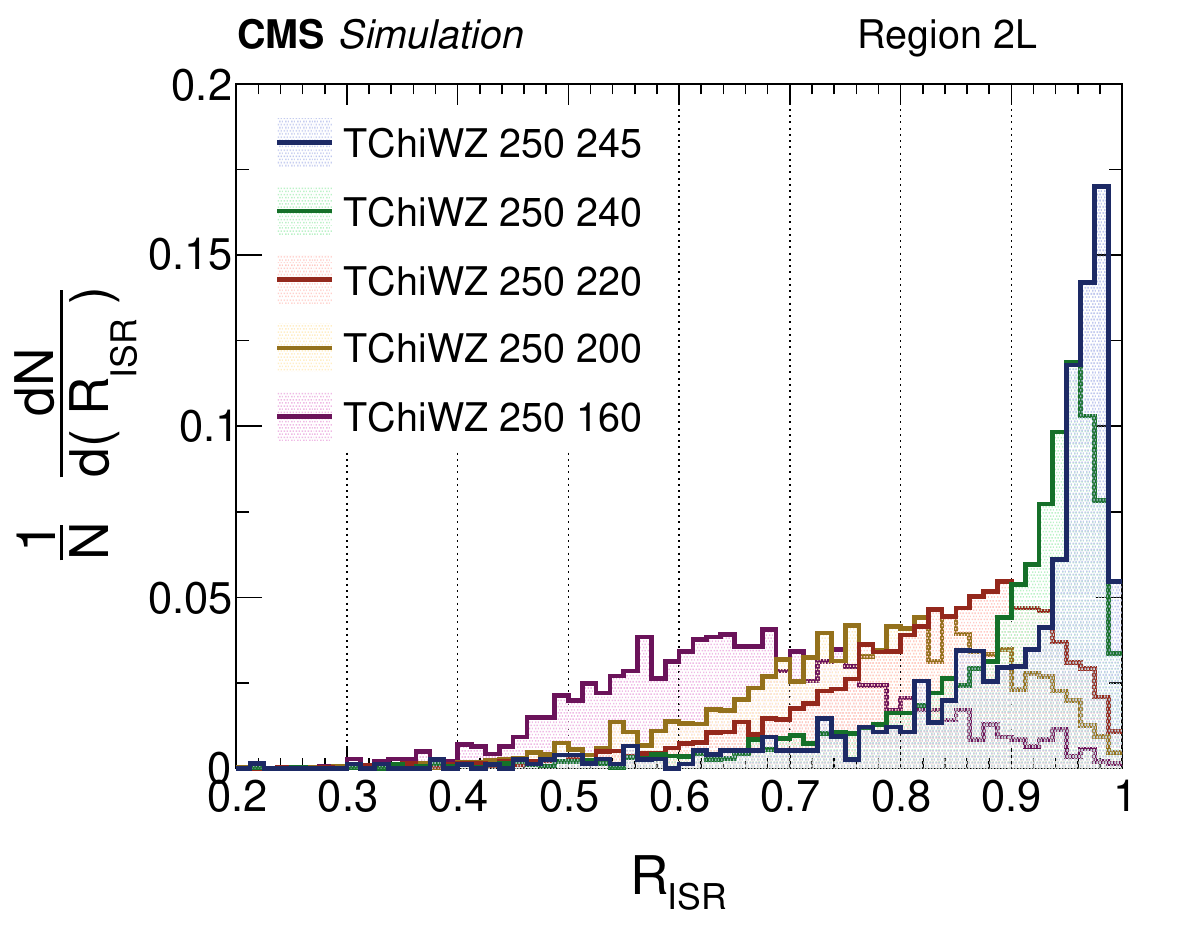}
\includegraphics[width=0.48\textwidth]{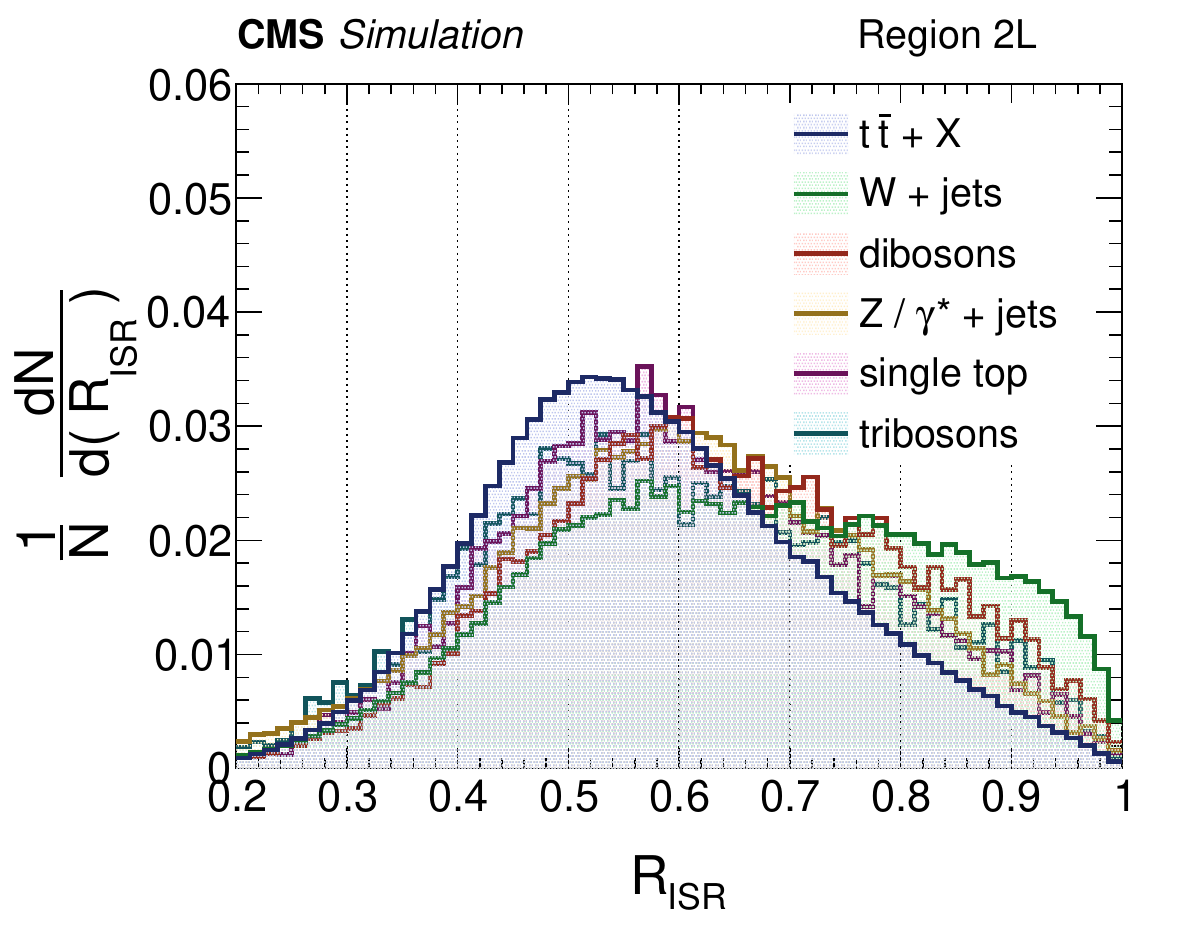}
\caption{Distributions of \risr for simulated events in the 2 lepton final state for 
TChiWZ signal models with 250\GeV parent mass and various LSP masses ranging from 160 to 245\GeV (\cmsLeft) 
and the SM backgrounds (\cmsRight).}
\label{fig:RISR}
\end{figure}

The distribution of \risr of a selection of compressed electroweakino TChiWZ model events is 
shown in Fig.~\ref{fig:RISR} (left), 
where it is observed to peak at $m_{\I{I}}/m_{\D{P}}$ as expected. With an absence of 
genuine, massive invisible particles, the SM backgrounds do not exhibit the same peaking behavior, 
with larger values of \risr suppressed, as apparent in Fig.~\ref{fig:RISR} (right). 
While observables sensitive to the absolute size of mass splittings in compressed scenarios can 
struggle to differentiate between signal and background, the resolution (and hence discriminating power) 
of \risr \textit{improves} with increasing compression. The peaking behavior of \risr depends predominantly 
on the event topology and particle masses, and is observed to be largely independent of the final state. 
Regardless of how particles in these events decay, \risr depends almost exclusively on the sparticle masses. 
The distribution of \risr for the SM backgrounds also behaves qualitatively similarly in different final states.

\begin{figure*}[!htbp]
\centering
\includegraphics[width=0.32\textwidth]{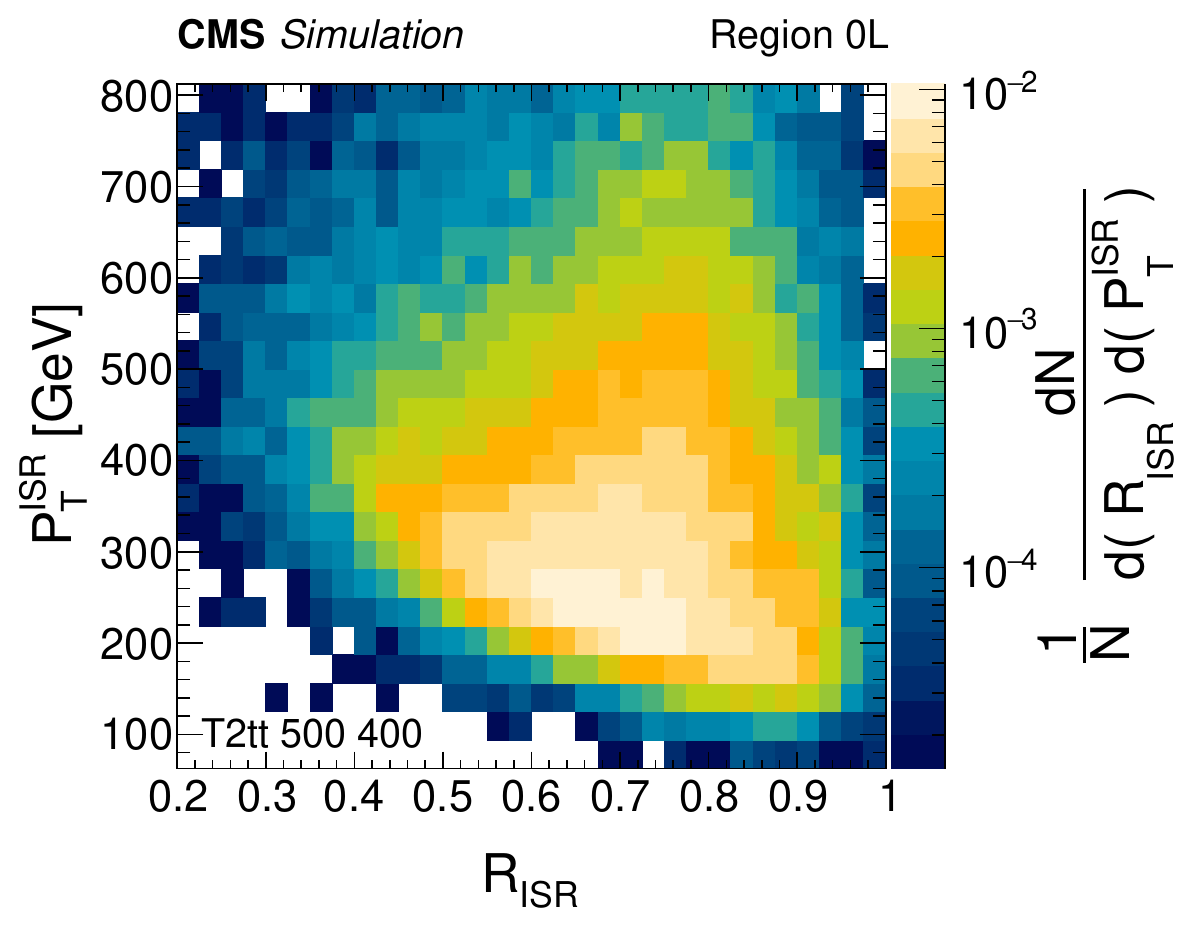}
\includegraphics[width=0.32\textwidth]{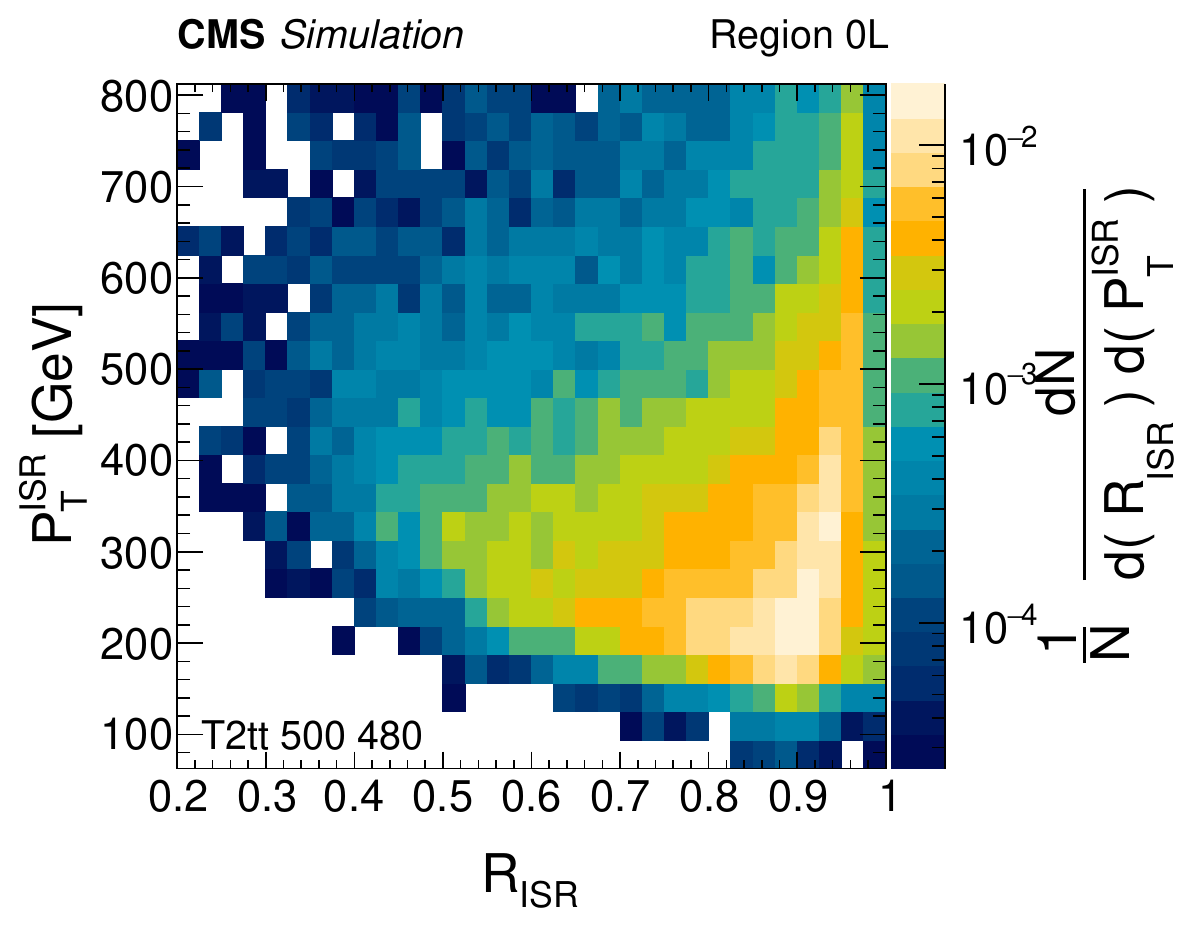}
\includegraphics[width=0.32\textwidth]{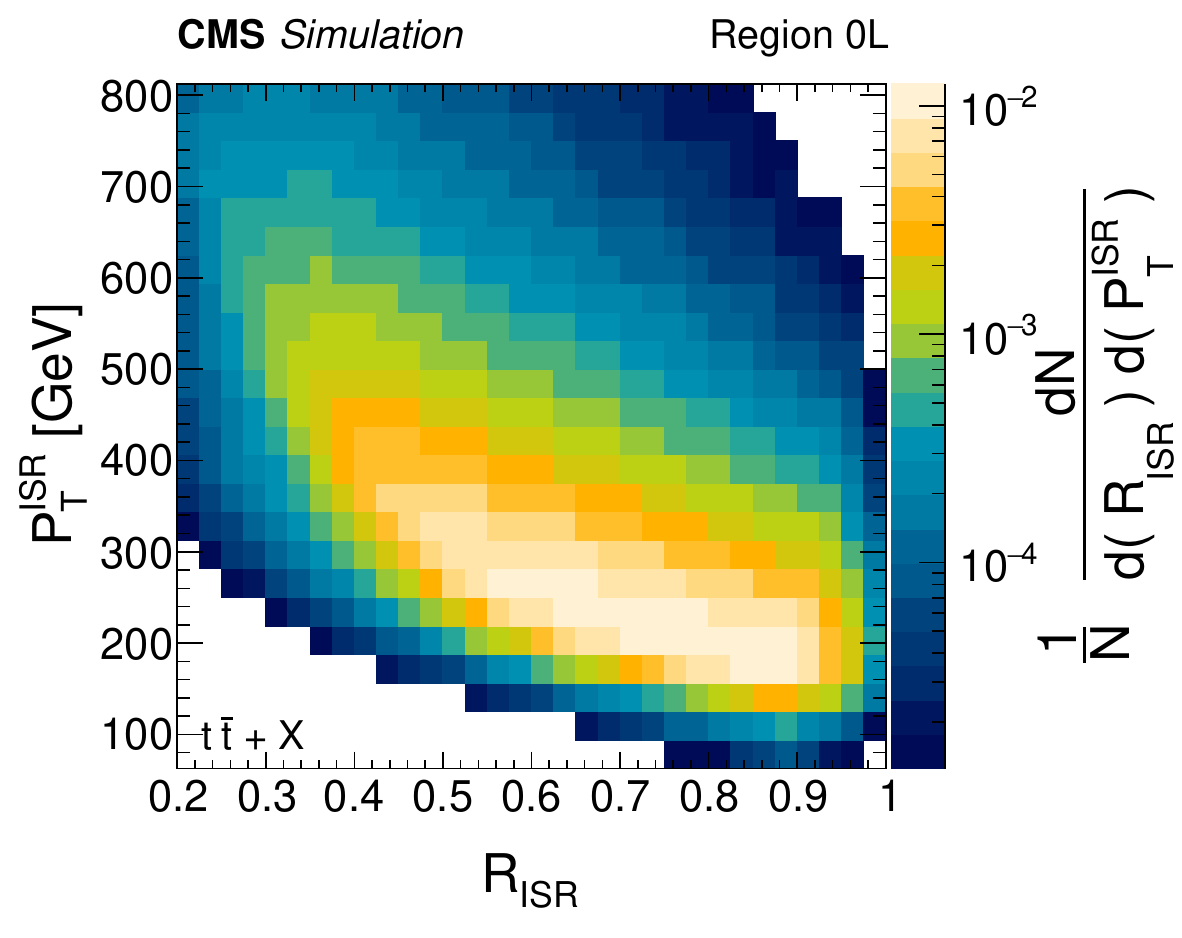}
\caption{ Distributions of \ptisr vs. \risr in events with 0 leptons 
for simulated top squark signals in the T2tt model with parent mass of 500\GeV and a LSP mass of 400\GeV (left), LSP mass of 480\GeV (center), 
and \ttjets background (right).}
\label{fig:PTISR_v_RISR}
\end{figure*}

The \ptisr observable 
quantifies the magnitude of the ISR kick to the sparticle system ($\ptisr = \abs{\pthree{ \V{ISR},\, \mathrm{T}}{\D{CM}}}$). 
As one might naively expect, the more kick, the more distinctive the peaking behavior of \risr for signals with massive invisible particles. This behavior is illustrated in Fig.~\ref{fig:PTISR_v_RISR}, where the \risr resolution improves for compressed signals with increasing \ptisr.
Conversely, the SM backgrounds have increasingly suppressed \risr distributions as \ptisr grows, as seen for the \ttjets background in Fig.~\ref{fig:PTISR_v_RISR}. In this analysis, \ptisr is used to define signal-enriched and control regions (CRs) 
by exploiting this behavior.

Additional observables sensitive to information independent from \risr and \ptisr, which reflect the $R$-parity-conserving decay topology, can be constructed by further resolving the decay kinematics inside the sparticle decay system \D{S} ``perpendicular'' to the axis of 
the ISR boost. Two additional approximate reference frames are defined relative to the \D{S} frame according to:
\begin{equation}
\sbeta{\V{V}}{\D{S}} = \argmin_{\sbeta{\V{V}}{\D{S}}} \left( \Lambda_{\sbeta{\V{V}}{\D{S}}} \pfour{\V{V}}{\D{S}} \right)_{0},~~
\sbeta{\I{I}}{\D{S}} = \argmin_{\sbeta{\I{I}}{\D{S}}} \left( \Lambda_{\sbeta{\I{I}}{\D{S}}} \pfour{\I{I}}{\D{S}} \right)_{0},
\end{equation}
where \sbeta{\V{V}}{\D{S}} and \sbeta{\I{I}}{\D{S}} are the velocities (restricted along the direction of the boost \vbeta{ \D{S}}{\D{CM}}) relating the \D{S} frame to the respective reference frames. The \V{V} and \I{I} systems are at rest along the \hbeta{ \D{S}}{\D{CM}} axis, and
$\left( \Lambda_{\sbeta{\V{V}}{\D{S}}} \pfour{\V{V}}{\D{S}} \right)_{0}$
is the energy of the visible system in this frame after the Lorentz transformation $\Lambda_{\sbeta{\V{V}}{\D{S}}}$, which is minimized. As opposed to simply projecting the four-vectors \pfour{\V{V_{a/b}}}{\D{S}} and \pfour{\I{I_{a/b}}}{\D{S}} individually into the plane perpendicular to the \hbeta{ \D{S}}{\D{CM}} boost direction, projecting the entire \V{V} and \I{I} groups maintains information along this axis and is insensitive to the previous inexact approximation.

From the ``perpendicular'' four-vectors in the two \D{S}-adjacent frames, defined as
\begin{equation}
\pfour{ \V{V_{a/b}}, \perp}{\D{S}} = \Lambda_{\sbeta{\V{V}}{\D{S}}} \pfour{\V{V_{a/b}}}{\D{S}},~~
\pfour{ \I{I_{a/b}}, \perp}{\D{S}} = \Lambda_{\sbeta{\I{I}}{\D{S}}} \pfour{\I{I_{a/b}}}{\D{S}},
\end{equation}
observables can be constructed:
\begin{align}
\begin{split}
\Mass{\D{P_{a/b}},\perp}{2} = \left(  \pfour{\V{V_{a/b}}, \perp}{\D{S}} + \pfour{ \I{I_{a/b}}, \perp}{\D{S}}   \right)^{2},\\
\Mass{\D{S},\perp}{2} = \left(  \pfour{\V{V_{a}}, \perp}{\D{S}} + \pfour{\V{V_{b}}, \perp}{\D{S}} + \pfour{ \I{I_{a}}, \perp}{\D{S}} + \pfour{ \I{I_{b}}, \perp}{\D{S}}  \right)^2.
\end{split}
\end{align}
While representing some independent information, the masses \Mass{\D{P_{a/b}},\perp}{} and \Mass{\D{S},\perp}{} are not entirely uncorrelated.
As the reconstruction in the \D{S} frame is executed by choosing several unknowns by minimizing \Mass{\D{P_{a}}}{2}+\Mass{\D{P_{b}}}{2}, a summary mass variable, \mperp, is defined from a combination of these masses as
\begin{equation}
\mperp = \sqrt{ \frac{\Mass{\D{P_{a}},\perp}{2} + \Mass{\D{P_{b}},\perp}{2}}{2} },
\end{equation}
such that it is related to the average (of squares) of the individual mass estimators for each sparticle parent.

\begin{figure}[!htbp]
\centering
\includegraphics[width=0.48\textwidth]{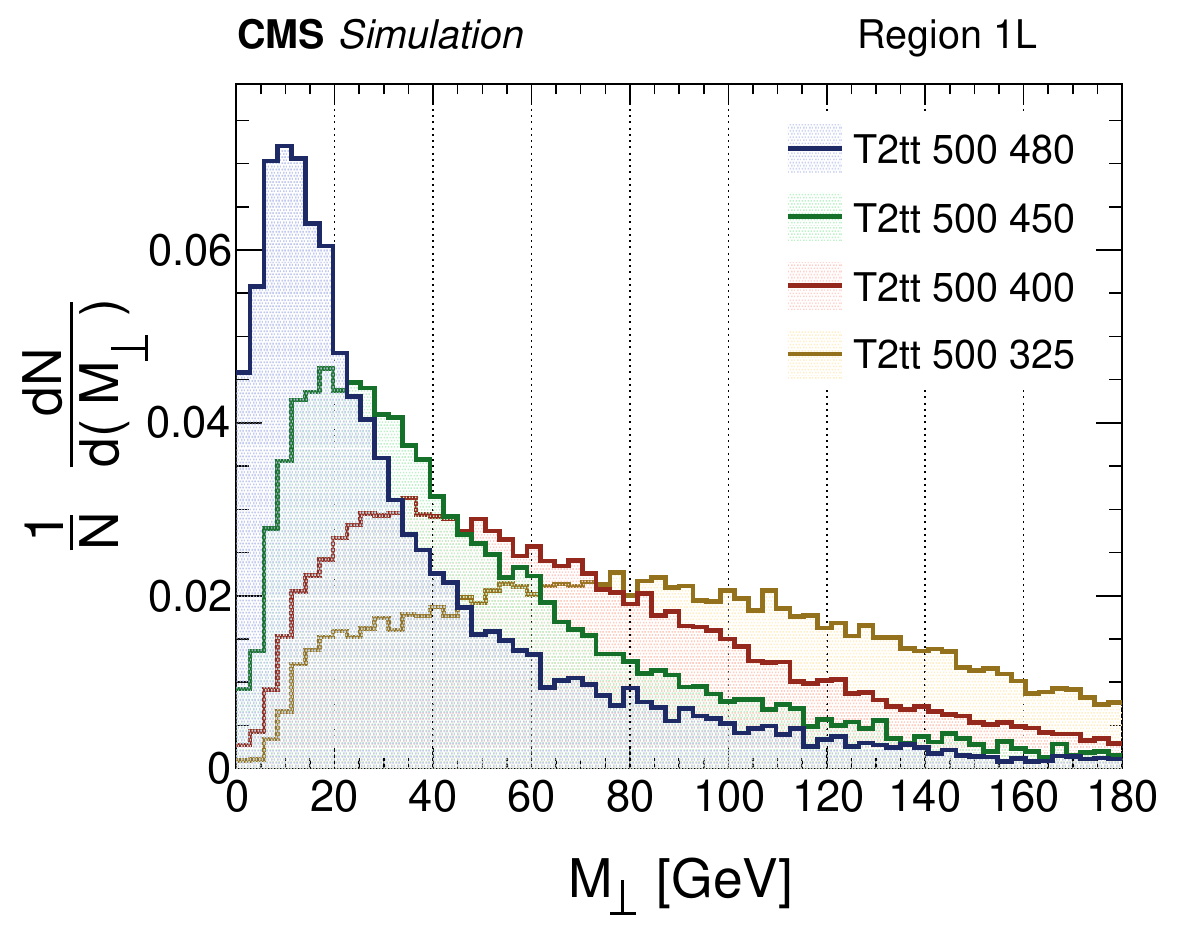}
\includegraphics[width=0.48\textwidth]{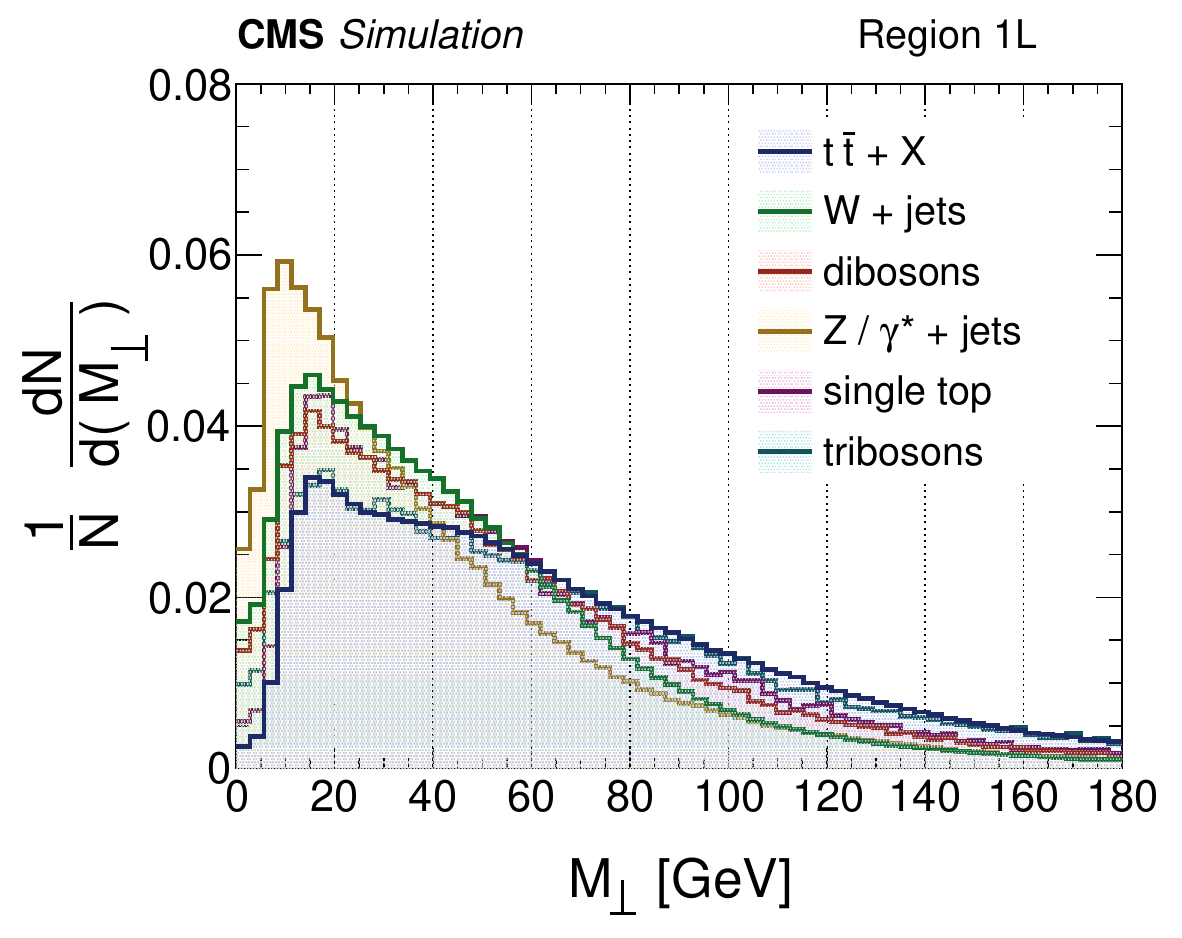}
\caption{Distributions of \mperp in one lepton final states for simulated events: compressed T2tt signal events 
with a parent top squark mass of 500\GeV and LSP masses ranging from 325 to 480\GeV (\cmsLeft) and  
the SM backgrounds (\cmsRight).}
\label{fig:Mperp}
\end{figure}

The distribution of \mperp is shown in Fig.~\ref{fig:Mperp} for compressed top squark 
signals and the SM backgrounds.
As the individual invisible particle masses are set to zero in the reconstruction, their expected contribution to \mperp is not accounted for, meaning that \mperp is not sensitive to the masses \mass{\V{P_{a/b}}}{} but rather to the mass splittings $\mass{\V{P_{a/b}}}{}-\mass{\I{I_{a/b}}}{}$.
The resulting \mperp distributions for signal events exhibit a kinematic endpoint at $\mass{\V{P_{a/b}}}{}-\mass{\I{I_{a/b}}}{}$ (modulo resolution effects), while backgrounds have falling distributions sensitive to the mass scale of particles appearing in the events.

\begin{figure*}[htbp]
\centering
\includegraphics[width=0.32\textwidth]{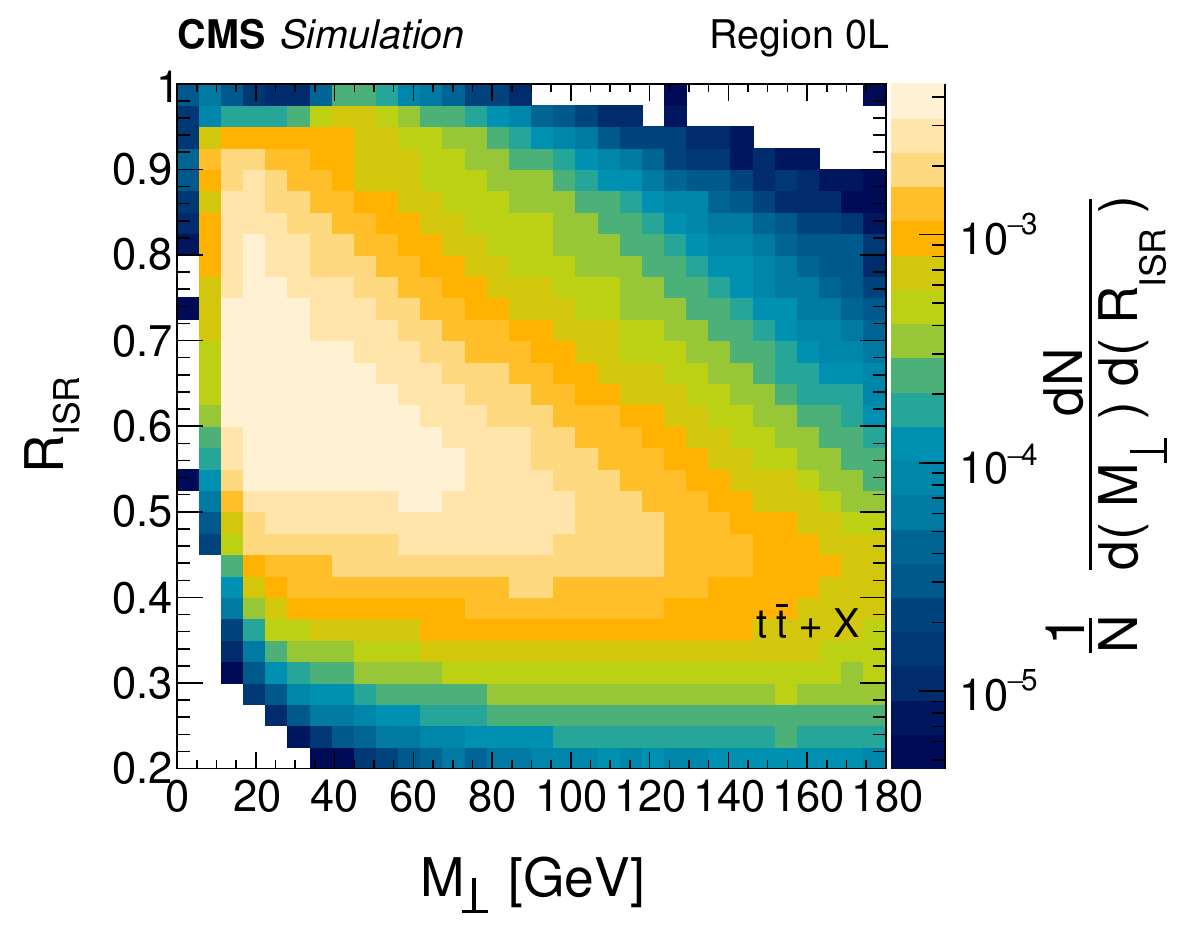}
\includegraphics[width=0.32\textwidth]{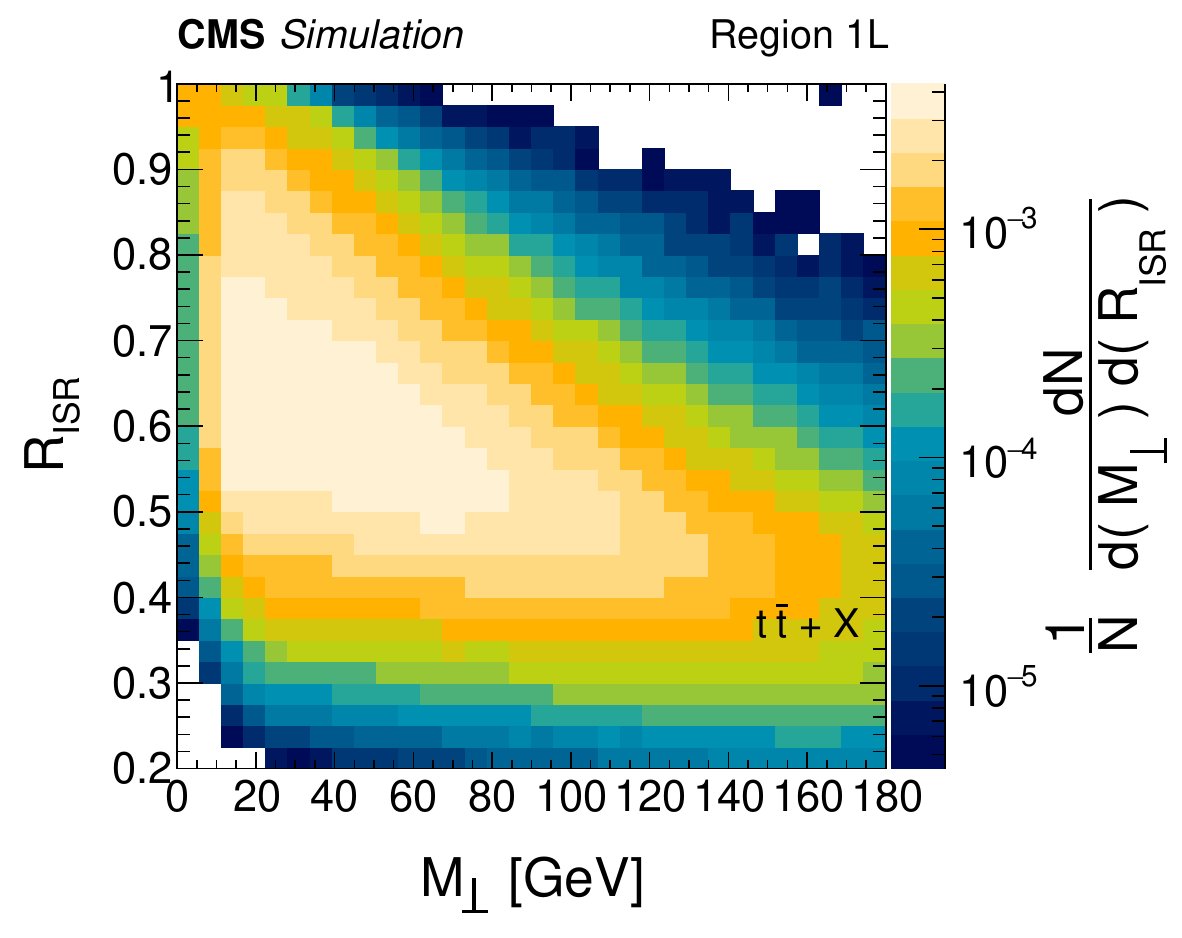}
\includegraphics[width=0.32\textwidth]{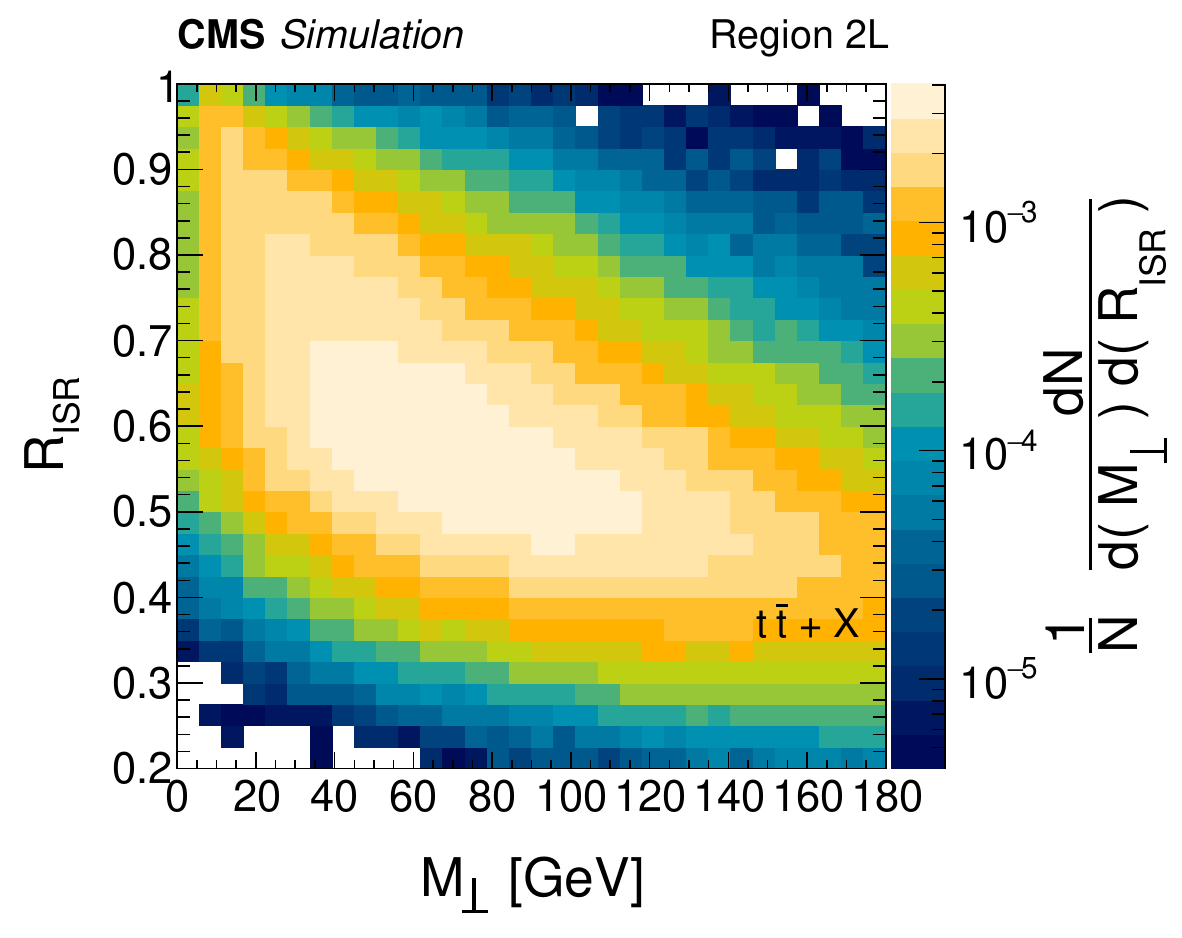}
\includegraphics[width=0.32\textwidth]{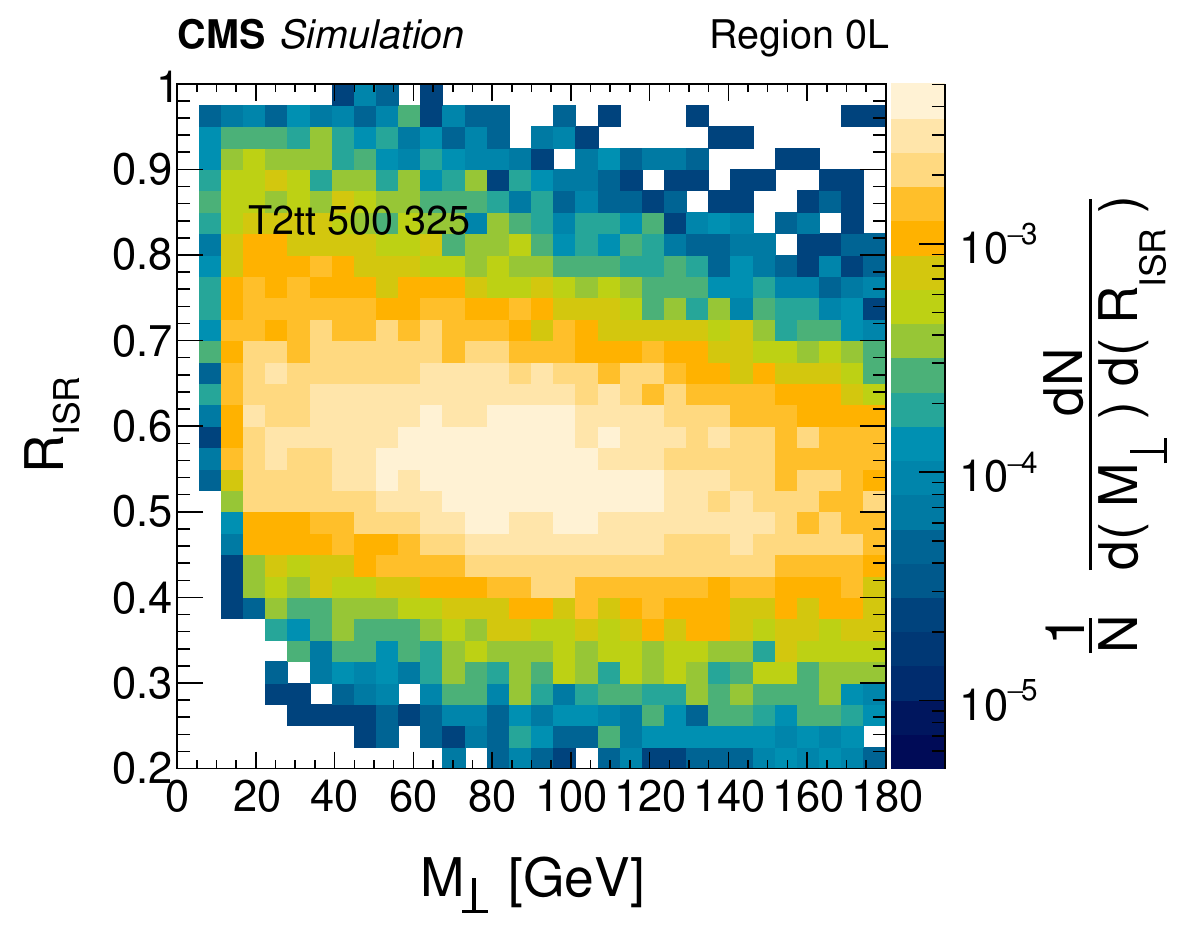}
\includegraphics[width=0.32\textwidth]{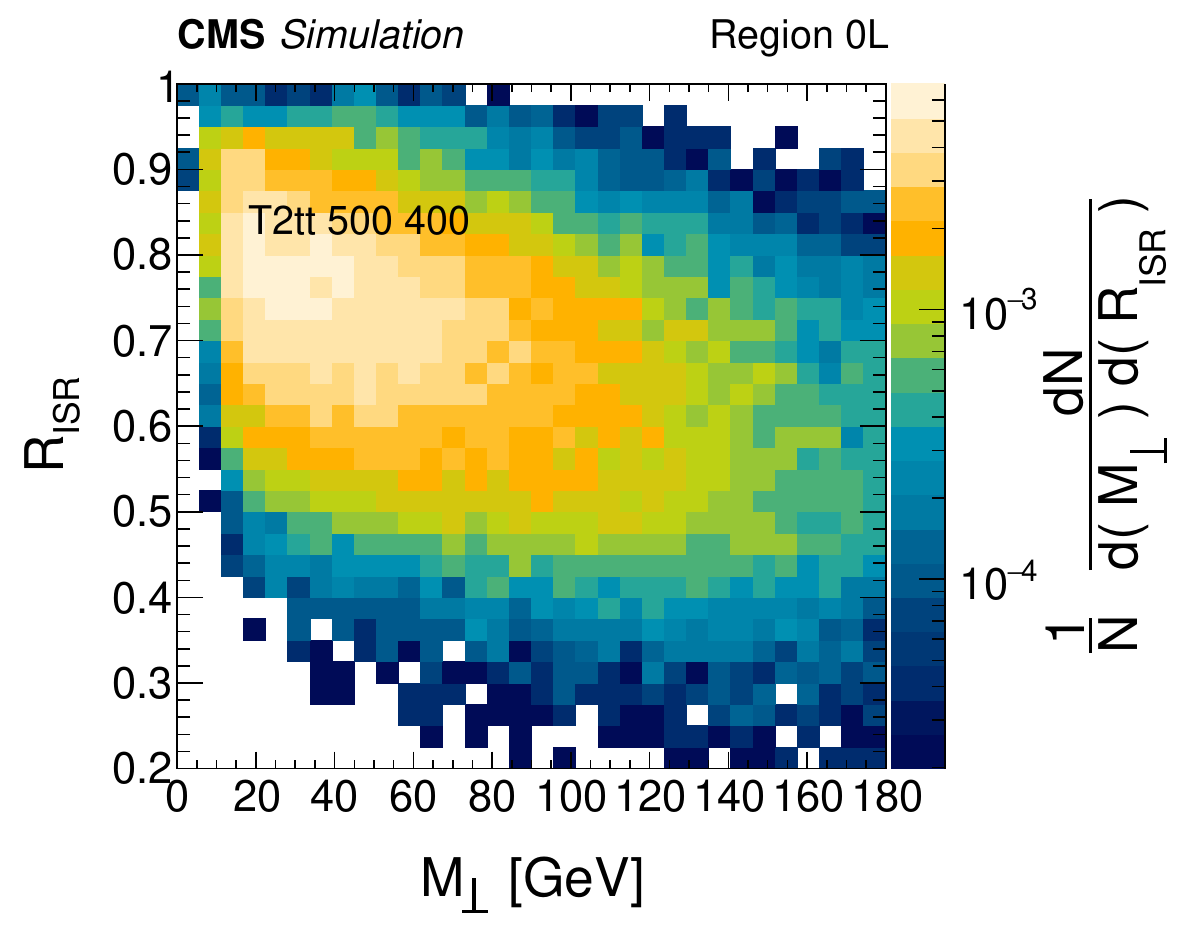}
\includegraphics[width=0.32\textwidth]{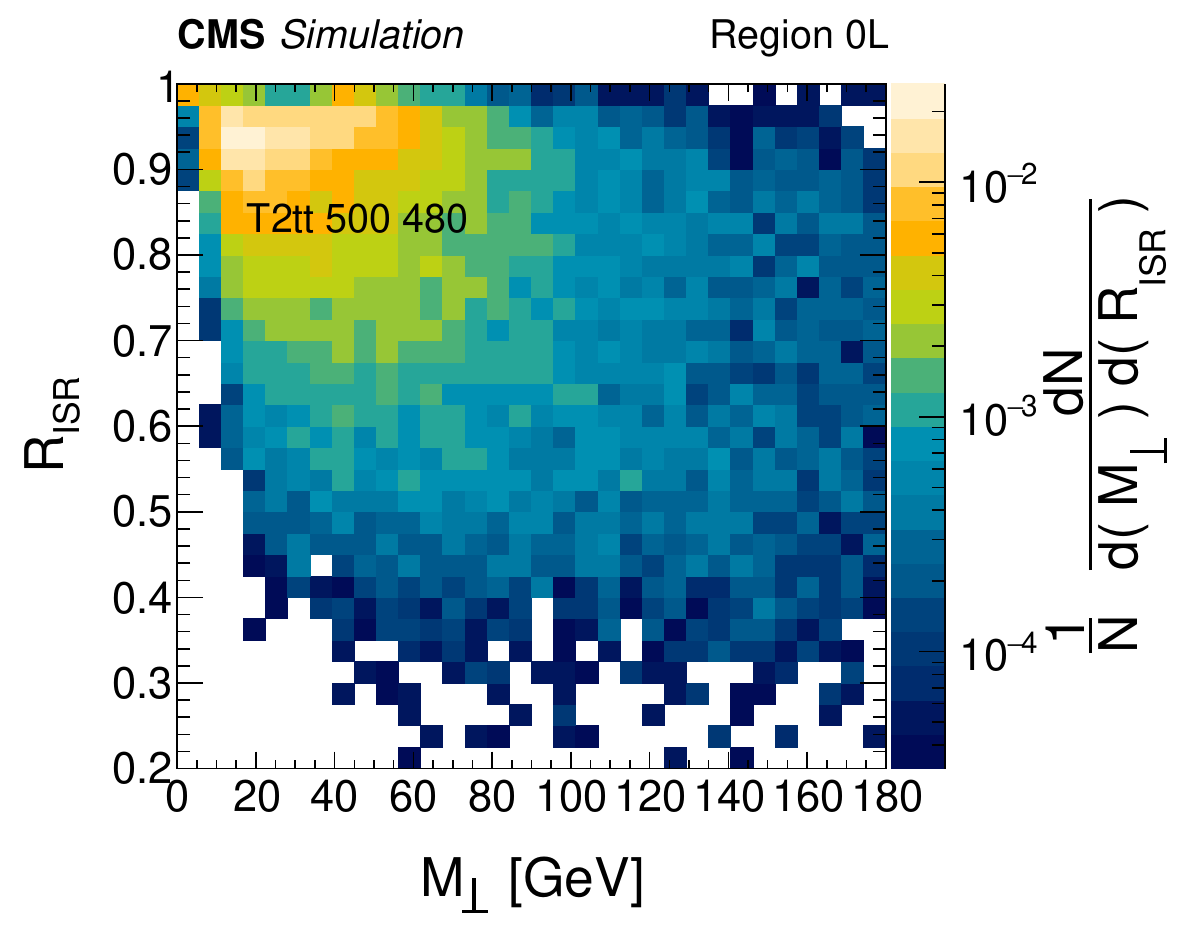}
\includegraphics[width=0.32\textwidth]{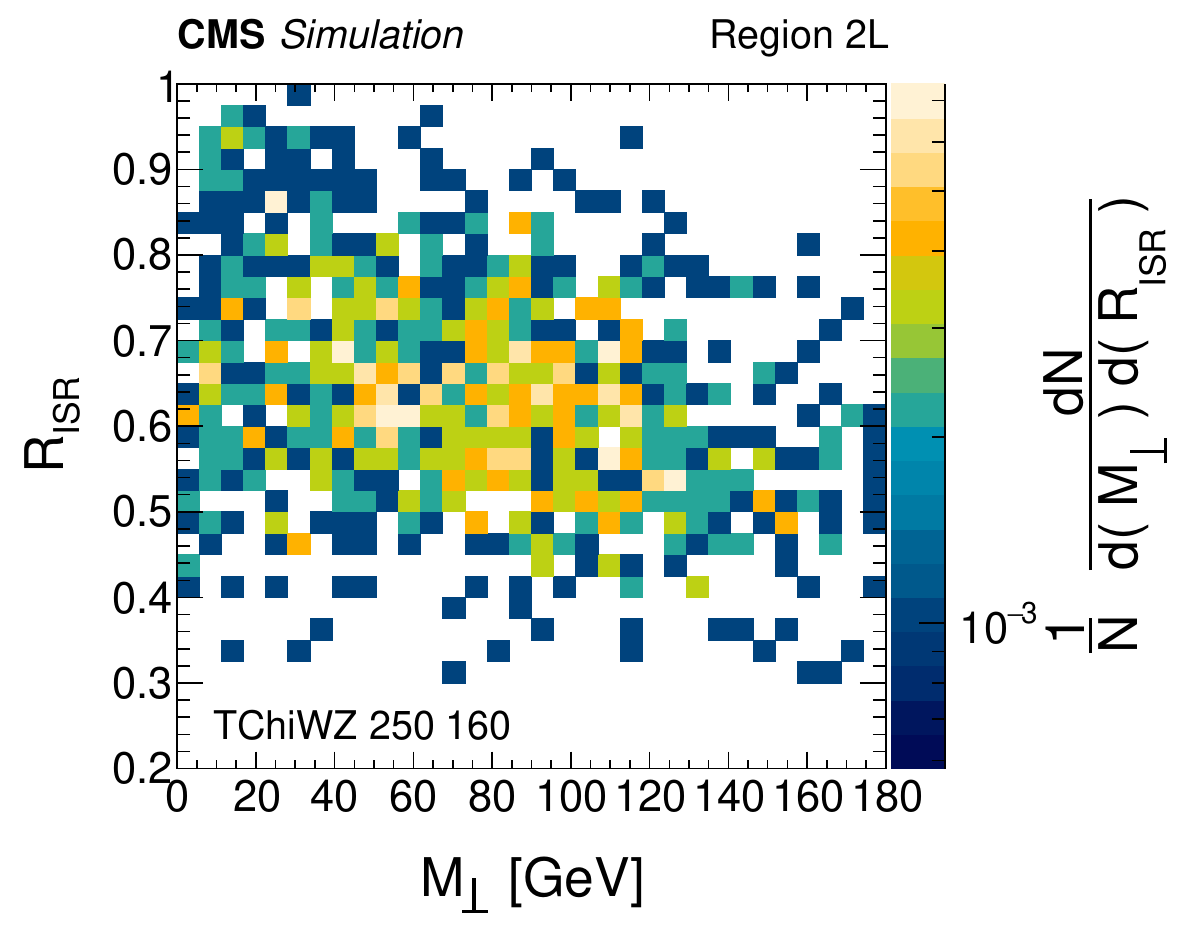}
\includegraphics[width=0.32\textwidth]{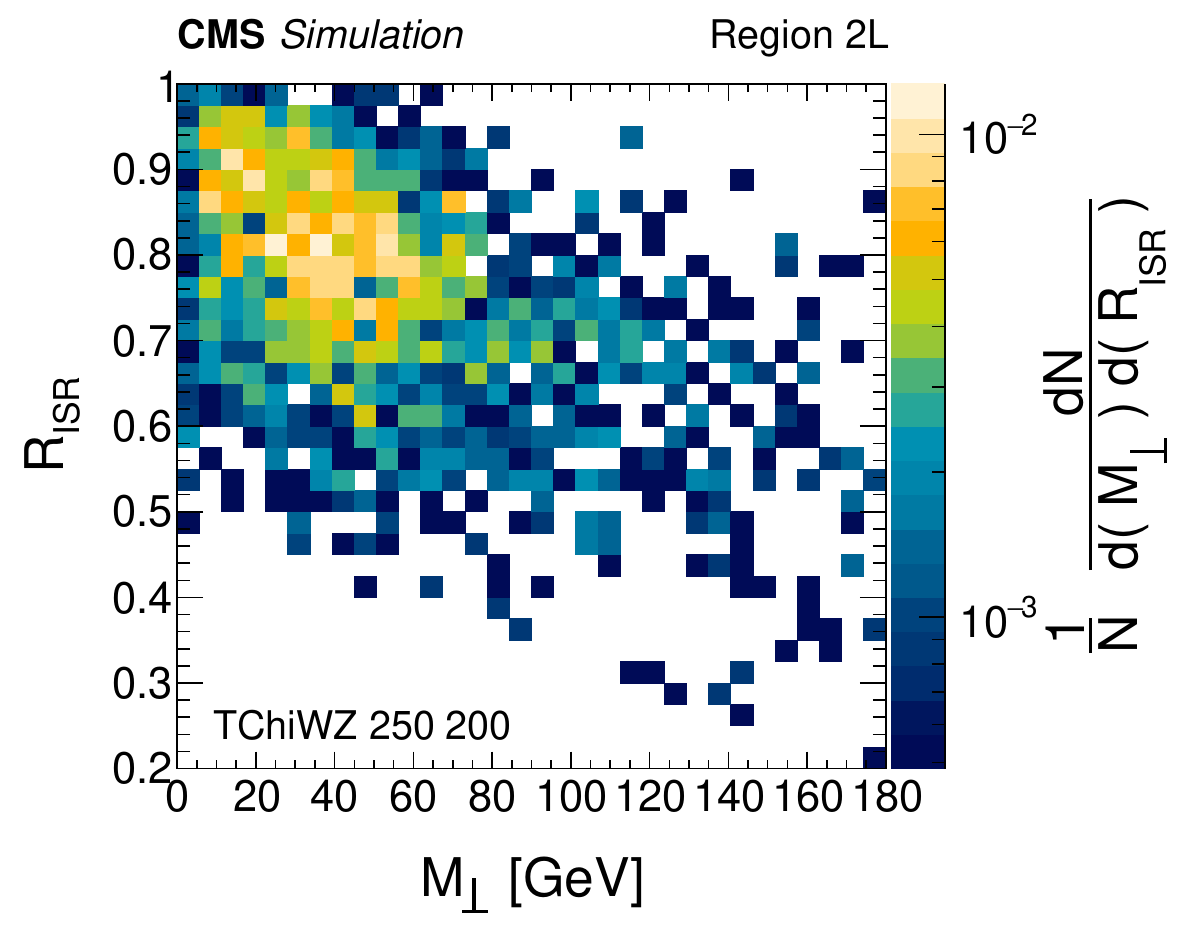}
\includegraphics[width=0.32\textwidth]{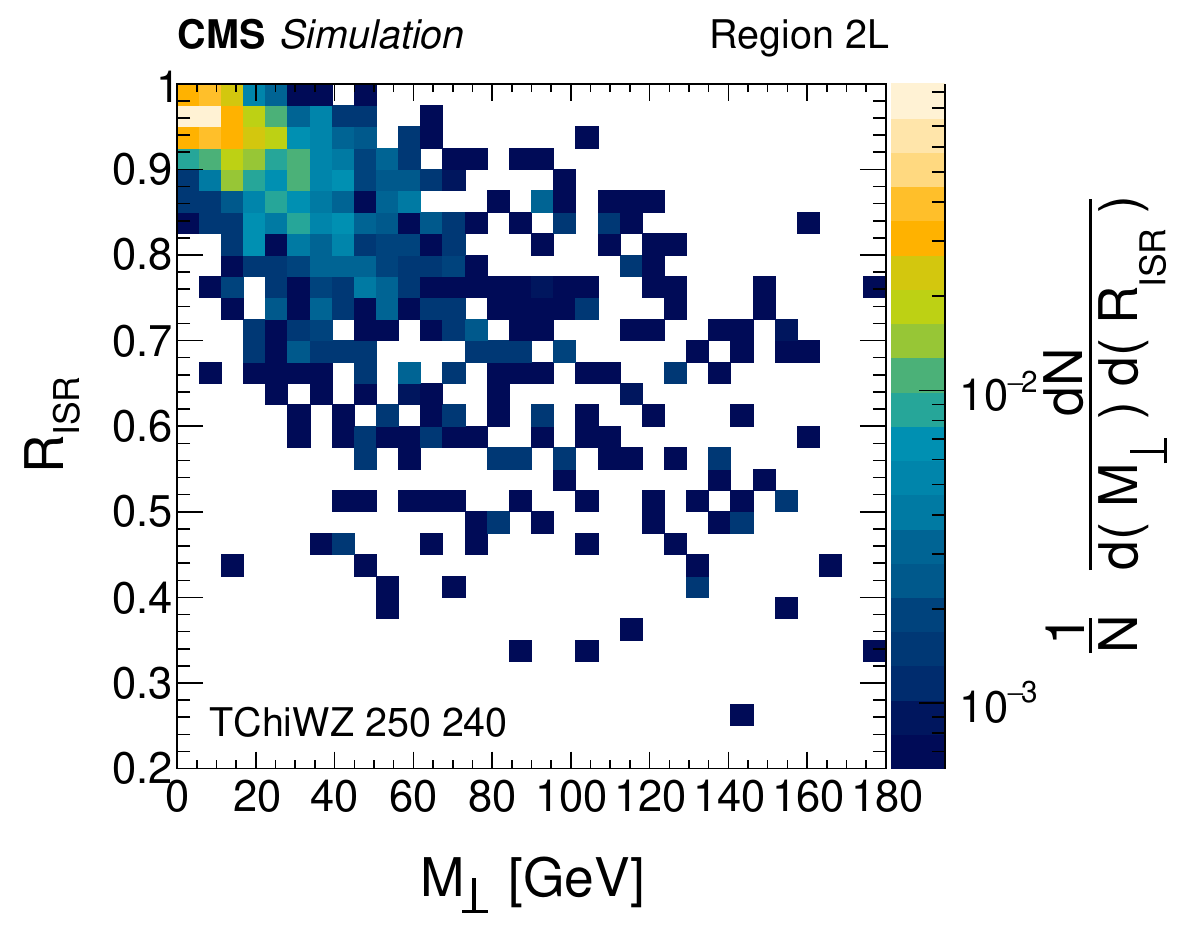}
\caption{Distributions of \risr vs. \mperp for simulated events in multiple final states. (Upper row) \ttjets background events in 0 lepton (0L), one lepton (1L), and two lepton (2L) final states. (Middle row) T2tt signals in 0L final states and (lower row) TChiWZ signals in 2L final states for various sparticle mass combinations.}
\label{fig:RISR_v_Mperp}
\end{figure*}

The \risr and \mperp variables are shown in Fig.~\ref{fig:RISR_v_Mperp} for various signal and background processes. The 
compressed SUSY signals appear as 2D ``bumps'' in the \risr vs. \mperp plane, with the location dictated by the sparticle masses.
Simultaneously, backgrounds are dispersed over the larger \risr and \mperp phase space, with larger values of \risr suppressed 
for increasingly large \mperp.

An additional variable, complementary to \mperp, is defined from a different combination of the ``perpendicular'' masses:
\begin{equation}
\gperp = \frac{2\mperp}{\Mass{\D{S},\perp}{}}.
\end{equation}
The observable \gperp is sensitive to the asymmetry of the \D{S} system decay, taking larger values (closer to 1) when the 
event is maximally imbalanced (invisible particles recoiling together against visible).
While the observable is not as powerful a discriminant as \risr or \mperp, it tends to larger values for signals 
relative to backgrounds and is also used in the event categorization. 
It is particularly effective against certain, otherwise difficult backgrounds, such as 
nonresonant SM $\PW\PW \to 2\Pell2\nu$ production, where the decay topology mimics that of $R$-parity-conserving SUSY.

The analysis proceeds by counting objects (leptons, jets,  \PQb-tagged jets, and  \PQb-tagged SVs) that were assigned to either the sparticle \D{S} system or the \V{ISR} system with the RJR reconstruction.  Any leptons reconstructed in the event are automatically assigned to the \D{S} system in the event interpretation, while jets and SVs can appear as either coming from sparticles or ISR.
One of the most important object counting observables is the multiplicity of jets assigned to the \D{S} system, \NjetS. While the distribution of \NjetS exhibits large variations depending on the level of compression of the signal model, the distribution of \NjetISR, the number of jets assigned to the \V{ISR} system in each event, is more uniform between differing signal masses and, more relevantly, is more similar to backgrounds.
Furthermore, \NjetS is a powerful discriminant when signals have mass splittings large enough to produce multiple above-threshold jets in sparticle decays.
For this analysis, the only requirement on \NjetISR is that there is at least one jet assigned to the ISR system.
For both signal and background, larger \risr and smaller \mperp are typically associated with lower \NjetS, with much weaker correlations for signals than for SM backgrounds. 

The analysis also categorizes events according to the number of jets that are tagged as coming 
from \PQb quarks in each of the \D{S} and \V{ISR} systems, \NbS and \NbISR, respectively. 
The \ttjets background often leads to at least one  \PQb-tagged jet in the \V{ISR} system; this arises from cases 
where  \PQb-tagged jets from the top quark decays get assigned erroneously to the \V{ISR} system.
This observation 
is used in the analysis by separating events with $\NbISR \geq 1$ from those with none 
in order to isolate \ttjets contributions. Also, requiring large \NbS and $\NbISR = 0$ selects a 
large fraction of top squark signal events, while rejecting most of the \zdy and \ttjets backgrounds.

\begin{figure}[htbp]
\centering
\includegraphics[width=0.48\textwidth]{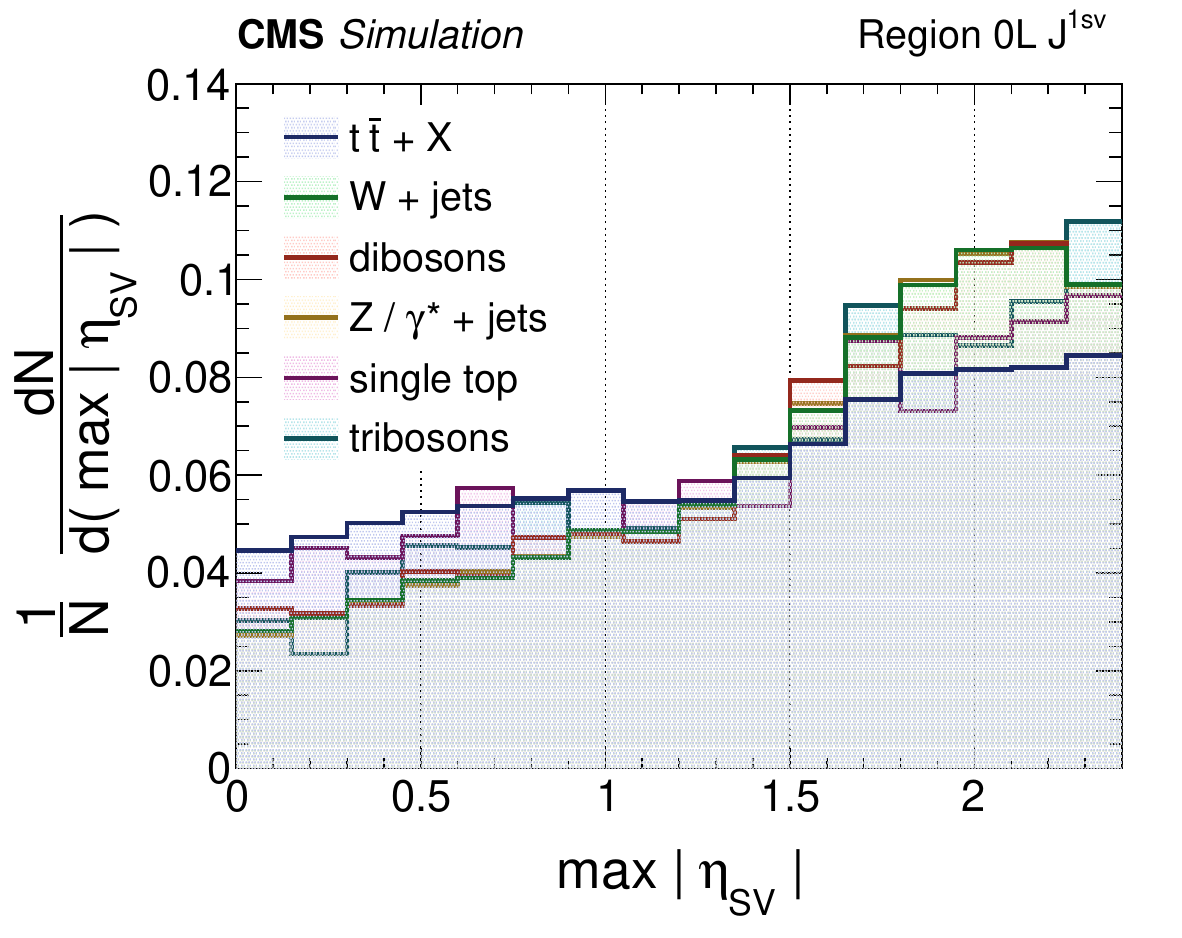}
\includegraphics[width=0.48\textwidth]{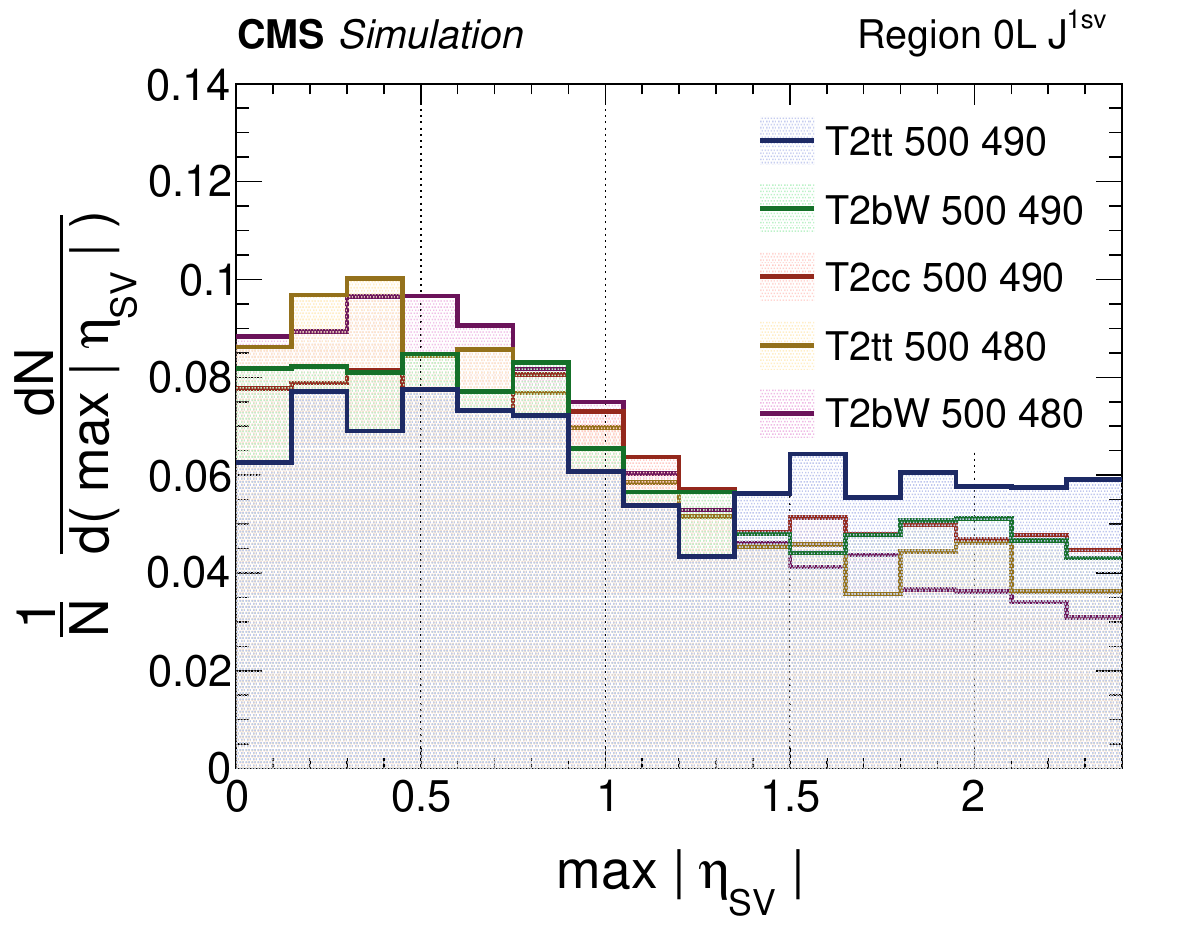}
\caption{Distributions of $\text{max} \: \abs{\etaSV}$ in final states with 0 leptons and ${\geq}1$ SVs associated with the \D{S} system, for simulated SM background events (\cmsLeft) 
and various top squark signal models (\cmsRight).}
\label{fig:etaSV}
\end{figure}

The final object multiplicity observable used in the analysis is the number of soft, stand-alone SVs assigned to the \D{S} system, \NsvS. The SV multiplicity is 
nearly identical among the different background events, 
with top squark signals having a higher probability of observing $\NsvS \geq 1$.  
In addition to categorizing based on the presence or absence of an SV in 
the \D{S} system, the $\abs{\eta}$ distribution of identified SVs also serves as a useful discriminant.
SVs associated with sparticle decays tend to be more central than those from background processes, especially backgrounds such as \Wjets 
where genuine bottom quarks usually arise from radiation or misidentification, as illustrated in Fig.~\ref{fig:etaSV}. 
The analysis uses the observable $\abs{\etaSV}$, defined as the maximum absolute value of the $\eta$ found for those SVs in the \D{S} system
to categorize events with SVs into central ($\abs{\etaSV} < 1.5$) and forward regions, with the latter acting as a CR for constraining the SV reconstruction efficiencies and kinematics with data.

\section{Event selection and categorization}\label{sec:analysis}
The analysis of data begins with preselection requirements that remove events that are not consistent with the compressed phase space of interest.
Events are then categorized into mutually exclusive analysis regions that are defined according to a combination of object multiplicities assigned to the \D{S} and \V{ISR} systems as well 
as categorization based on kinematic variables. The events falling in each region are further binned in 2D in the primary sensitive variables, \risr and \mperp, with bin boundaries common among regions with the same lepton and \NjetS multiplicities. 
This analysis implements a fitting approach based on control samples in data (described in Section~\ref{sec:fit}) to model background, 
accounting for differences in lepton flavor and source in the rates of their contributions, while also considering potential kinematic data/simulation deviations associated with the leptons, jets, and SVs. 
Some of the categories effectively act as background-dominated 
CRs used to constrain both the normalization 
and shape of the \risr and \mperp bin yields in signal-sensitive
ones. A subset of these analysis bins (over many categories) are identified 
as having negligible expected signal yields (for all of the signal
models considered in this analysis, through explicit evaluation) and are designated 
as control regions for use in blinded fits to data and model independent interpretations. 

Events are selected from the \ptmiss trigger data sets for each year by requiring 
that the offline $\ptmiss > 150$\GeV and 
applying the \ptmiss related event filters~\cite{ref:metperf}.
These events must have at least one visible object assigned to the \D{S} system and at least one jet assigned to the \V{ISR} system. To provide a moderate ISR kick, only events with $\ptisr > 250$\GeV are retained 
and it is further required that $\risr > 0.5$ thus 
primarily targeting signals with LSP masses exceeding about half the parent sparticle mass. 
A $\abs{\Delta \phi_{\ptvecmiss, V}}<\pi/2$ requirement ensures that the visible and invisible systems associated with sparticle decays are pointing in the same transverse hemisphere. 
A further event filter requirement is used to remove events that are poorly modeled in simulation. 
The magnitude of the vector sum of the transverse momentum of the center of mass frame (\ptcm) and the azimuthal angle between the center of mass system and the invisible system (\dphicmi) are defined. Larger values of \ptcm tend to come from out-of-acceptance noise 
or misreconstruction of physics objects, including \ptmiss, while \dphicmi will peak near 0 or $\pi$ if there 
are misreconstructed events in data. Events are removed if $\ptcm > 200$\GeV or \dphicmi is 
near 0 or $\pi$ (with a \ptcm-dependent requirement). 
To account for inoperable HCAL endcap sectors during 2018 data taking, events are also discarded from the 2018 data set, 
if there are any leptons or jets passing the respective object selections and within the ($-3.2<\eta<-1.2$, $-1.77<\phi<-0.67$) region.
Events with lepton pairs having invariant masses consistent with \JPsi meson decays are rejected.

{\tolerance=800
Categorization of events can be imagined with a hierarchical ordering using 
a notation, $\mathrm{NL}^{\Pell~\text{type}}_{\Pell~\text{qual}}~\mathrm{NJ}_{\mathrm{NbS}}^{\mathrm{NsvS}}~\PX_\mathrm{NbISR}^\text{kin}$, 
that indicates the reconstruction category of the event,
with NL corresponding to the number of reconstructed leptons; 0, 1, 2,
and 3, with events containing more than three reconstructed leptons
discarded. These categories are further subdivided by lepton flavor, charge, and reconstruction quality (gold, silver, bronze), appearing as superscripts and subscripts of NL, respectively.  The next tier of categorization is by \NjetS (NJ), the multiplicity of reconstructed jets assigned to the S system in each event.
Depending on the region, there are then further subdivisions according to the number of  \PQb-tagged jets observed in the S and ISR systems (NbS and NbISR), or the number of SVs assigned to S (NsvS), which are further split into SV central and forward categories (svc and svf, respectively). 
Finally, some categories have additional subdivisions according to the kinematic variables \ptisr and \gperp, with each variable being used to define a ``low'' signal-depleted region (denoted by $\mathrm{p}-, \gamma-$ respectively) and a ``high'' 
signal-enriched region (denoted by $\mathrm{p}+, \gamma+$ respectively), leading 
to potentially four such separate subdivisions ( $\mathrm{p}-$ $\gamma-$, $\mathrm{p}-$ $\gamma+$, $\mathrm{p}+$ $\gamma-$, and $\mathrm{p}+$ $\gamma+$ ). The cases where the four kinematic subdivisions are employed are 
denoted by ``$\mathrm{p}\pm \; \gamma\pm$'' 
in the category definition tables. In all cases, $\gamma-$ corresponds to $\gperp <0.5$ and $\gamma+$ corresponds to $\gperp > 0.5$.
\par}

There are 84 exclusive zero lepton (0L) categories defined as outlined in Table~\ref{tab:cat0L}. Within each of these categories, events are counted in bins of \risr and \mperp, with the definition of the \NjetS-dependent bin boundaries for all of the 0L regions listed in Table~\ref{tab:bin0L}. 
In the limit of extreme mass spectrum compression it is still possible to reconstruct soft SVs, therefore, 
dedicated 0L categories that require the presence of soft SVs associated with the S system are introduced for regions with lower S jet multiplicities. 
For larger \NjetS, there are regions ranging from 2 jets (2J) to 5 jets (5J) that target intermediate mass splittings in models such as T2tt and T2bW.  The 0L categories with higher S object multiplicities and bins with lower \risr values are very good at providing constraints for QCD multijet and other backgrounds.

\begin{table}[!ht]
  \centering
  \topcaption{Category definitions for 0L regions for 
  each \NjetS multiplicity. The highest (5J) is inclusive ($\NjetS \geq 5$). There are 84 exclusive categories in 
  total for the 0L regions.}
  \label{tab:cat0L}
  \renewcommand{\arraystretch}{1.1}
  \begin{scotch}{ c c c c c c }
    \NjetS & \NbS & \NbISR & \NsvS & kin & \ptisr [GeV] \\ [\cmsTabSkip]
	\hline
    0J                        &    &    &  1 or $\geq 2$   & svc or svf  & $[350, \infty)$ \\ [\cmsTabSkip]
    \multirow{2}*{1J} &    &    &  ${\geq}1$   & svc or svf & $[400, \infty)$ \\
           &  0 or 1  &  0 or ${\geq}1$  &  0  & $\mathrm{p}-$ or $\mathrm{p}+$ & $[400, 550]$ or $[550, \infty)$ \\ [\cmsTabSkip]
    \multirow{2}*{2J}  &  0 or 1  &  0 or ${\geq}1$  &    & \multirow{2}*{$\mathrm{p}\pm \; \gamma\pm$} & \multirow{2}*{$[350, 500]$ or $[500, \infty)$} \\
           &  $\geq$2  &                       &    &                                    &     \\ [\cmsTabSkip]
    \multirow{2}*{3J}  &  0 or 1  &  0 or ${\geq}1$  &    & \multirow{2}*{$\mathrm{p}\pm \; \gamma\pm$} & \multirow{2}*{$[350, 500]$ or $[500, \infty)$} \\
           &  $\geq$2  &                       &    &                                    &     \\ [\cmsTabSkip]
    \multirow{2}*{4J}  &  0 or 1  &  0 or ${\geq}1$  &    & \multirow{2}*{$\mathrm{p}\pm \; \gamma\pm$} & \multirow{2}*{$[350, 500]$ or $[500, \infty)$} \\
           &  $\geq$2  &                       &    &                                    &     \\ [\cmsTabSkip]
    \multirow{2}*{5J}  &  0 or 1  &  0 or ${\geq}1$  &    & \multirow{2}*{$\mathrm{p}-$ or $\mathrm{p}+$} & \multirow{2}*{$[350, 500]$ or $[500, \infty)$} \\
           &  $\geq$2  &                       &    &                                    &     \\ 
  \end{scotch}
\end{table}

\begin{table}
  \centering
  \topcaption{The \risr and \mperp bin definitions for 0L regions for each \NjetS multiplicity. The highest (5J) is inclusive ($\NjetS \geq 5$). The lower \risr bins denoted as "CR" are used as control regions.}
  \label{tab:bin0L}  
  \renewcommand{\arraystretch}{1.1}
  \begin{scotch}{ c  l  l  c }
    \NjetS & \risr & \mperp [GeV] & \Nbins \\ [\cmsTabSkip]
    \hline
	\multirow{2}*{0J} &  $[0.95, 0.985]$ CR & $[0, \infty)$ & \multirow{2}*{4}\\
           &  $[0.985, 1]$ & $[0, 5]$ or $[5, 10]$ or $[10, \infty)$ & \\
    [\cmsTabSkip]
    \multirow{4}*{1J} &  $[0.8, 0.9]$ CR & $[0, \infty)$ & \multirow{2}*{6} \\
           &  $[0.9, 0.93]$ CR & $[0, \infty)$ &\\
           &  $[0.93, 0.96]$ & $[0, 20]$ or $[20, \infty)$ &\\
           &  $[0.96, 1]$ & $[0, 15]$ or $[15, \infty)$ &\\
    [\cmsTabSkip]
    \multirow{5}*{2J} &  $[0.65, 0.75]$ CR &  $[0, \infty)$ & \multirow{5}*{6}\\
           &  $[0.75, 0.85]$ CR & $[0, \infty)$ &\\
           &  $[0.85, 0.9]$ & $[0, \infty)$ &\\
           &  $[0.9, 0.95]$ & $[0, 20]$ or $[20, \infty)$ &\\
           &  $[0.95, 1]$ & $[0, \infty)$ & \\
    [\cmsTabSkip]
    \multirow{5}*{3J} &  $[0.55, 0.65]$ CR & $[0, \infty)$ & \multirow{5}*{6}\\
           &  $[0.65, 0.75]$ CR & $[0, \infty)$ &\\
           &  $[0.75, 0.85]$ & $[0, \infty)$ &\\
           &  $[0.85, 0.9]$ & $[0, 50]$ or $[50, \infty)$ &\\
           &  $[0.9, 1]$ & $[0, \infty)$ &\\
    [\cmsTabSkip]
    \multirow{4}*{4J} &  $[0.55, 0.65]$ CR & $[0, \infty)$ & \multirow{4}*{5}\\
           &  $[0.65, 0.75]$ CR & $[0, \infty)$ &\\
           &  $[0.75, 0.85]$ & $[0, 80]$ or $[80, \infty)$ &\\
           &  $[0.85, 1]$ & $[0, \infty)$ &\\
    [\cmsTabSkip]
    \multirow{4}*{5J} &  $[0.5, 0.6]$ CR & $[0, \infty)$ & \multirow{4}*{5}\\
           &  $[0.6, 0.7]$ CR & $[0, \infty)$ &\\
           &  $[0.7, 0.8]$ & $[0, 150]$ or $[150, \infty)$ &\\
           &  $[0.8, 1]$ & $[0, \infty)$ &\\
  \end{scotch}
\end{table}

The categorization of the one-lepton (1L) regions is the most expansive in the analysis with 178 exclusive regions, 
as they are applicable to a wide range of signals. 
Additional CRs are defined based on the lepton reconstruction quality, allowing for further background constraints among regions, particularly allowing for the shapes and normalization of various types of lepton contributions to be measured from data. Depending on \NjetS, 1L events 
can be categorized by either the lepton flavor, or charge (to better control $\PW(\Pell\nu)$ + jets backgrounds), or both. As in 0L, the 1L categories include dedicated regions requiring tagged SVs, which are most relevant for the most compressed signals also having soft heavy-flavor decays. 
The 1L category definitions are listed in Table~\ref{tab:cat1L}. The \risr-\mperp bin definitions within each of these regions are shown in Table~\ref{tab:bin1L}.

\begin{table*}[t]
  \centering
  \topcaption{Category definitions for 1L regions for each \NjetS multiplicity. The highest (4J) is inclusive ($\NjetS \geq 4$). There are a total of 178 categories for the 1L regions.}
  \label{tab:cat1L}
  \cmsTable{
  \begin{scotch}{ c  c  c  c  c  c  c  c }
    \NjetS & lep qual & lep cat &  \NbS & \NbISR & \NsvS & kin & \ptisr [GeV] \\[\cmsTabSkip]
    \hline
    \multirow{4}*{0J}  & gold                  &  \multirow{1}*{$\Pellp$ or $\Pellm$}  & & & \multirow{2}*{${\geq}1$}   & \multirow{2}*{svc or svf}  & \multirow{2}*{$[350, \infty)$} \\
           & silver or bronze & $\Pe$ or $\PGm$ & & &    &  &  \\
	[\cmsTabSkip]
           & \multirow{2}*{gold}                   &  \multirow{1}*{$\Pep$ or $\Pem$ or}  & & \multirow{2}*{0 or ${\geq}1$} & \multirow{3}*{0}   & \multirow{3}*{$\mathrm{p}-$ or $\mathrm{p}+$}  & \multirow{3}*{$[350, 500]$ or $[500, \infty)$ } \\
           & & \multirow{1}*{$\PGmp$ or $\PGmm$} & & & & & \\
           & silver or bronze & $\Pe$ or $\PGm$ & & &    &  &  \\
    [\cmsTabSkipLarge]
    \multirow{4}*{1J}  & gold                  &    & & & \multirow{2}*{${\geq}1$}   & \multirow{2}*{svc or svf}  & \multirow{2}*{$[350, \infty)$} \\
           & silver or bronze & $\Pe$ or $\PGm$  & & &    &  &  \\
	[\cmsTabSkip]
           & gold                   &  \multirow{1}*{$\Pellp$ or $\Pellm$}  & 0 or 1 & 0 or ${\geq}1$ & \multirow{2}*{0}   & \multirow{2}*{p$\pm \; \gamma\pm$}  & \multirow{2}*{$[350, 500]$ or $[500, \infty)$ } \\
           & silver or bronze & $\Pe$ or $\PGm$ & & &    &  &  \\
    [\cmsTabSkipLarge]
    \multirow{3}*{2J}   & \multirow{2}*{gold} &  & 0 or 1 & 0 or ${\geq}1$ & \multirow{3}*{}   & \multirow{3}*{p$\pm \; \gamma\pm$}  & \multirow{3}*{$[350, 500]$ or $[500, \infty)$ } \\
           &                                &  & ${\geq}2$ &  &   &   &  \\
           & silver or bronze & $\Pe$ or $\PGm$ & & &    &  &  \\
    [\cmsTabSkipLarge]
    \multirow{3}*{3J}   & \multirow{2}*{gold} &  & 0 or 1 & 0 or ${\geq}1$ & \multirow{3}*{}   & \multirow{3}*{p$\pm \; \gamma\pm$}  & \multirow{3}*{$[350, 500]$ or $[500, \infty)$ } \\
           &                                &  & ${\geq}2$ &  &   &   &  \\
           & silver or bronze & $\Pe$ or $\PGm$ & & &    &  &  \\
    [\cmsTabSkipLarge]
    \multirow{3}*{4J}   & \multirow{2}*{gold} &  & 0 or 1 & 0 or ${\geq}1$ & \multirow{3}*{}   & \multirow{3}*{p$\pm \; \gamma\pm$}  & \multirow{3}*{$[350, 500]$ or $[500, \infty)$ } \\
           &                                &  & ${\geq}2$ &  &   &   &  \\
           & silver or bronze & $\Pe$ or $\PGm$ & & &    &  &  \\
  \end{scotch}
}
\end{table*}

\begin{table}
  \centering
  \topcaption{The \risr and \mperp bin definitions for 1L regions for each \NjetS multiplicity. The highest (4J) is inclusive ($\NjetS \geq 4$). The lower \risr bins denoted as "CR" are used as control regions.}
  \label{tab:bin1L}
  \renewcommand{\arraystretch}{1.1}
  \begin{scotch}{ c  l  l  c }
    \NjetS & \risr & \mperp [GeV] & \Nbins \\ [\cmsTabSkip]
    \hline
    \multirow{2}*{0J} &  $[0.96, 0.98]$ CR & $[0, 10]$ or $[10, \infty)$ &  \multirow{2}*{5}\\
           &  $[0.98, 1]$ & $[0, 5]$ or $[5, 10]$ or $[10, \infty)$ & \\
    [\cmsTabSkip]
    \multirow{3}*{1J}  &  $[0.85, 0.9]$ CR & $[0, 30]$ or $[30, \infty)$ & \multirow{3}*{5} \\ 
           &  $[0.9, 0.95]$ & $[0, 20]$ or $[20, \infty)$ &\\
           &  $[0.9, 0.95]$ & $[0, \infty)$ &\\
    [\cmsTabSkip]
    \multirow{3}*{2J}  &  $[0.8, 0.85]$ CR & $[0, 70]$ or $[70, \infty)$ & \multirow{3}*{5}\\
           &  $[0.85, 0.9]$ & $[0, 50]$ or $[50, \infty)$ &\\
           &  $[0.9, 1]$ & $[0, \infty)$ & \\
    [\cmsTabSkip]
    \multirow{3}*{3J}  &  $[0.65, 0.75]$ CR & $[0, 100]$ or $[100, \infty)$ &  \multirow{3}*{5}\\
           &  $[0.75, 0.85]$ & $[0, 80]$ or $[80, \infty)$ &\\
           &  $[0.85, 1]$ & $[0, \infty)$ &\\
    [\cmsTabSkip]
    \multirow{3}*{4J} &  $[0.6, 0.7]$ CR & $[0, 180]$ or $[180, \infty)$ &  \multirow{3}*{5}\\
           &  $[0.7, 0.8]$ & $[0, 150]$ or $[150, \infty)$ &\\
           &  $[0.8, 1]$ & $[0, \infty)$ &\\
  \end{scotch}
\end{table}

\begin{figure}[ht]
  \centering
  \includegraphics[width=0.49\textwidth]{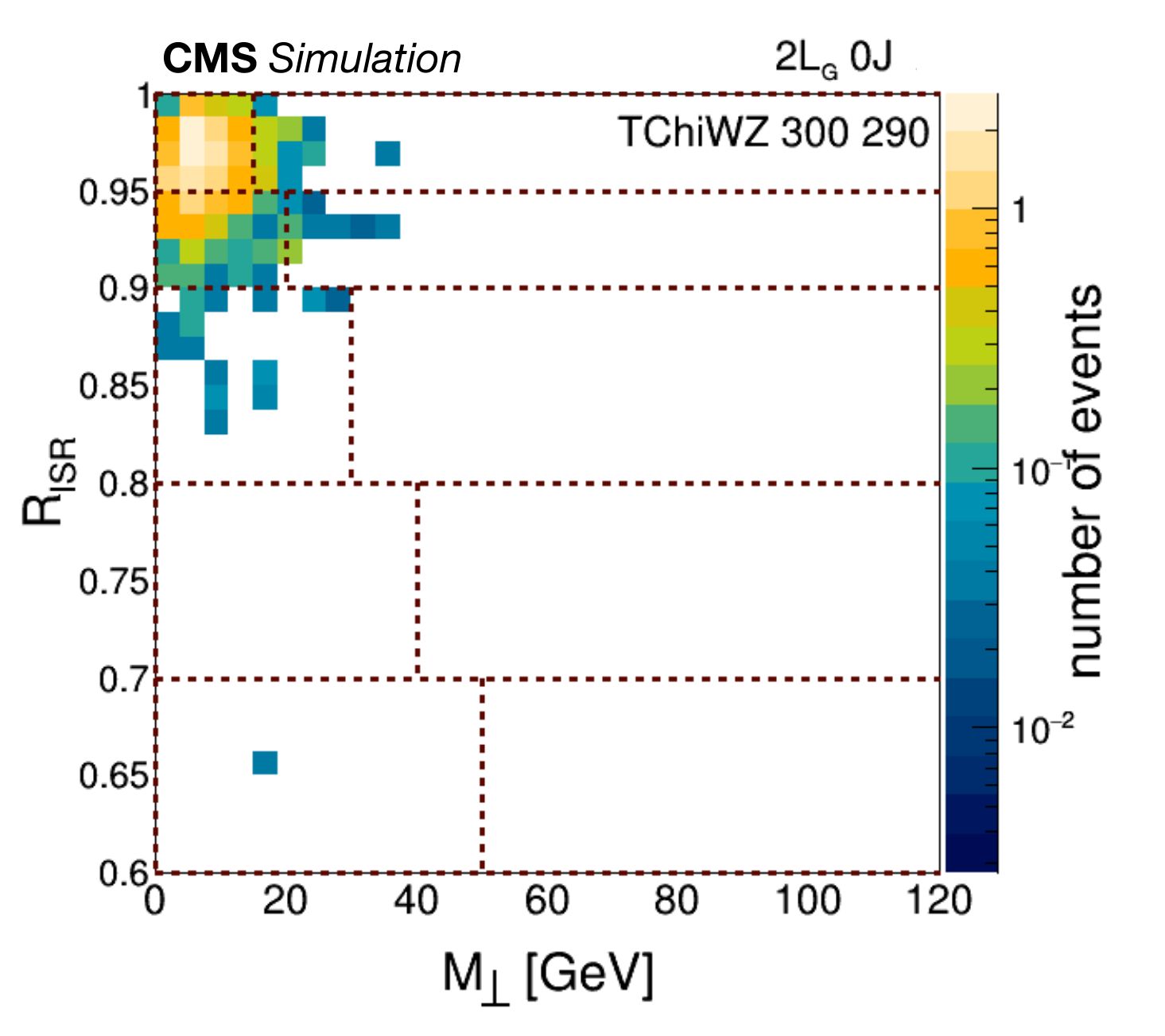}
  \includegraphics[width=0.49\textwidth]{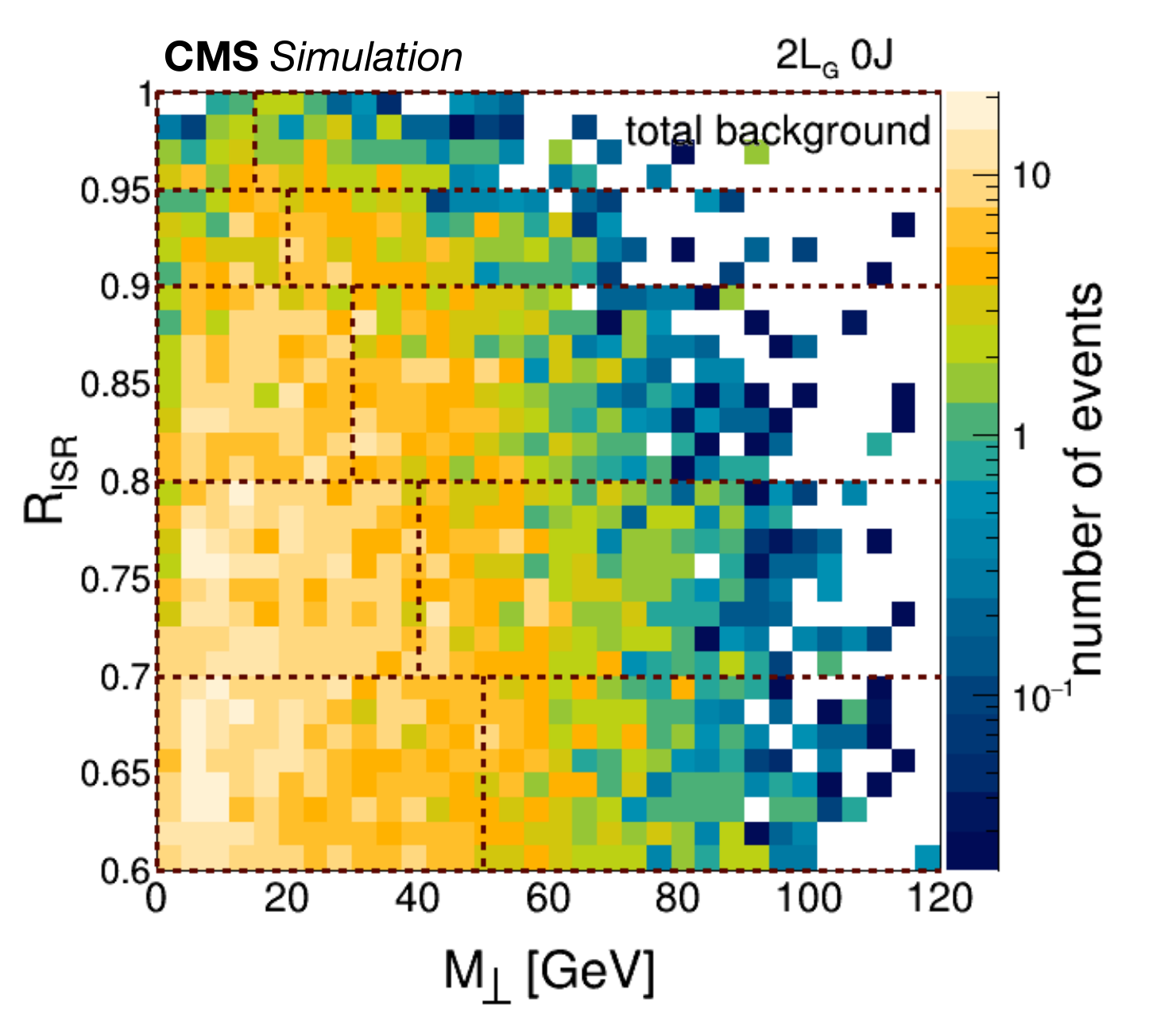}
  \caption{Distributions of \risr vs. \mperp for a TChiWZ signal sample with a parent mass of 300\GeV and a LSP mass of 290\GeV (\cmsLeft) and the corresponding total SM background (\cmsRight) for the 2L, 0~S-jet category. 
  The dashed lines show the bin edges for this particular jet multiplicity.}
  \label{fig:RvMp2L0J300290}
\end{figure}

For the two lepton (2L) final-state categorization, the gold regions are defined as having two gold 
leptons. Similarly, silver categorization requires one gold and one silver or both silver, while bronze only 
includes the gold-bronze and silver-bronze cases. 
The 2L categories with gold leptons are signal rich, while the silver and bronze ones  
are important for constraining nonprompt- and misidentified-lepton backgrounds.
Figure~\ref{fig:RvMp2L0J300290} shows an example \risr-\mperp distribution with the bin boundaries overlaid for the 
2L, 0~S-jet category.
The 2L category definitions are presented in Table~\ref{tab:cat2L}. Depending on the \NjetS multiplicity, different combinations 
of lepton charge and flavor are used to further split categories. In the 2L category with 0~S-jets, the gold category includes 
separate regions for each of the lepton flavor combinations, while the region requiring a 
soft SV tag integrates over lepton flavor and charge. There is also a same-sign (SS) lepton category, which is integrated over the lepton flavor. 

Categories are established for cases where the dilepton pair is consistent with a ``$\PZst$'' candidate, specifically when 
there are one or two \D{S} jets present. Here, these $\PZst$ candidates, have opposite-sign (OS) 
leptons with same flavor (SF) also in the same S-system hemisphere. The ``no$\PZ$'' categorization is for OS, 
different-flavor (DF) events, or OS-SF events with the leptons appearing in different hemispheres. 
These regions also have an SS dilepton category, with all three of these classes integrated over lepton flavor. For all of the 2L silver and bronze regions, leptons are separated by 
flavor (\elel, \MUMU, or \emu) in order 
to better serve as CRs capable of disentangling the contributions from different nonprompt-lepton sources and processes. 
Bin boundaries in \risr and \mperp for all the 2L regions are summarized in Table~\ref{tab:bin2L}.
 
\begin{table*}[t]
  \centering
  \topcaption{Category definitions for 2L regions for each \NjetS multiplicity. The highest (2J) is inclusive ($\NjetS \geq 2$). There is a total of 115 exclusive 2L categories. }
  \label{tab:cat2L}  
  \cmsTable{
  \begin{scotch}{ c c c c c c c c }
    \NjetS & lep qual & lep cat &  \NbS & \NbISR & \NsvS & kin & \ptisr [GeV] \\ [\cmsTabSkip]
    \hline
    \multirow{5}*{0J}  & \multirow{1}*{gold or silver}   &    & & & \multirow{2}*{${\geq}1$}   & \multirow{2}*{svc or svf}  & \multirow{2}*{$[250, \infty)$} \\
                                & \multirow{1}*{ or bronze}  & & & & & & \\
    [\cmsTabSkip]
           & \multirow{3}*{gold}  &  $\Pep\Pem$ or $\PGmp\PGmm$   & & \multirow{2}*{0 or ${\geq}1$} & \multirow{4}*{0}   & \multirow{2}*{p$\pm \; \gamma\pm$}  & \multirow{2}*{$[250, 350]$ or $[350, \infty)$ } \\
           &  & or $\Pepm\PGmmp$ & & & & & \\
           &                                                &  SS   & & &   &   & \multirow{2}*{$[250, \infty)$} \\
           & silver or bronze & $\Pe\Pe$ or $\PGm\PGm$ or $\Pe\PGm$ & & &    &  &  \\
    [\cmsTabSkipLarge]
    \multirow{4}*{1J}   & \multirow{2}*{gold}                   &  $\PZst$ or no$\PZ$ (OS)   & 0 or 1 & 0 or ${\geq}1$ & \multirow{3}*{}   & \multirow{1}*{p$\pm \; \gamma\pm$}  & \multirow{1}*{$[250, 350]$ or $[350, \infty)$ } \\
           &                                                &  SS   & & &   &   & \multirow{2}*{$[350, \infty)$} \\
           & silver or bronze & $\Pe\Pe$ or $\PGm\PGm$ or $\Pe\PGm$ & & &    &  &  \\
    [\cmsTabSkipLarge]
    \multirow{4}*{2J}   & \multirow{2}*{gold}                   &  $\PZst$ or no$\PZ$ (OS)   & 0 or ${\geq}1$ & 0 or ${\geq}1$ & \multirow{3}*{}   & \multirow{1}*{p$\pm \; \gamma\pm$}  & \multirow{1}*{$[250, 350]$ or $[350, \infty)$ } \\
           &                                                &  SS   & & &   &   & \multirow{2}*{$[350, \infty)$} \\
           & silver or bronze & $\Pe\Pe$ or $\PGm\PGm$ or $\Pe\PGm$ & & &    &  &  \\
  \end{scotch}
}
\end{table*}

\begin{table}
  \centering
  \topcaption{The \risr and \mperp bin definitions for 2L regions for each \NjetS multiplicity. The highest (2J) is inclusive ($\NjetS \geq 2$). The lower \risr bins denoted as "CR" are used as control regions.}
  \label{tab:bin2L}  
  \renewcommand{\arraystretch}{1.1}
  \begin{scotch}{ c l l  c }
    \NjetS & \risr & \mperp [GeV] & \Nbins \\ [\cmsTabSkip]
    \hline
    \multirow{5}*{0J} &  $[0.6, 0.7]$ CR & $[0, 50]$ or $[50, \infty)$ & \multirow{5}*{10}\\
           &  $[0.7, 0.8]$ CR & $[0, 40]$ or $[40, \infty)$ &\\
           &  $[0.8, 0.9]$ & $[0, 30]$ or $[30, \infty)$ &\\
           &  $[0.9, 0.95]$ & $[0, 20]$ or $[20, \infty)$ &\\
           &  $[0.95, 1]$ & $[0, 15]$ or $[15, \infty)$ &\\
    [\cmsTabSkip]
    \multirow{5}*{1J} &  $[0.5, 0.6]$ CR & $[0, 100]$ or $[100, \infty)$ & \multirow{5}*{10}\\
           &  $[0.6, 0.7]$ CR & $[0, 80]$ or $[80, \infty)$ &\\
           &  $[0.7, 0.8]$ & $[0, 60]$ or $[60, \infty)$ &\\
           &  $[0.8, 0.9]$ & $[0, 40]$ or $[40, \infty)$ &\\
           &  $[0.9, 1]$ & $[0, 30]$ or $[30, \infty)$ &\\
    [\cmsTabSkip]
    \multirow{4}*{2J} &  $[0.5, 0.65]$ CR & $[0, 100]$ or $[100, \infty)$ & \multirow{4}*{7}\\
           &  $[0.65, 0.75]$ CR & $[0, 80]$ or $[80, \infty)$ &\\
           &  $[0.75, 0.85]$ & $[0, 60]$ or $[60, \infty)$ &\\
           &  $[0.85, 1]$ & $[0, \infty)$ &\\
  \end{scotch}
\end{table}

For the three-lepton (3L) categories, the gold categories are defined such that all three leptons must have gold quality (GGG). The sub-splittings for these include events with 0 or ${\geq}1$ S-jet, the presence of a $\PZst$ candidate, no \PZ boson, or three same-sign leptons.  Throughout the 3L category there is no  \PQb-tagged jet
counting and no other splitting of categories based on \ptisr, \gperp, or the presence of SVs. The 3L silver category includes GGS and GSS combinations for the individual lepton quality, while the 3L bronze category includes GGB and GSB quality combinations. 
The sub-categorization for all lepton quality criteria in 3L is the same. The definitions of the 3L analysis regions are presented in Table~\ref{tab:cat3L}. There are no \mperp bins in the 3L categories (one integrated bin), only bins in \risr, with bin boundaries 
for 3L with 0 S jet and 3L with 1 S jet categories summarized in Table~\ref{tab:bin3L}.

\begin{table}
  \centering
  \caption{Category definitions for the 3L regions for each \NjetS multiplicity. The highest (1J) is inclusive ($\NjetS \geq 1$). There is a total of 15 exclusive 3L categories.}
  \label{tab:cat3L}  
  \renewcommand{\arraystretch}{1.1}
  \begin{scotch}{ c  c  c  c }
    \NjetS & lep qual & lep cat  &  \ptisr [GeV] \\ [\cmsTabSkip]
    \hline
    \multirow{1}*{0J}   & gold or silver or bronze  &  $\PZst$ or no$\PZ$ or $\Pellpm\Pellpm\Pellpm$   & \multirow{1}*{$[250, \infty)$ } \\
    \multirow{1}*{1J}   & gold or silver or bronze  &  $\PZst$ or no$\PZ$ or $\Pellpm\Pellpm\Pellpm$   & \multirow{1}*{$[250, \infty)$ } \\
  \end{scotch}
\end{table}

\begin{table}
  \centering
  \topcaption{The \risr and \mperp bin definitions for 3L regions for each \NjetS multiplicity. The highest (1J) is inclusive ($\NjetS \geq 1$). The lower \risr bins denoted as "CR" are used as control regions. An additional control region with $0.5\leq \risr <0.6$ is also used with the 0~S-jet region.}
  \label{tab:bin3L}  
  \renewcommand{\arraystretch}{1.1}
  \begin{scotch}{ c  l  l  c }
    \NjetS & \risr & \mperp [GeV] & \Nbins \\ [\cmsTabSkip]
    \hline
    \multirow{4}*{0J} &  $[0.6, 0.7]$ CR & $[0, \infty)$ & \multirow{4}*{4}\\
           &  $[0.7, 0.8]$ CR & $[0, \infty)$ &\\
           &  $[0.8, 0.9]$ & $[0, \infty)$ &\\
           &  $[0.9, 1]$ & $[0, \infty)$ &\\
    [\cmsTabSkip]
    \multirow{3}*{1J} &  $[0.55, 0.7]$ CR & $[0, \infty)$ & \multirow{4}*{3}\\
           &  $[0.7, 0.85]$ & $[0, \infty)$ & \\
           &  $[0.85, 1]$ & $[0, \infty)$ &\\
  \end{scotch}
\end{table}

While the signal model-dependent results involve a simultaneous fit to all of the bins and categories included in this analysis, 
a subset of CR bins are identified 
in order to both study the fit model in data prior to the unblinding of signal-sensitive regions, and also to derive model-independent
upper limits on a subset of bin/category combinations.  
To be included as a CR bin, there must be less than 1\% signal contamination (relative to expected backgrounds) for any of the signals considered in the analysis. 
The CR bins account 
for 648 out of the total, 2443 bins, encompassing 62\% of the expected SM background events. 
In general, the chosen CR bins 
are at lower values of \risr, where expected signal-to-background is significantly less favorable. A simultaneous fit of 
the CR bins is able to constrain all of the nuisance parameters with data.  

To derive model-independent 
limits, combinations 
of the previously defined ``signal'' bins are combined into seven
``superbins'', defined in Table~\ref{tab:modind}. Five of the superbins were derived by systematically examining the expected signal significance for clusters of bins from admixtures of signals corresponding to five groupings of top squark, electroweakino, and slepton pair models with low and intermediate $\Delta m$. 
To look for  \PQb-enriched signals similar to those from top squark pair decays, there are three superbins for low $\Delta m$ and one for medium $\Delta m$. 
One of these low-$\Delta m$ superbins was defined with only low-momentum SV candidates. 
One superbin is defined targeting signal models with sparticle decays to \PW or \PZ bosons, such as TChiWZ and TChiWW. For signals consistent with compressed slepton decays with $\Delta m < 10\GeV$, another superbin is similarly defined to feature two leptons with OS-SF. An additional superbin is defined for 3L final states. 

\begin{table*}
\centering
\topcaption{List of categories and \mperp/\risr bins corresponding to each model-independent superbin.}\label{tab:modind}
\cmsTable{
\begin{scotch}{lccccccccc}
Label & Lepton ID & $N_{\text{lep}}$ & \NjetS & $N^{\mathrm{S}}_{\mathrm{b/SV}}$ & \ptisr & \gperp & \NbISR & \mperp [GeV] & \risr \\ [\cmsTabSkip]
\hline
  \PQb jets & \multirow{2}*{$\PGmm$}    & \multirow{2}*{1} &  \multirow{2}*{0}       &  \multirow{2}*{0}       & \multirow{2}*{high} & \multirow{2}*{incl} & \multirow{2}*{0}     & \multirow{2}*{$[0,10]$}     & \multirow{2}*{$[0.98,1]$} \\
  low-$\Delta m$ 1L &  &  &  &  & & &  &  & \\ 
[\cmsTabSkip]
\PQb jets&
OS-DF    & 2 &  0       &  0       & high & high &  0     & $[0,15]$     & $[0.95,1]$ \\
 low-$\Delta m$ 2L  & OS-SF    & 2 &  0       &  0       & high & high & 0     & $[0,15]$     & $[0.95,1]$ \\ 
 & $\PZ$/no$\PZ$  & 2 &  1       &  0       & high & high & 0     & $[0,30]$     & $[0.85,1]$ \\ 
 & \PZ       & 2 & ${\geq}2$ &  0       & high & high & 0     & $[0,\infty]$ & $[0.9,1]$ \\ 
[\cmsTabSkipLarge]
SV  &
-            & 0 &  0       & ${\geq}2$ & low  & low  & ${\geq}1$ & $[5,\infty]$ & $[0.985,1]$ \\ 
 & $\Pellm$   & 1 &  0       & ${\geq}1$ & low  & low  & ${\geq}1$ & $[0,10]$     & $[0.98,1]$ \\ 
 & $\Pellpm$ & 1 &  1       & ${\geq}1$ & low  & low  & ${\geq}1$ & $[0,\infty]$ & $[0.95,1]$ \\ 
 & $\Pell\Pell$   & 2 &  0       & ${\geq}1$ & low  & low  & ${\geq}1$ & $[0,15]$     & $[0.95,1]$ \\ 
[\cmsTabSkipLarge]
\PQb jets & -      & 0 & 4       & 1     & high & incl & 0    & $[0,\infty]$ & $[0.85, 1]$ \\
moderate-$\Delta m$ & -      & 0 & 4       & 2     & high & incl & ${\geq}1$ & $[0,\infty]$ & $[0.85, 1]$ \\
 & -      & 0 & ${\geq}5$ & 1     & high & low  & 0    & $[0,\infty]$ & $[0.8, 1]$ \\ 
 & -      & 0 & ${\geq}5$ & ${\geq}2$ & high & low  & ${\geq}1$ & $[0,\infty]$ & $[0.8, 1]$ \\ 
 & $\Pell$ & 1 & 2       & 1     & high & high & 0    & $[0,\infty]$ & $[0.9, 1]$ \\
 & $\Pell$ & 1 & 2       & 2     & high & high & ${\geq}1$ & $[0,\infty]$ & $[0.9, 1]$ \\ 
 & $\Pell$ & 1 & 3       & 1     & high & high & 0    & $[0,\infty]$ & $[0.85, 1]$ \\ 
 & $\Pell$ & 1 & 3       & ${\geq}2$ & high & high & ${\geq}1$ & $[0,\infty]$ & $[0.85, 1]$ \\ 
 & $\Pell$ & 1 & ${\geq}4$ & 1     & high & high & 0    & $[0,\infty]$ & $[0.8, 1]$ \\ 
 & $\Pell$ & 1 & ${\geq}4$ & ${\geq}2$ & high & high & ${\geq}1$ & $[0,\infty]$ & $[0.8, 1]$ \\ 
 & \PZ      & 2 & 1       & ${\leq}1$ & high & high & 0    & $[0,\infty]$ & $[0.9, 1]$ \\ 
 & \PZ      & 2 & ${\geq}2$ & ${\leq}1$ & high & high & 0    & $[0,\infty]$ & $[0.85, 1]$ \\ 
 & no$\PZ$    & 2 & 1       & ${\leq}1$ & high & high & 0    & $[0,\infty]$ & $[0.9, 1]$ \\ 
 & no$\PZ$    & 2 & ${\geq}2$ & ${\leq}1$ & high & high & 0    & $[0,\infty]$ & $[0.85, 1]$ \\ 
 [\cmsTabSkipLarge]
Electroweak & OS-SF & 2 & 0       & 0    & high & high & 0    & $[0,15]$     & $[0.95,1]$ \\ 
 & \PZ    & 2 & 1       & 0 & high & incl & 0    & $[0,\infty]$ & $[0.9,1]$ \\ 
 & \PZ    & 2 & ${\geq}2$ & 0 & high & incl & 0    & $[0,\infty]$ & $[0.85,1]$ \\ 
 & \PZ    & 3 & 0       & 0    & low  & incl & ${\geq}1$ & $[0,\infty]$ & $[0.9,1]$ \\ 
 & \PZ    & 3 & ${\geq}1$ & 0    & low  & incl & ${\geq}1$ & $[0,\infty]$ & $[0.85,1]$ \\ 
[\cmsTabSkipLarge]
2L OS-SF & OS-SF & 2 & 0       & 0    & high & incl & 0    & $[0,15]$     & $[0.95,1]$ \\ 
[\cmsTabSkipLarge]
3L & \PZ   & 3 & 0       & 0 & low & incl & ${\geq}1$ & $[0,\infty]$  & $[0.9,1]$  \\
 & \PZ   & 3 & ${\geq}1$ & 0 & low & incl & ${\geq}1$ & $[0,\infty]$  & $[0.85,1]$ \\ 
 & no$\PZ$ & 3 & 0       & 0 & low & incl & ${\geq}1$ & $[0,\infty]$  & $[0.9,1]$  \\
 & no$\PZ$ & 3 & ${\geq}1$ & 0 & low & incl & ${\geq}1$ & $[0,\infty]$  & $[0.85,1]$ \\
 & SS  & 3 & incl & 0 & low & incl & ${\geq}1$ & $[0,\infty]$  & $[0.9,1]$  \\ 
\end{scotch}
}
\end{table*}

\section{Background and signal fit model}\label{sec:fit}
A maximum likelihood fit is performed simultaneously to all 
of the 2443 \risr/\mperp bins in the 392 categories, as defined in the previous section. The fit model is implemented with 
the \textsc{Combine} statistical analysis tool~\cite{COMBINE} based on the \textsc{RooFit}~\cite{Verkerke:2003ir} and \textsc{RooStats} frameworks~\cite{Moneta:2010pm}. 
In the fit model, the likelihood is modeled as a product of Poisson probability distributions, one for each bin, with a rate parameter equal to the total expected event yield in that bin, calculated as the sum of contributions from different processes. 
The primary dependence on simulation in the background modeling is associated with 
the nominal initial values of these process-dependent bin yields (accounting for all 
previously described scale factors and corrections). 

The modeling of different background contributions in the fit (\eg, \Wjets, \ttx) is further subdivided by number and type of 
nonprompt leptons in defining the individual modeled processes.
The nominal normalization for each process is modified by a collection of nuisance parameters, with each applied to one or (typically) more processes in a given category or bin, multiplying yields by a scale factor 
that is profiled in the maximum likelihood fit. This means that, for example, dibosons with one electron coming 
from a heavy-flavor decay and dibosons with two nonprompt muons are modeled as different processes with 
separate bin-dependent yields. This allows for a single nuisance parameter to modify the event yield 
of all diboson-associated processes or, independently, the yield of all events with nonprompt electrons from a single nonprompt source.  

The nuisance parameters modeled in the fit are either \textit{externally constrained} or \textit{constrained with data}. 
Externally constrained parameters are associated with auxiliary measurements, performed using data sets independent 
of those  considered in this analysis, from which prior uncertainties are correspondingly derived~\cite{ref:JES,ref:bjetcor}. These prior uncertainties, 
modeled as lognormal distributions, further multiply the likelihood such that externally constrained parameters are informed by both 
auxiliary measurements and the data set selected in this analysis. Alternatively, parameters are freely-floating 
in the fit, constrained instead using categorization and kinematic sidebands corresponding to CRs in data. 

Data from all three years (2016--2018) are fit simultaneously in all categories and bins, with nuisances modeled as either common to all years or independent. There are numerous systematic uncertainties for 
which the parameters are determined in the fit. Each of the systematic uncertainty 
contributions, along with details of their number, year-by-year implementation, and 
size range, are summarized in Table~\ref{tab:syst}.  

\begin{table*}[ht]
  \centering
  \topcaption{Summary of systematic uncertainties for the full fit. The number of nuisance parameters is listed, with details as to how they are partitioned by data-taking period. The range of the parameter impact variation post-fit is given in the final column.}
  \label{tab:syst}  
  \begin{scotch}{ l  c  c }
    \multicolumn{1}{ c }{Source} & Number of parameters & Uncertainty (\%)\\
    \hline 
    \multicolumn{1}{ c }\textbf{Externally constrained} &  &  \\
    Integrated luminosity & 1 / year (3) & 1.5-2.5 \\
    Factorization/normalization scales, PDF, and $Q^2$ & 8 & 0-11 \\
    Pileup & 1 / year (3) & 1-7\\
    \ptmiss trigger & 2 / year (6) & 1-3 \\
    Electron \& muon efficiency & 8 / year (24) & 1 \\
    Jet energy scale and resolution  & 2 / year (6) & 0-10\\
    \ptmiss unclustered energy & 1 / year (3) & 0-6\\
    \ptmiss trigger & 2 / year (6) & 1-3 \\
    \PQb jet efficiency & 2 + 2 / year (8) & 0-3\\
    Fast simulation corrections & 7 / year (21) & 1-10 \\
    Monte Carlo event count & 1 / bin & 1-15 \\ [\cmsTabSkip]
    \multicolumn{1}{ c }\textbf{Constrained with data} &  &  \\
    \Wjets normalization & 16 & 1-12\\
    \ttx normalization & 16 & 2-20 \\
    QCD multijet normalization & 15 & 5-30 \\
    \zdy, diboson + Higgs boson normalization & 5 & 2-10\\
    Single top quark, triboson/rare normalization & 4 & 5-30\\   
    \ptisr and \gperp  & 28 & 1-10\\
    Lepton category normalization & 21 & 5-10 \\
    Misidentified and nonprompt leptons & 36 & 3-12\\
    \PQb-tagged jet category normalization & 68 & 1-10\\ 
    SV tagging efficiency & 3 & 1-10 \\
  \end{scotch}
\end{table*}

\subsection{Externally constrained systematic uncertainties}
Uncertainties in the collected integrated luminosity are split by year, with correlations corresponding to common 
systematic uncertainty sources~\cite{ref:lum1, ref:lum2, ref:lum3}. 
The uncertainty in the simulations of background processes resulting from missing higher-order corrections is estimated by varying the renormalization and factorization scales by a factor of two, with each of the two scales taken to be the same in each variation~\cite{ref:renandfac1, ref:renandfac2}. 
Systematic uncertainties associated with the modeling of parton distribution functions are estimated using 100 variations provided with the NNPDF sets, while the effect of the uncertainty in the value of the strong coupling constant is estimated by varying the value $\alpha_S(m_{\PZ}) = 0.1180$ by $\pm0.0015$~\cite{ref:renandfac3}. Simulated events are reweighted such that the distribution of the number of additional pileup interactions matches that observed over the different data-taking periods, with associated systematic uncertainties evaluated by varying the total inelastic cross section within measured uncertainties~\cite{ref:pileup}.

Differences between simulation and data in the efficiencies of lepton identification requirements for prompt leptons are evaluated using the tag-and-probe method 
applied to \PZ boson and \JPsi meson events, as described in Section~\ref{sec:reco}. 
Independent scale factors and corresponding uncertainties are derived for electrons and muons, for each of the three data-taking years, and are separated by loose selection, identification, isolation, and impact parameter requirements. Uncertainties in the jet energy scale and resolution are modeled 
independently for each of the three data-taking periods~\cite{ref:jes1, ref:jes2}, with the effects of changes to the momentum of jets 
propagated to the \ptmiss. Similar year-independent nuisance parameters associated with the effects of unclustered energy on \ptmiss are included. 
The scale factors associated with differences for \PQb jet tagging between simulation and data have corresponding nuisance parameters separated by source (heavy-flavor 
and light-flavor quark or gluon jets) and by year, with an additional parameter for each source accounting for correlations between years. 

The efficiency as a function of \ptmiss to pass the trigger requirements has been measured with data collected using single-lepton reference triggers, and is compared to that found using background sample simulations to derive scale factors and corresponding uncertainties. 
These factors include variations depending on $\HT$ (the scalar sum of the \pt of all jets with $\pt > 20$\GeV and $\abs{\eta} < 5.0$ in the event), the number and flavor of leptons, as well as the S jet multiplicity.  

Uncertainties arising from differences between the fast simulation used for signal processes 
and the full \GEANTfour-based simulation used for background processes are accounted for through a set of additional nuisance parameters. These uncertainties 
cover reconstructed leptons, \PQb-tagged jets, SVs, and \ptmiss. 

\subsection{Systematic uncertainties determined from data}
In order to account for data/simulation modeling differences that are not covered by the externally constrained systematic uncertainties 
included in the fit model, a collection of scale factor parameters obtained from control samples in data is included. 
Mismodeling can result from unaddressed topology-dependent effects in the derivation of prior constraints or from 
inherent shortcomings in the simulation. 
The large number and types of event categories included in the fit allows for these factors to be constrained directly from data, using a high-dimensional collection of sidebands associated with kinematic, object quality, and multiplicity categorizations.   

Background processes are separated into several groups that are treated with common 
systematic uncertainties: \Wjets, \ttx, \zdy, single top quark, diboson (including Higgs to diboson final states), rare backgrounds, and QCD multijets. 
Each of these groups of processes is associated with a set of normalization parameters, which are determined directly from data, 
primarily using the effectively signal-free CR bins. 
The number of parameters for each group of processes depends on their relative importance to the analysis, 
and how well they can be constrained by CRs. 
The dominant backgrounds, \Wjets and \ttx, have a separate scale factor for each lepton and S jet multiplicity category. 
The QCD background also has normalization factors, which are split by number of leptons and S jets. Intermediate 
backgrounds, \zdy, diboson + Higgs boson, and single top quark, have one factor per lepton multiplicity. 
The triboson/rare backgrounds group has a single scale factor that maps to all subprocesses. 
The result is that major backgrounds have independent normalizations, constrained by data control regions, for 
each lepton and S jet multiplicity, allowing for data/simulation discrepancies specific to the modeling of that process to be evaluated from data without assumptions about how they could appear in different categories.  

Kinematic requirements on \ptisr and \gperp are used to define categorizations expected to result 
in signal enriched and background dominated regions at higher and lower values, respectively. The 
analysis applies data-driven, kinematic category scale factors which are common to all background processes, with independent factors for \ptisr and \gperp categories that are also independently determined for different lepton and S jet multiplicities.

Lepton flavor, charge, quality, and configuration provide 
some of the most powerful types of categorization included in the analysis.
There are scale factors modeled as common to all the processes in the analysis, which can account for data/simulation discrepancies in their relative rates that affect all processes in similar ways. This could follow from, for example, higher-order correlations between hadronic activity and lepton isolation, or mismodeling of event kinematics that modifies how leptons are clustered into hemispheres. 

The analysis additionally measures scale factors for both electrons and muons 
that modify the rates of lepton candidates coming from either heavy-flavor (HF) decays or misidentified/nonprompt leptons from non-heavy-flavor (LF) decays, and also 
for changes in the \mperp/\risr distributions for processes associated with these lepton sources. One set of factors scales all processes with an associated lepton, while two additional sets are specific to the lepton isolation and impact parameter requirements that define the various regions.  In addition to normalization parameters, the fit also includes nuisance parameters that can change distribution shapes for nonprompt lepton backgrounds. Generic variations in  \risr and \mperp are parametrized in ``up'' and ''down'' templates  calculated separately for \risr and \mperp bins, where each bin is multiplied by a sliding fraction, which has the effect of skewing the kinematic distributions higher or lower.  A nominal prior of 5\% maximum variation is applied, leading to up to 10\% relative variations in the highest and lowest \mperp/\risr bins in a given category. These shape variations are implemented with separate factors for each lepton flavor and independently for each lepton and S jet multiplicity category. These constraints rely on the assumption that any such kinematic discrepancies in modeling of these background processes should be largely independent of the lepton quality, such that the bronze regions are used to predominantly constrain shape parameters common with gold for a given process.

To take into account uncertainties appearing due to the assignment of  \PQb-tagged jets to the S or the ISR system, there are scale factors to specifically account for data/simulation differences in relative categorization frequencies, in addition to the constrained efficiency scale factors previously described. 
Top quark and non-top-quark backgrounds are modeled with separate scale factors. For the SVs, signal events will tend to have more central ($\abs{\eta}\leq 1.5$) SV candidates resulting from real bottom or charm quarks, so scale factors are defined independently for central and forward SV categories. 

The background fit model was studied using a full CR fit (also split by data-taking period), a fit 
also including all bronze lepton categories, and 
finally in a fit to all of the bins and categories included in the analysis over 
the entire data set. 
The quality of these background-only fits was primarily assessed by considering 
the distribution of data/fit-model residuals in each bin in 
the analysis. Each fit is observed to give a reasonably consistent description of event yields in data.

The fit quality for the full fit to the 
entire dataset is assessed more quantitatively by considering the 
distribution of the post-fit tail probability computed for each fit bin.
The significance of 
an excess or deficit per bin is evaluated using the one-sided 
upper-tail Poisson probability. This probability is corrected\footnote{The corrected upper-tail probability sum reads 
as 
$p(n_{\text{obs}}; \mu_{\mathrm{b}}) = \sum_{n=n_{\text{obs}}}^{\infty} \mathrm{Po}(n; \mu_{\mathrm{b}} ) - \frac{1}{2} \mathrm{Po}(n_{\text{obs}}; \mu_{\mathrm{b}})$, where $n_{\text{obs}}$ events are observed, $\mu_{\mathrm{b}}$ is the expected number of background events, and Po represents the Poisson distribution.} for bias associated with double-counting $n=n_{\text{obs}}$  
such that the average expected probability for background-only is 0.5 and the corresponding 
one-sided upper- and lower-tail probabilities sum to unity.
The pseudo-data based evaluation integrates 
the upper-tail probability over the post-fit uncertainty in the background event 
yield per bin using a Gaussian posterior model; the method is closely related to the $Z_{\mathrm{N}}$ procedure 
of Ref.~\cite{Cousins:2007yta}. 
The resulting background-averaged probability (pseudo $p$-value) for 
consistency with the background-only hypothesis is then expressed in signed 
Gaussian quantiles as a z-score.
 
\begin{figure}[ht]
\centering
\includegraphics[width=0.49\textwidth]{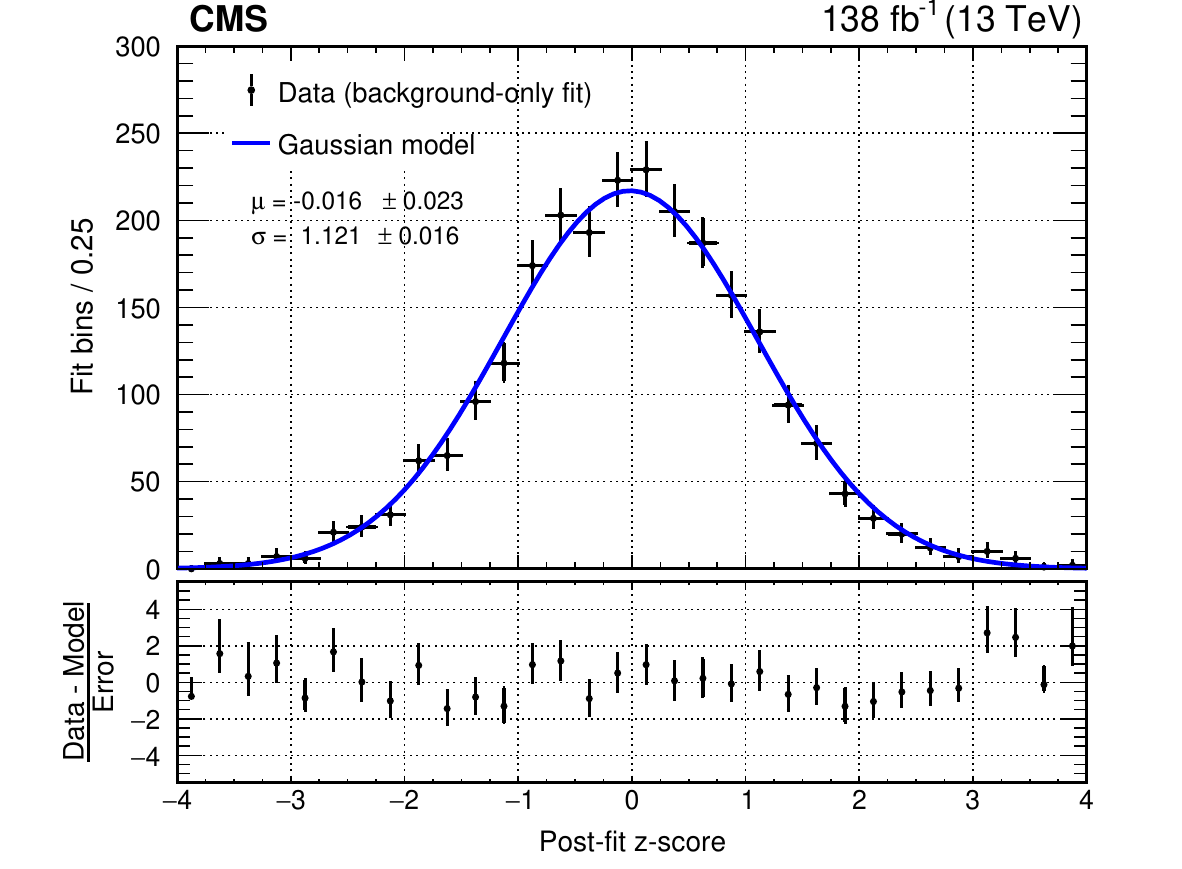}
\caption{Distribution of post-fit z-scores for the full data set 
background-only fit. The superimposed Gaussian model uses the observed mean and 
standard deviation.}
\label{fig:unblind_summ_pull}
\end{figure}

For the considered data set, under the simplifying assumptions of independent 
bins and neglecting the small expected reduction in variance associated with 
the fit to data, this post-fit z-score distribution is expected to 
be approximately Gaussian with a mean of zero and a standard deviation of 1.0.
The distribution of the observed post-fit z-scores per bin is shown 
in Fig.~\ref{fig:unblind_summ_pull} and is compared with the Gaussian 
distribution inferred from the sample mean and standard deviation.
The observed distribution is consistent with a Gaussian having a mean of zero 
and a standard deviation 12\% larger than unity. The characteristics 
of the ten most outlying bins have been examined; 
it is found that all but one is a CR bin.

\section{Results}\label{sec:results}
\begin{figure*}
  \centering
  \includegraphics[width=0.95\textwidth]{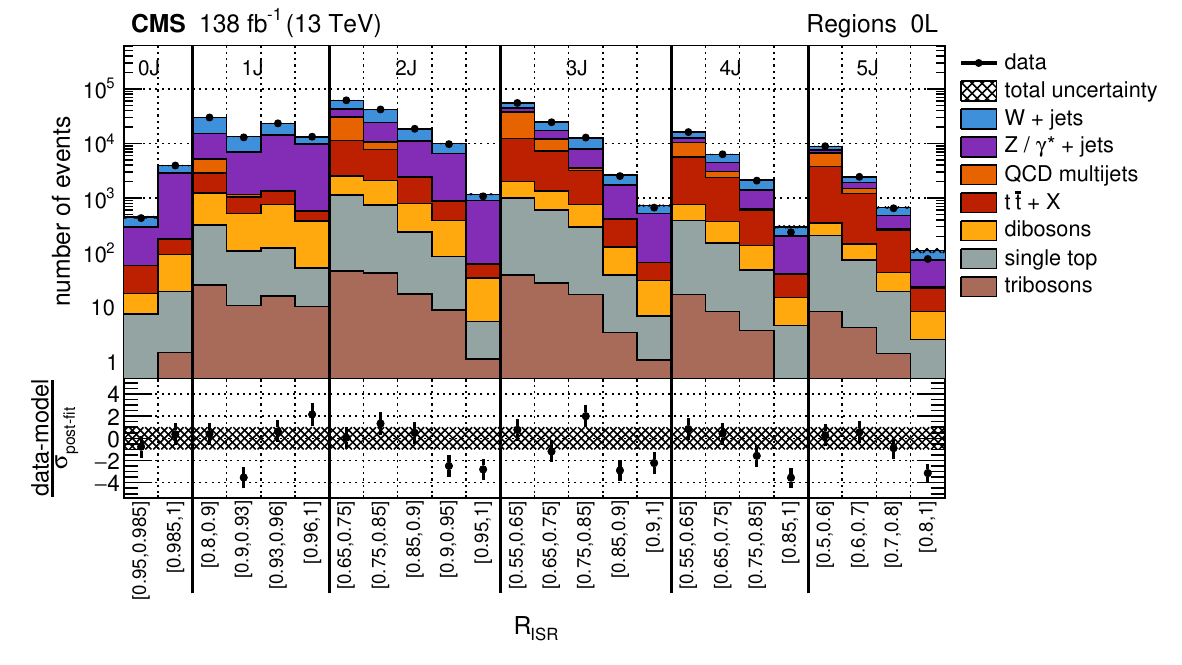}
  \includegraphics[width=0.95\textwidth]{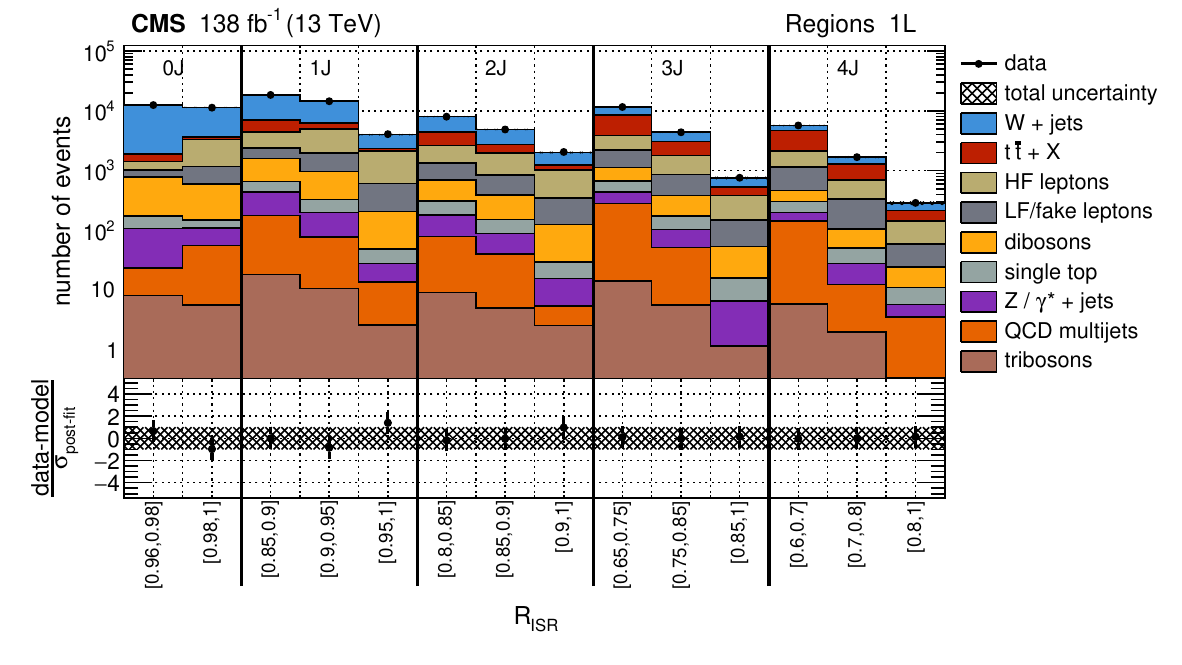}
  \caption{Post-fit distributions of data with the background-only fit model for the full data set in the 0L region (upper) and 1L region (lower). 
  Bins are split by \risr along with \NjetS.  
  Yields are integrated over all other sub-categorizations and \mperp. The sub-panels below the panels show the data minus fit model scaled by the post-fit model uncertainty. This uncertainty neglects correlations among the individual fitted bin event yields.}
  \label{fig:0L1LJetSummary}
\end{figure*}

\begin{figure*}
	\centering
  \includegraphics[width=0.95\textwidth]{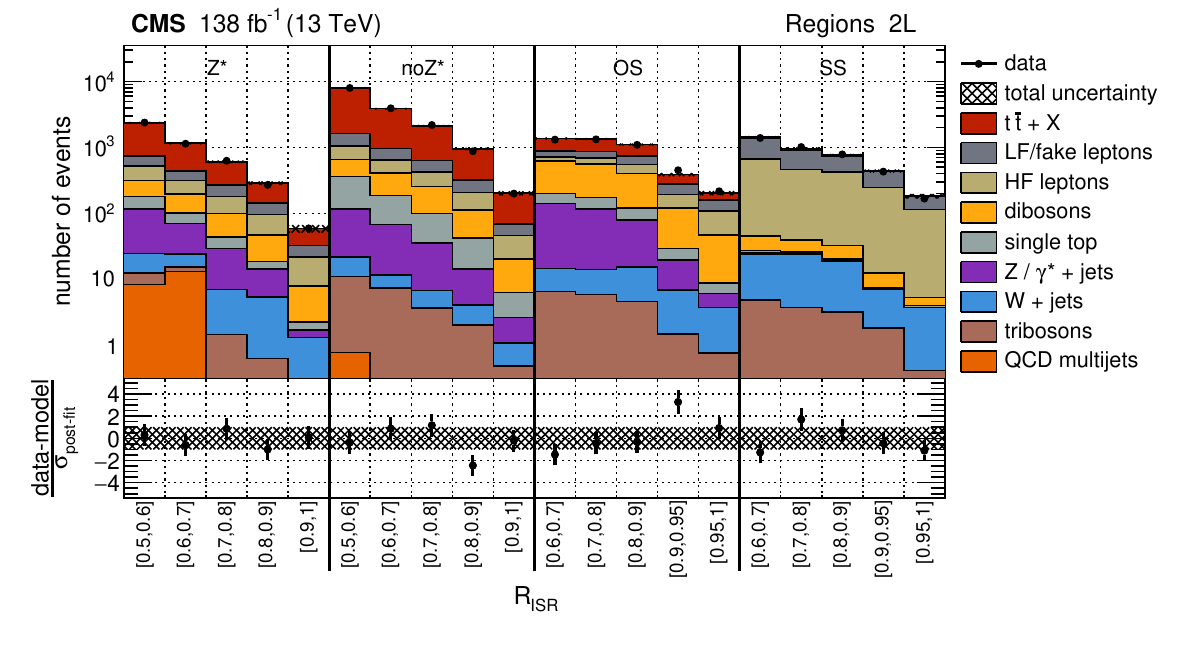}
	\includegraphics[width=0.95\textwidth]{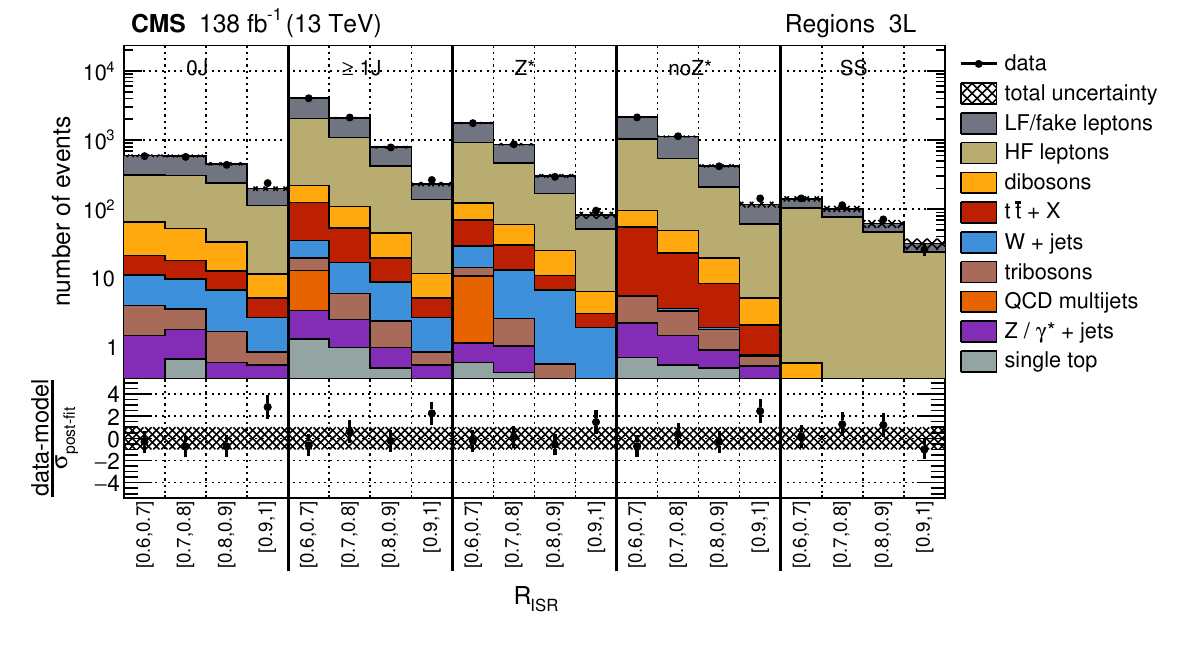}
	\caption{Post-fit distributions of data with the background-only fit model for the full data set in the 2L region (upper) and 3L region (lower). Bins are split by \risr along with lepton categorization. Yields are integrated over all other sub-categorizations and \mperp. The sub-panels below the panels show the data minus fit model scaled by the post-fit model uncertainty. This uncertainty neglects correlations among the individual fitted bin event yields.}
  \label{fig:2L3LJetSummary}
\end{figure*}

The maximum likelihood fit over 2443 bins in 392 categories is performed using the full data set.
Event yields in data are observed to be in statistically good agreement with the background-only fit model 
within the uncertainties included in the fit model.
Summaries of these data yields, integrated over categories and bins of \mperp, are shown in Figs.~\ref{fig:0L1LJetSummary} and~\ref{fig:2L3LJetSummary} for the 0L, 1L, 2L, and 3L regions.

\begin{figure*}
  \centering
  \includegraphics[width=0.95\textwidth]{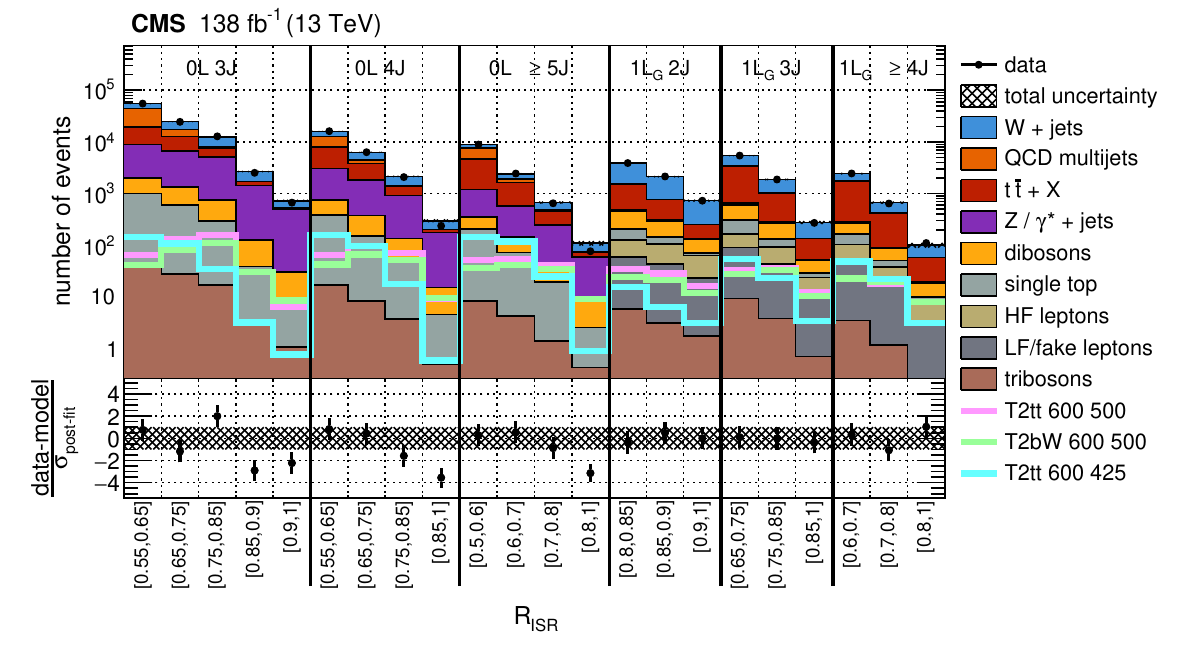}
  \includegraphics[width=0.95\textwidth]{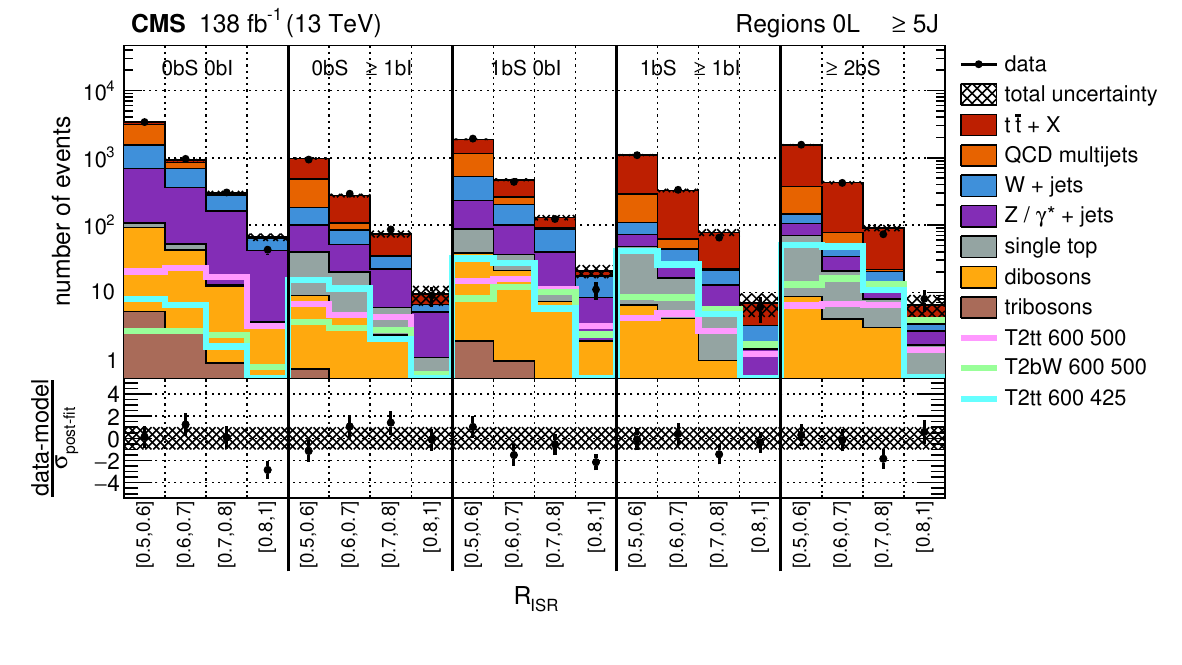}
  \caption{Post-fit distributions of data with the background-only fit model for the full data 
set. (Upper) 0L and 1L gold regions with larger jet multiplicities. (Lower) 0L 5J regions separated by $b$-tagged jet multiplicities in the S and ISR systems. Bins are split by \risr with yields integrated over all other sub-categorizations and \mperp.  The sub-panels below the panels show the data minus fit model scaled by the post-fit model uncertainty. This uncertainty neglects correlations among the individual fitted bin event yields. Expected yields for 
example signal models are superimposed.}
  \label{fig:Summary_1}
\end{figure*}

Event yields in data are compared to the background-only fit model for 0L and 1L final states in Fig.~\ref{fig:Summary_1} for categories with higher S jet multiplicities and also separated by  \PQb-tagged jet categorization.
These categories are designed to be particularly sensitive to signal models
with larger numbers of jets following from sparticle decays,
including heavy-flavor quarks following from intermediate top quark decays,
as is the case for sparticle production of top squarks and gluinos.
Such signals would appear as excesses in the high-\risr bins;
no such excesses are observed in this data set.

\begin{figure*}
  \centering
  \includegraphics[width=0.95\textwidth]{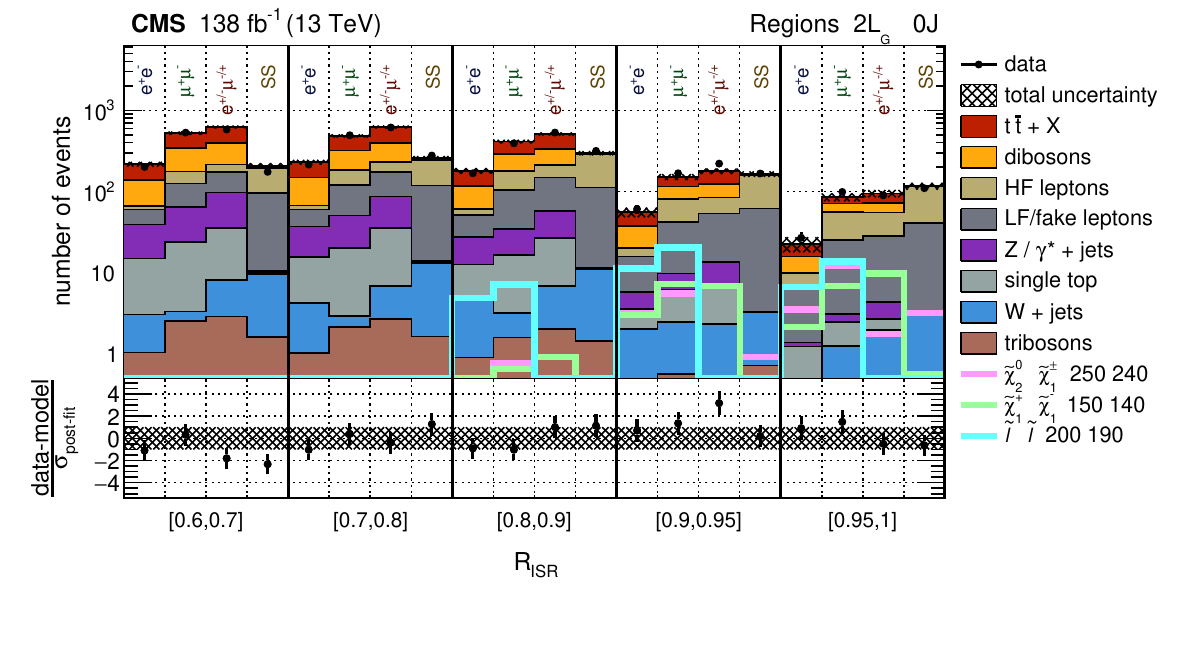}
  \includegraphics[width=0.95\textwidth]{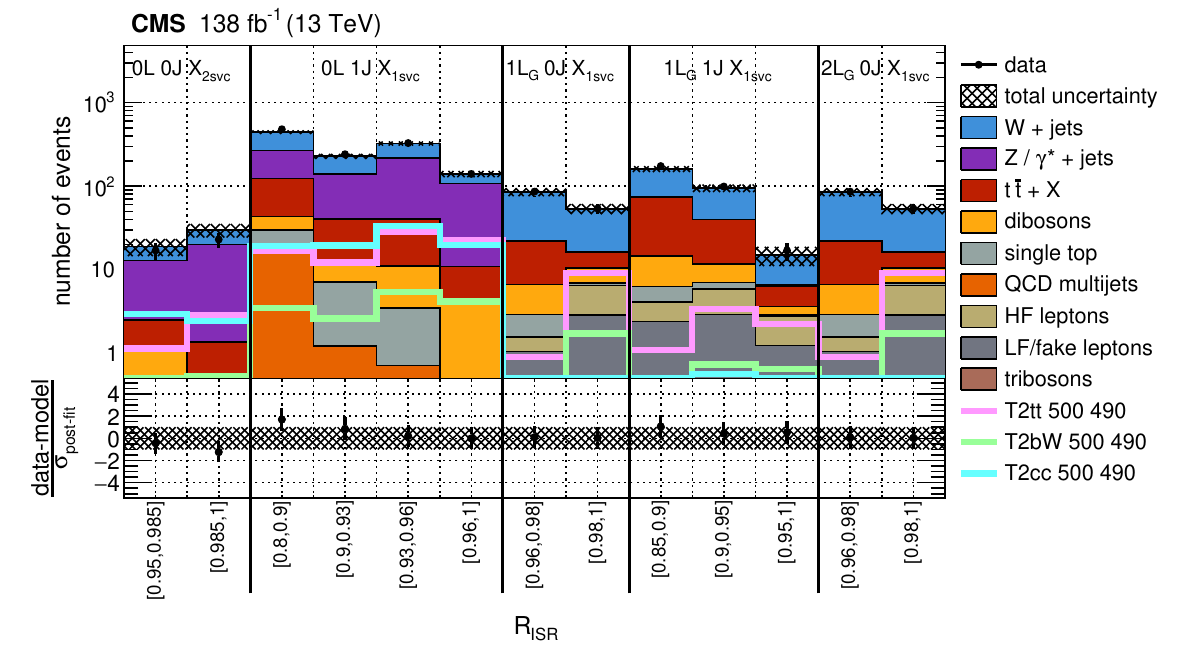}
  \caption{Post-fit distributions of data with the background-only model for the full data set. (Upper) 2L 0J gold regions separated 
  by lepton flavor and charge. (Lower) Central  \PQb-tagged SV regions in 0L, 1L, and 2L final states. Bins are split by \risr with yields integrated over all other sub-categorizations and \mperp. The sub-panels below the panels show the data minus fit model scaled by the post-fit model uncertainty. This uncertainty neglects correlations among the individual fitted bin event yields. Expected yields for example signal models are superimposed.}
  \label{fig:Summary_2}
\end{figure*}

Similarly, comparisons of data and background-only fit model event yields for regions with two gold leptons and no S jets and categories containing one or more central SV candidates are shown in Fig.~\ref{fig:Summary_2}.
Overall, no significant deviations between the background model and data are observed, particularly those consistent with patterns expected from the presence of signals with electroweakinos, sleptons, or top squarks. 
 
\begin{figure*}
  \centering
  \includegraphics[width=0.95\textwidth]{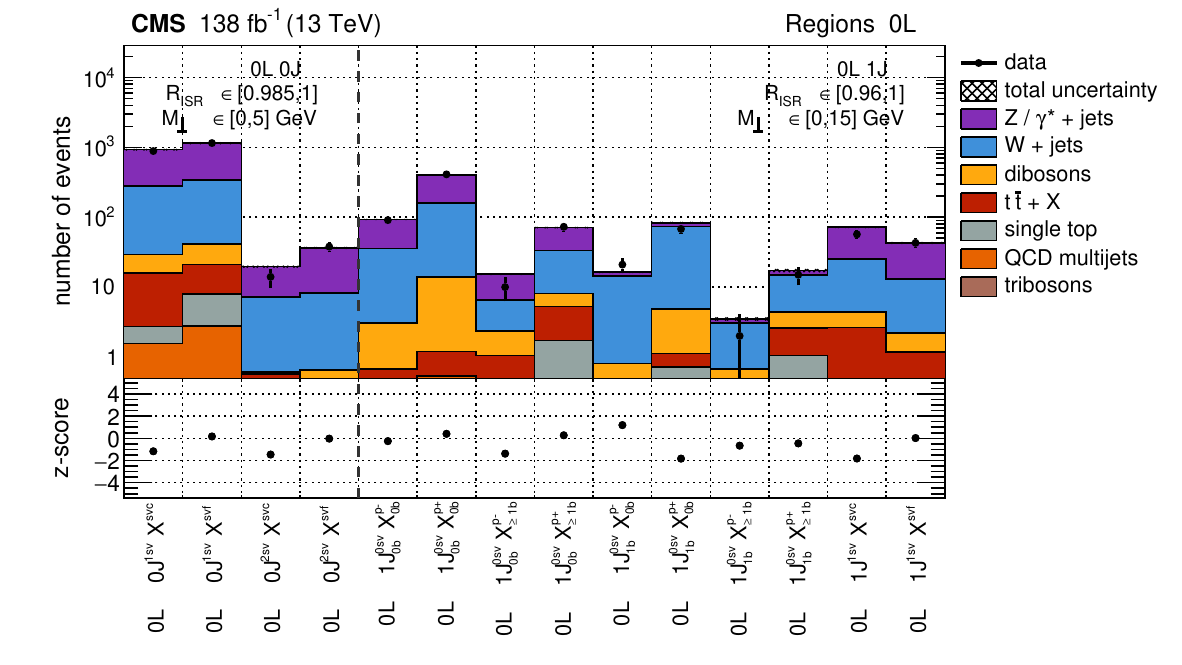}
  \includegraphics[width=0.95\textwidth]{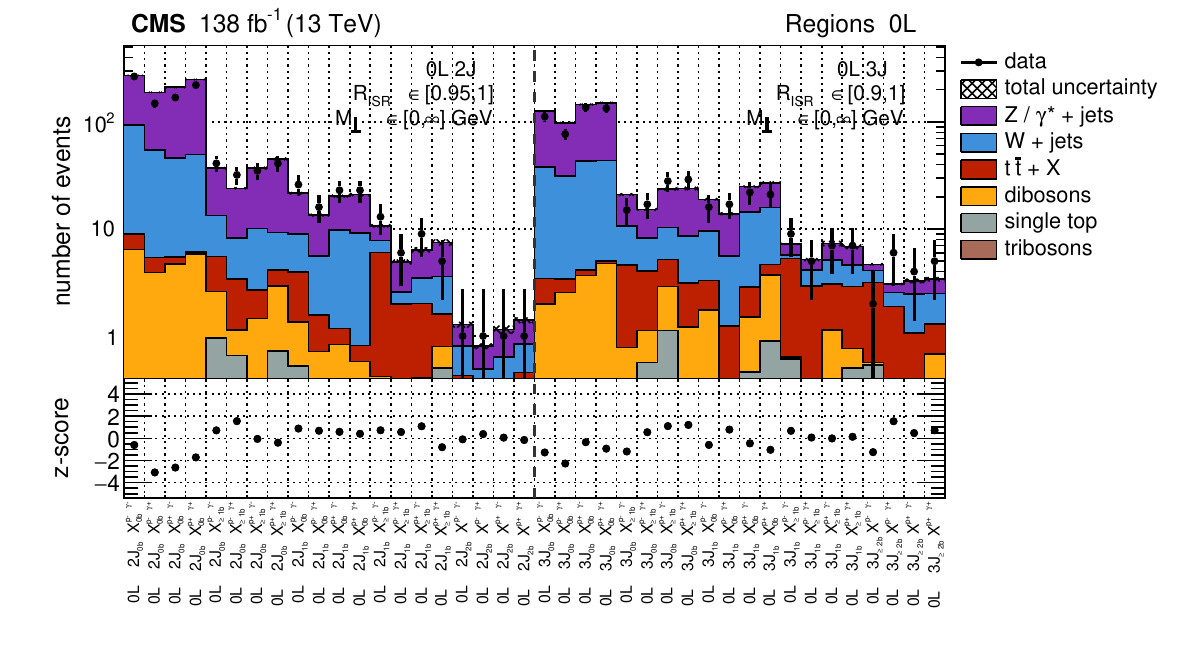}
  \caption{Post-fit distributions of data with the background-only
    model for the full data set for the highest \risr bin in
    each analysis category.  (Upper) 0L 0J and 1J regions. (Lower) 0L
    2J and 3J regions. The sub-panels below the panels indicate the
    post-fit z-score for each bin.}
  \label{fig:stackZscore0}
\end{figure*}

\begin{figure*}
  \centering
  \includegraphics[width=0.95\textwidth]{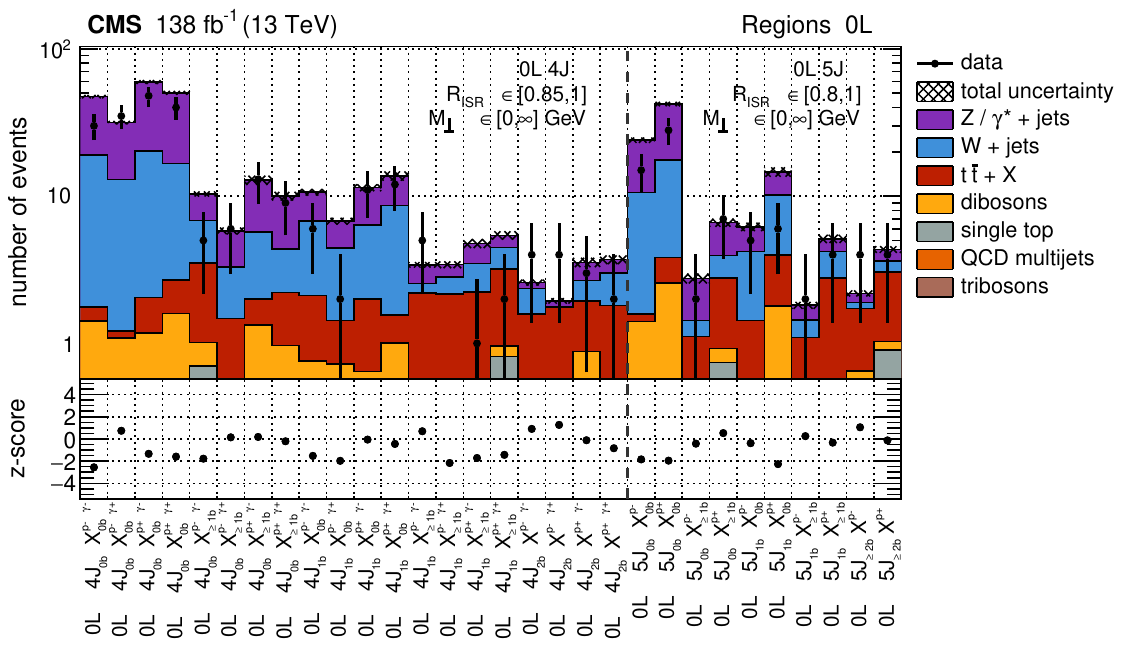}
  \includegraphics[width=0.95\textwidth]{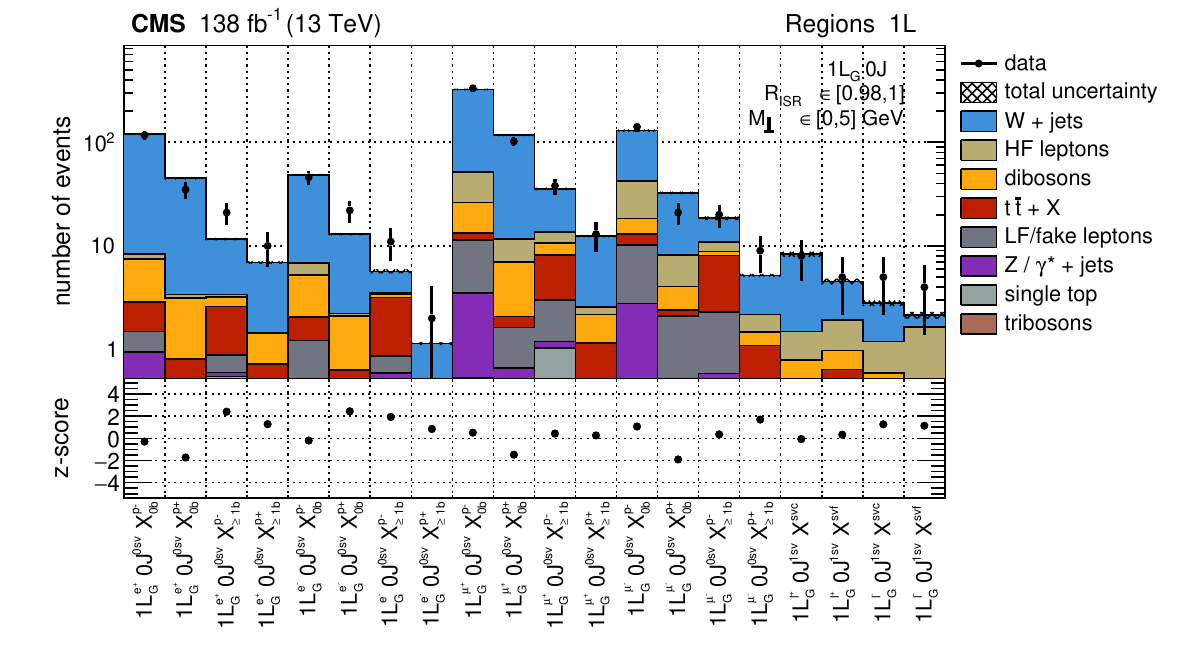}
  \caption{Post-fit distributions of data with the background-only
    model for the full data set for the highest \risr bin in
    each analysis category. (Upper) 0L 4J and $\ge$5J
    regions. (Lower) 1L 0J regions with a gold lepton. The sub-panels below the panels indicate the
    post-fit z-score for each bin.}
  \label{fig:stackZscore1}
\end{figure*}

\begin{figure*}
  \centering
  \includegraphics[width=0.95\textwidth]{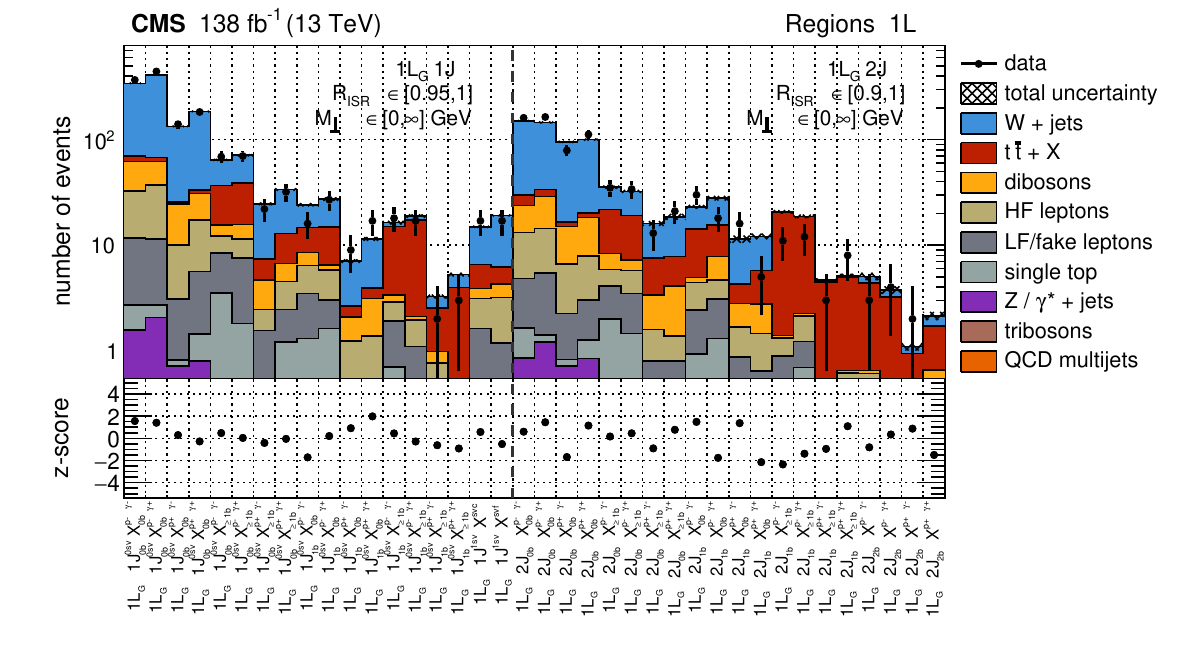}
  \includegraphics[width=0.95\textwidth]{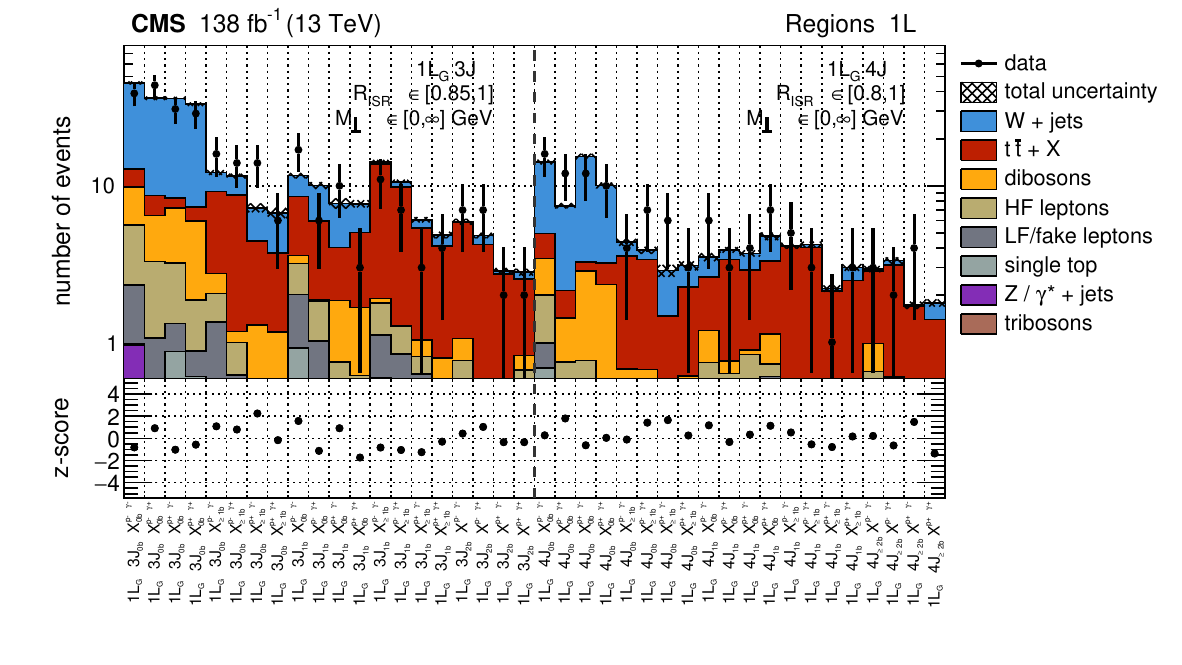}
  \caption{Post-fit distributions of data with the background-only
    model for the full data set for the highest \risr bin in
    each analysis category. (Upper) 1L 1J and 2J regions with a gold lepton. (Lower) 1L
    3J and $\ge$4J regions with a gold lepton. The sub-panels below the panels indicate the
    post-fit z-score for each bin.}
  \label{fig:stackZscore2}
\end{figure*}

\begin{figure*}
  \centering
  \includegraphics[width=0.95\textwidth]{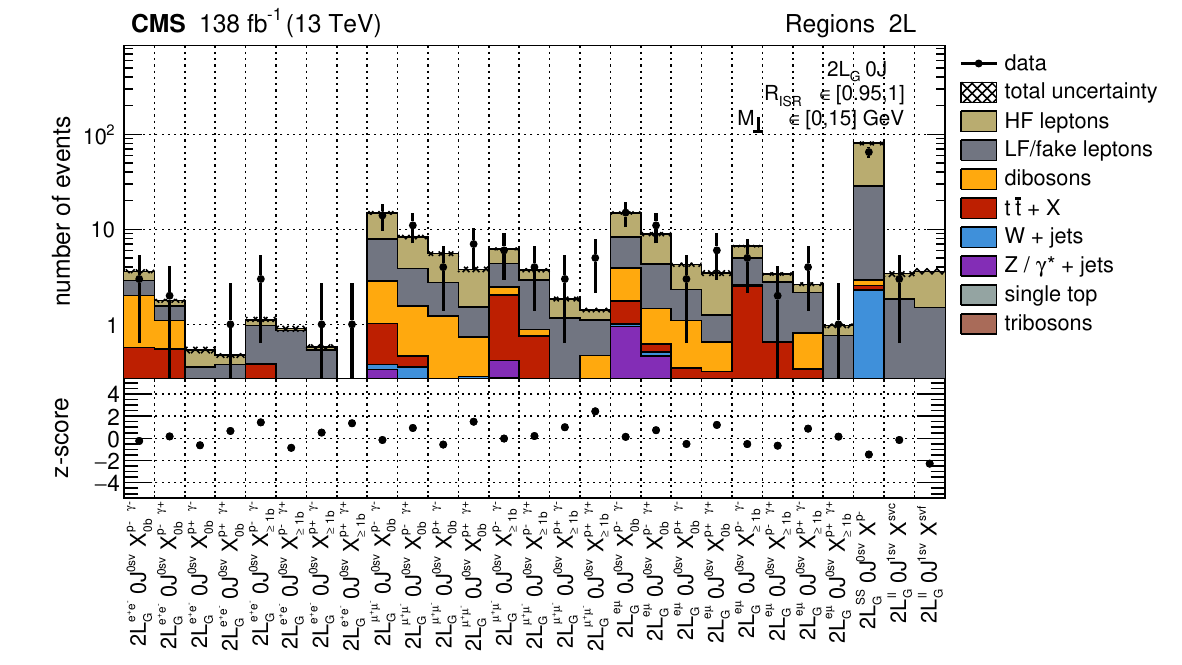}
  \caption{Post-fit distributions of data with the background-only
    model for the full data set for the highest \risr bin in
    each analysis category of the 2L 0J regions with gold leptons. The sub-panels below the panels indicate the
    post-fit z-score for each bin.}
  \label{fig:stackZscore3}
\end{figure*}

\begin{figure*}
  \centering
  \includegraphics[width=0.95\textwidth]{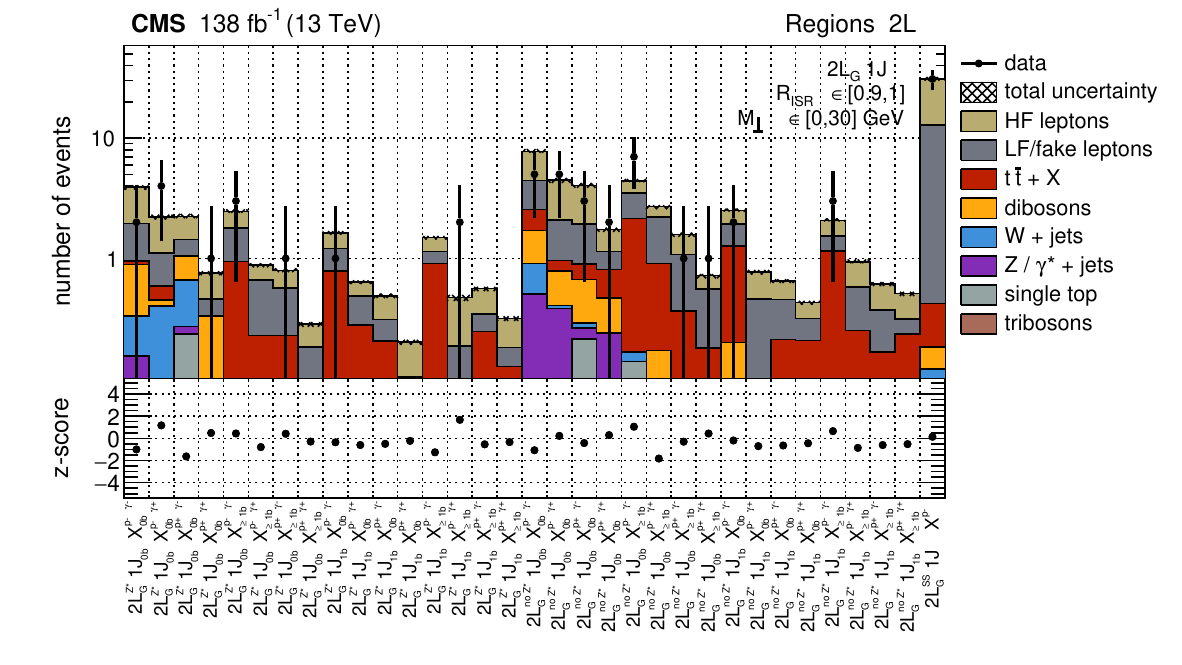}
  \includegraphics[width=0.95\textwidth]{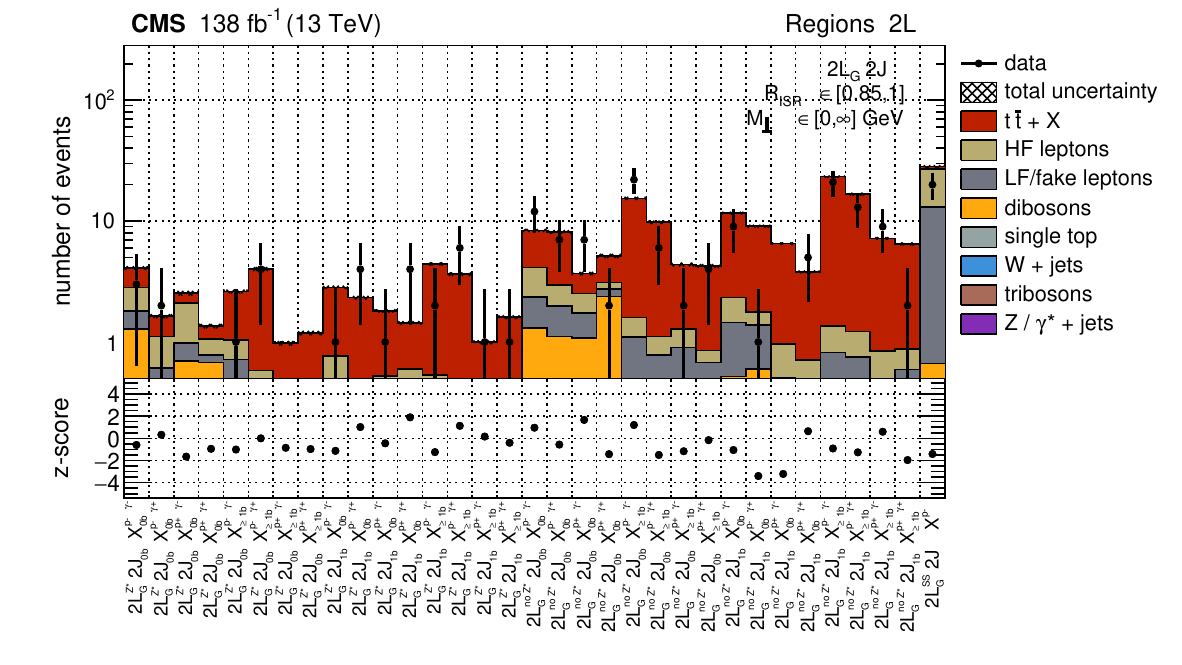}
  \caption{Post-fit distributions of data with the background-only
    model for the full data set for the highest \risr bin in
    each analysis category. (Upper) 2L 1J regions with gold
    leptons. (Lower) 2L $\ge$2J regions with gold leptons. The sub-panels below the panels indicate the
    post-fit z-score for each bin.}
  \label{fig:stackZscore4}
\end{figure*}

Event yields in data are also compared to the background-only fit model for all 0L, 1L, and 2L categories for the most compressed bin (the one with the 
highest \risr and lower \mperp) for the gold regions in Fig.~\ref{fig:stackZscore0} (0L), Fig.~\ref{fig:stackZscore1} (0L \& 1L), Fig.~\ref{fig:stackZscore2} (1L) 
and Figs.~\ref{fig:stackZscore3} and ~\ref{fig:stackZscore4} (2L)  
amounting to 294 independent bins. Each panel also shows the observed post-fit z-score for the bin. Again,
the data and the background-only fit model are generally in reasonable agreement and 
with no significant excess over the background-only fit model that could be an indication for a compressed signal.

We proceed to interpret results as constraints on potential signals using both model-dependent and model-independent interpretations.
In order to constrain signal models outside of the collection considered in this analysis,
aggregations of signal-sensitive bins in different final state categories, or superbins (defined in Section~\ref{sec:analysis}) are considered.
Upper limits at 95\% confidence level (\CL) for the signal strength
(\SUL) are calculated for each of these superbins using the modified
frequentist \CLs 
method~\cite{ref:ffit2, ref:ffit1}.
The distribution of the expected number of events in each superbin is evaluated from the generation of pseudo-experiments from the background-only fit model,
taking into account the posterior covariance of all nuisance parameters.
From these distributions,
the mean expected background, $N^{\text{pred}}_{\text{bkg}}$, 
standard deviation, $\sigma(N^{\text{pred}}_{\text{bkg}})$,
and signal event number upper limits, \SUL,
are calculated,
as summarized in Table~\ref{table-yields} for each of the eight superbins.

\begin{table}[h!t]
\centering
\topcaption{Event counts observed in data, $N_{\text{obs}}$, in each of the model-independent bins, compared with the mean expected background predictions 
from the CR fit, $N^{\text{pred}}_{\text{bkg}}$, their 
corresponding standard deviations, $\sigma(N^{\text{pred}}_{\text{bkg}})$, and the upper limits at 95\%~\CL 
on the signal strength (\SUL) in event counts. 
All superbins are mutually exclusive except the \PQb jets low-$\Delta m$ case which aggregates the \PQb jets low-$\Delta m$~1L, \PQb jets low-$\Delta m$~2L, and SV superbins.}
\begin{scotch}{l r r r r }
Region & $N_{\text{obs}}$ & $N^{\text{pred}}_{\text{bkg}}$ & $\sigma(N^{\text{pred}}_{\text{bkg}})$  & \SUL \\[\cmsTabSkip]
\hline 
\PQb jets low-$\Delta m$~1L   &  50 &  65.8 & 21.4 & 9.5\\ 
\PQb jets low-$\Delta m$~2L   &  16 &  10.3 &  3.9 &  14.6\\ 
SV                        &  38 &  37.2 &  8.5 & 17.8\\ [\cmsTabSkip]
\PQb jets low-$\Delta m$      & 104 & 115.5 & 22.3 & 16.2\\ [\cmsTabSkip]
\PQb jets moderate-$\Delta m$ &  83 & 108.1 & 18.2 & 9.9\\ 
Electroweak               &  26 &  30.2 &  5.5 & 9.6\\ 
2L OSSF                   &  12 &  10.2 &  4.5 &  9.9\\ 
3L                        &  21 &  25.2 &  5.0 &  9.0\\ 
\end{scotch}
\label{table-yields}
\end{table}
 
Model-dependent interpretations are calculated by performing signal-plus-background model 
fits for each considered scenario, independently for each simulated combination of sparticle 
masses. Using these signal-plus-background fits, along with the background-only fit to data, 
a profile likelihood ratio test statistic is used to evaluate upper limits on each of these model points. 
These are then interpolated among model points to fill in the complete sparticle mass plane.
The model-dependent results represent upper exclusion limits at 95\%~\CL on the product of the cross section 
and branching fractions for top squark, neutralino/chargino, and slepton processes. The asymptotic approximation 
of the modified frequentist approach is used to calculate these confidence levels 
with the profile likelihood ratio test statistic~\cite{ref:ffit2, ref:ffit1, ref:ffit3}.  

\begin{figure}[!ht]
  \centering
  \includegraphics[width=0.48\textwidth]{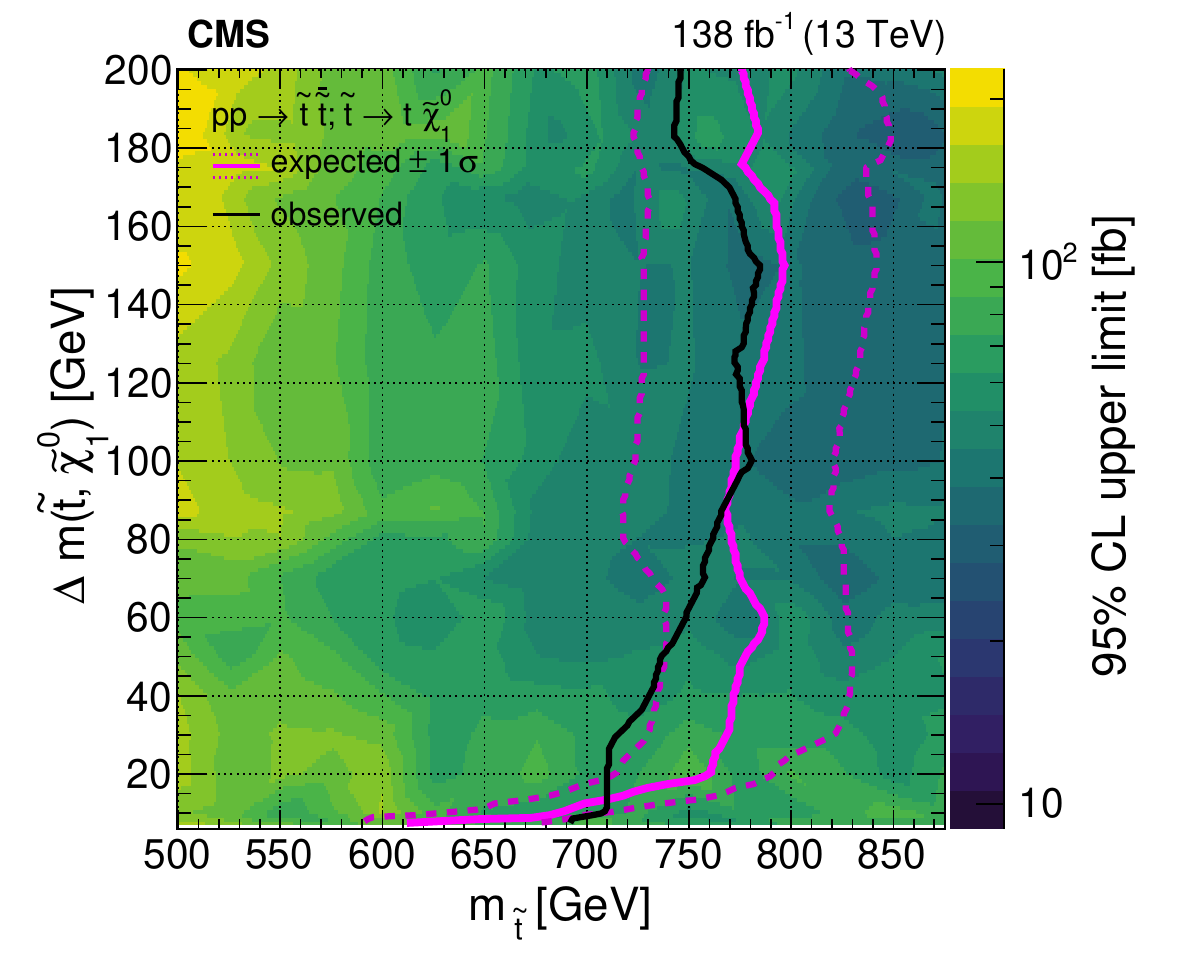}
  \includegraphics[width=0.48\textwidth]{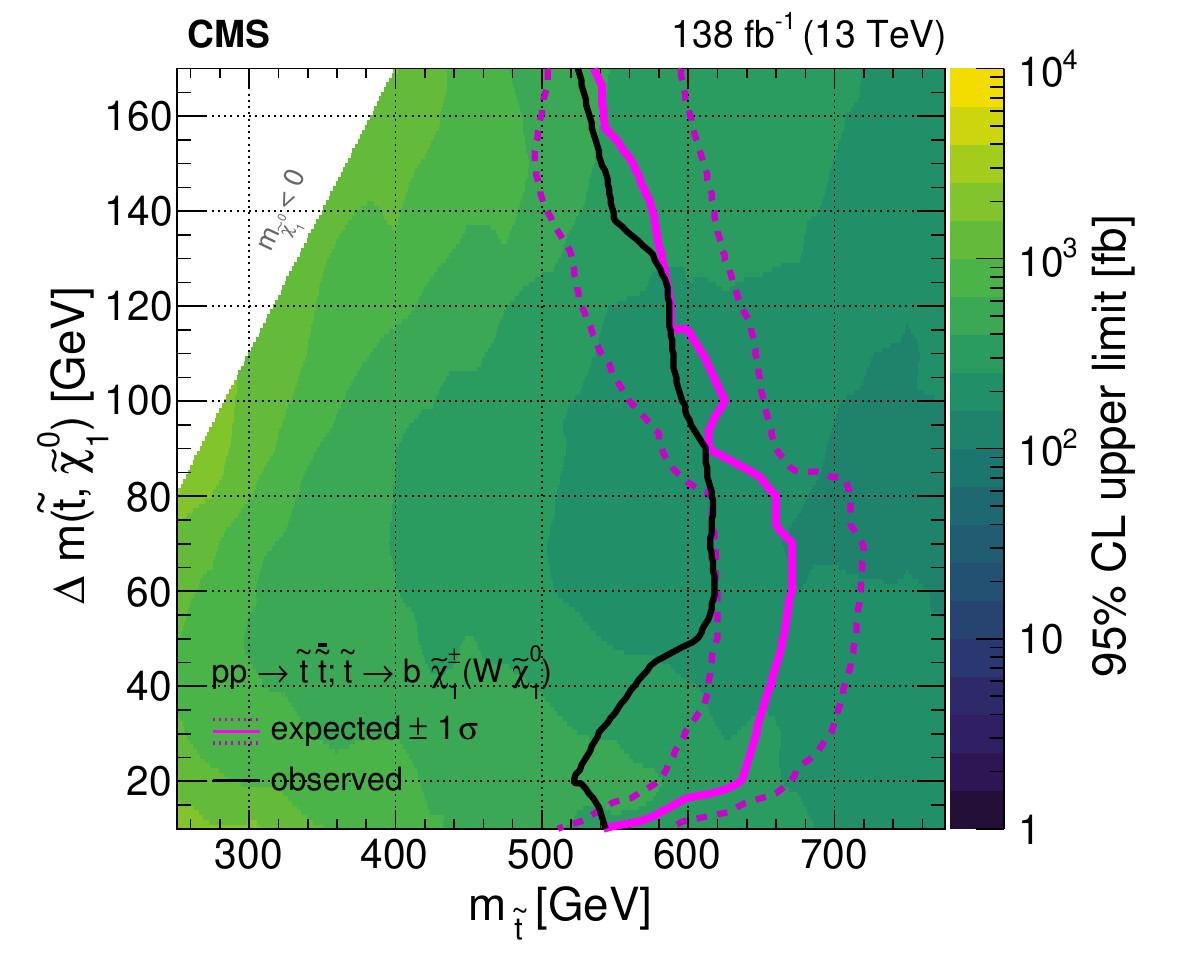}
  \caption{Top squark pair production. Observed upper limits at 95\%~\CL on the product of the cross section and relevant branching fractions 
 are shown using the color scale where the $\PSQt$ mass is on the $x$-axis and 
 the mass difference between the $\PSQt$ and the LSP 
 is on the $y$-axis. 
 The expected lower mass limits (magenta line) together with their $\pm 1\sigma$ uncertainties (magenta dashed lines) and 
 the observed lower mass limits (black line) are indicated for 100\% branching fractions. 
 The \cmsLeft panel shows the results for the T2tt model with limits on $\sigma (\PSQt \, \PASQt) \, \mathcal{B}^{2} (\PSQt \to \PQt \PSGczDo)$.
 The \cmsRight panel shows the results for the T2bW model with limits on $\sigma (\PSQt \, \PASQt) \, \mathcal{B}^{2} (\PSQt \to \PQb \PSGcpDo)  \mathcal{B}^{2} ( \PSGcpDo\to\PWp \PSGczDo ) $. 
  }
  \label{fig:t2tt_limits_2017v4}
\end{figure}

\begin{figure}[!ht]
  \centering
  \includegraphics[width=0.48\textwidth]{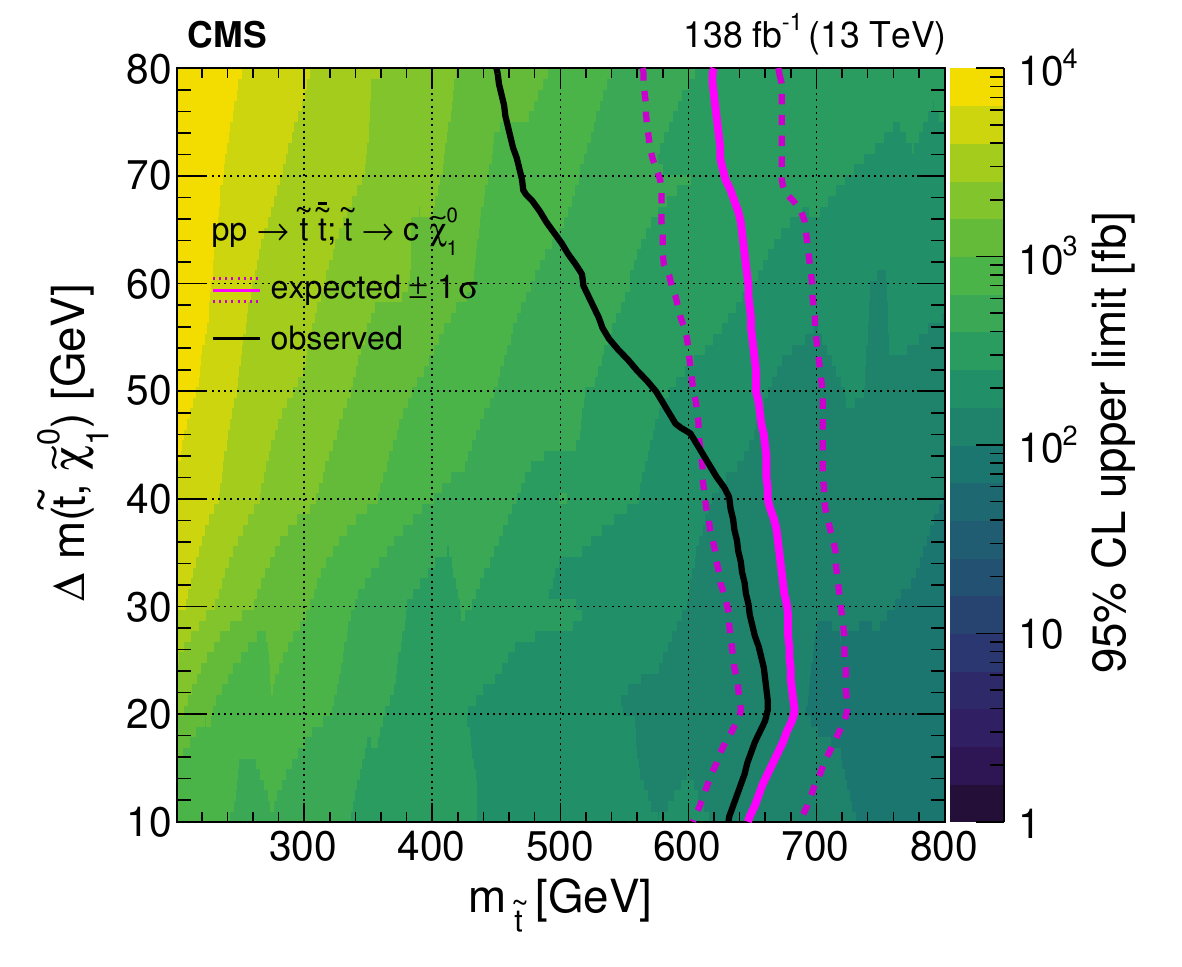}
  \includegraphics[width=0.48\textwidth]{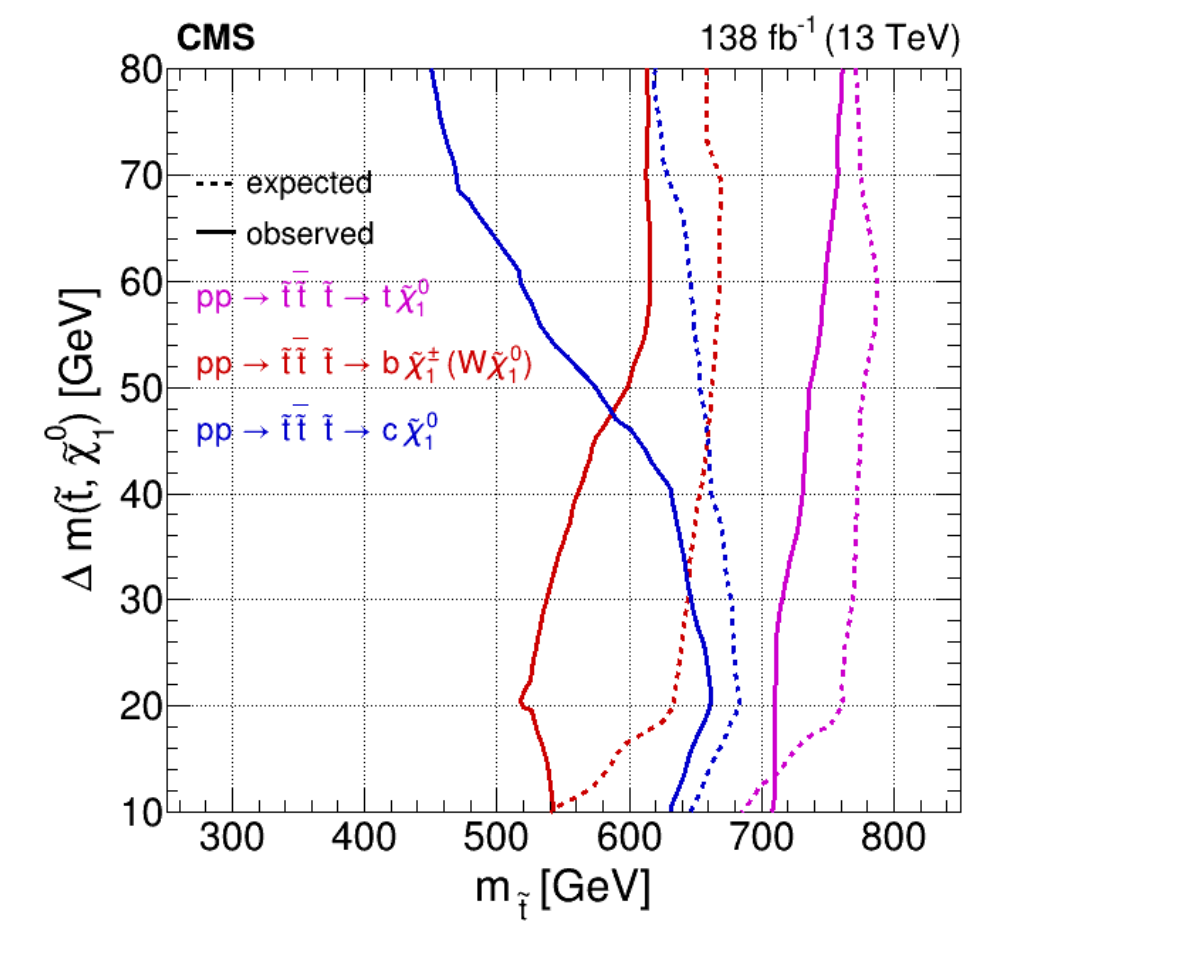}
  \caption{Top squark pair production. 
  Observed upper limits at 95\% \CL on the product of the cross section and branching fraction squared, $\sigma (\PSQt \, \PASQt) \, \mathcal{B}^{2} (\PSQt\to\PQc \PSGczDo )$ (\cmsLeft), 
 are shown using the color scale where the $\PSQt$ mass is on the $x$-axis and 
 the mass difference between the $\PSQt$ and the LSP 
 is on the $y$-axis. 
 The expected lower mass limits (magenta line) together with their $\pm 1\sigma$ uncertainties (magenta dashed lines) and 
 the observed lower mass limits (black line) are indicated for 100\% branching fractions. 
 Observed and median expected limits for top squark pair production at 95\%~\CL (\cmsRight) for the three decay modes investigated.
  }
  \label{fig:tt_limits_Summary}
\end{figure}

Figure~\ref{fig:t2tt_limits_2017v4} 
shows the limits found for  
top squark pair production with the decay $\PSQt \to \PQt \PSGczDo$ (the T2tt model) in the left panel 
and for $\PSQt \to \PQb \PSGcpDo$ with $\PSGcpDo \to \PWp \PSGczDo$ decay (the T2bW model) in the right panel.
Figure~\ref{fig:tt_limits_Summary} shows the limits found for top squark pair production with 
the decay $\PSQt \to \PQc \PSGczDo$ (the T2cc model). 
The T2tt limits are generally stronger than the previous limits from CMS and ATLAS for these compressed models 
with $\Delta m < 200$\GeV~\cite{Sirunyan:2020tyy,CMS:2021edw, CMSPRD.104.052001, Atlas2004.14060, Atlas2012.03799, ATLAS:2024rcx}.

\begin{figure}[!ht]
  \centering
  \includegraphics[width=0.48\textwidth]{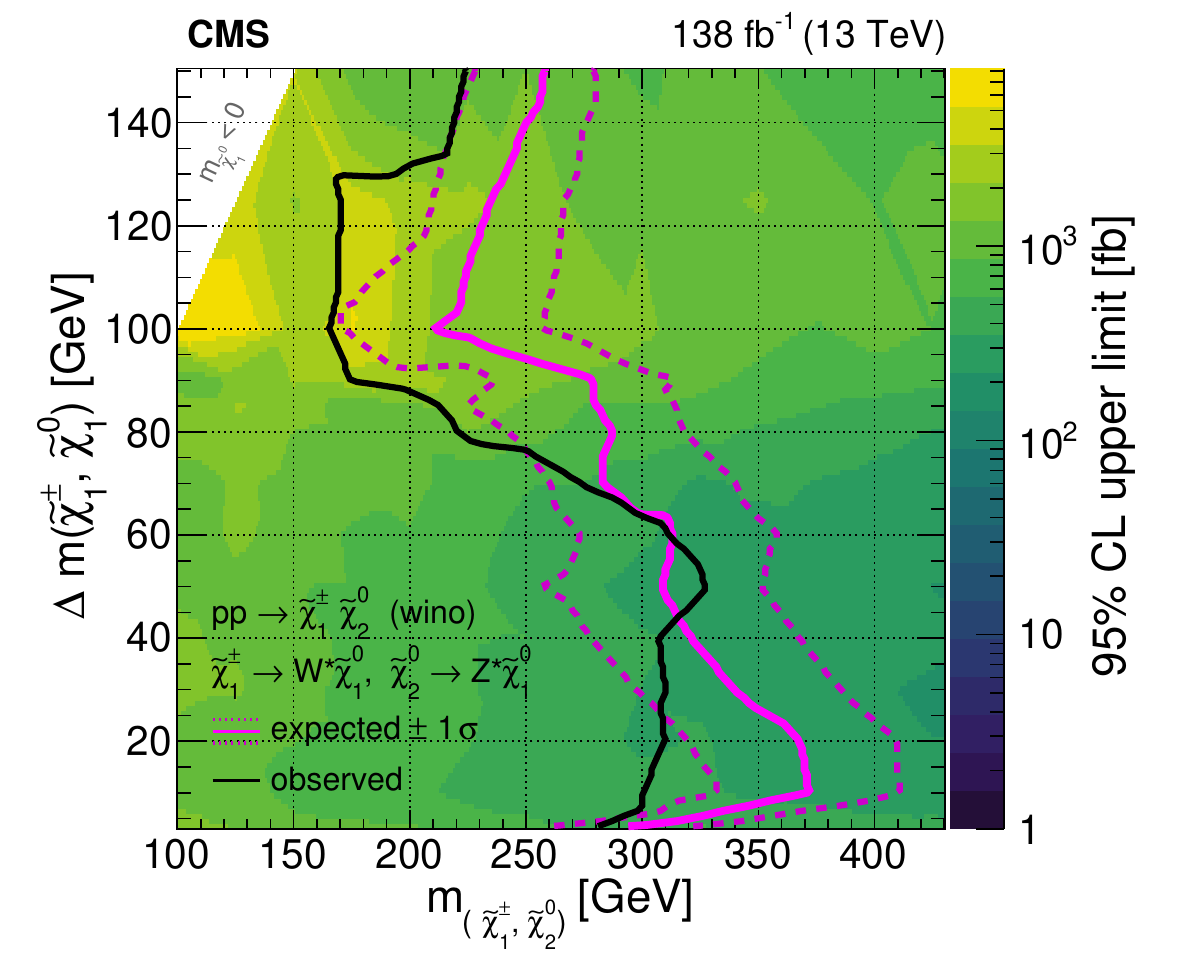}
  \includegraphics[width=0.48\textwidth]{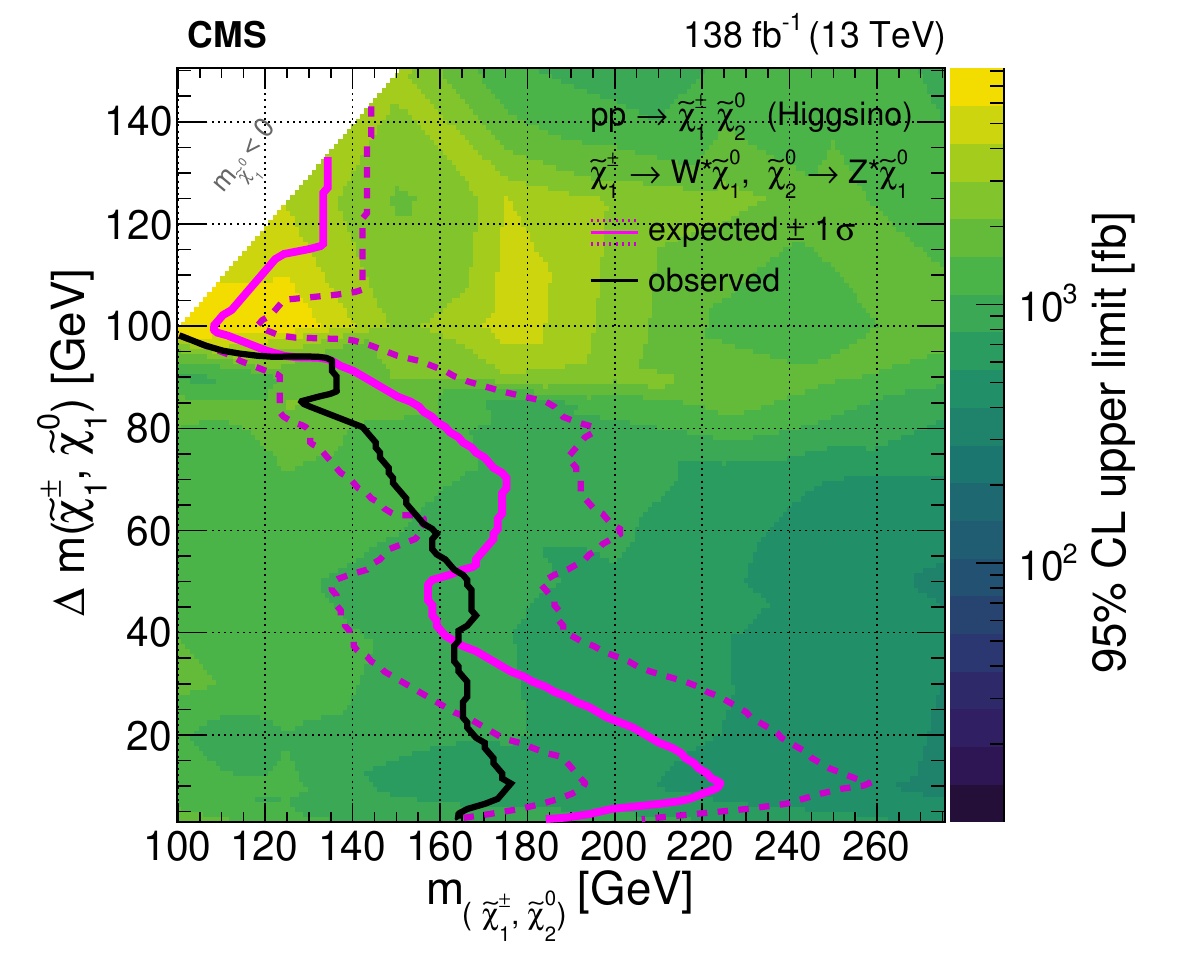}
  \caption{
  Chargino-neutralino production. Observed upper limits at 95\% \CL on the product of the cross section and
 the two branching fractions, $\sigma (\PSGcpmDo \PSGczDt) \, \mathcal{B} ( \PSGcpmDo \to \PWpm \PSGczDo ) \,  
  \mathcal{B} ( \PSGczDt \to \PZ \PSGczDo )$,  
 are shown using the color scale where the $\PSGcpmDo/\PSGczDt$ mass is on the $x$-axis and 
 the mass difference between the $\PSGcpmDo/\PSGczDt$ and the LSP 
is  on the $y$-axis. For these results, based on the TChiWZ simplified model, 
 the $\PSGcpmDo$ and $\PSGczDt$ masses are set equal.
 The expected lower mass limits (magenta line) together with their $\pm 1\sigma$ uncertainties (magenta dashed lines) and 
 the observed lower mass limits (black line) are indicated for 100\% branching fractions for 
 wino-like cross-sections (\cmsLeft) and for higgsino-like cross-sections (\cmsRight). 
 }
  \label{fig:tchiwz_limits_2017v4}
\end{figure}

Limits on the TChiWZ model are presented in Fig.~\ref{fig:tchiwz_limits_2017v4}, using both wino-like and higgsino-like cross sections. 
These limits are generally stronger than the previous limits from CMS and ATLAS for these compressed models with $\Delta m < 80$\GeV~\cite{Aaboud:2017leg, Sirunyan:2019zfq, Aad:2019qnd, ATLASPRD.103.112006, CMS:2024gyw}. 
For $3 < \Delta m < 50$\GeV, the observed 95\% \CL lower mass limit on higgsino-like 
chargino-neutralino production exceeds 163\GeV. For $8 < \Delta m < 65$\GeV, the observed 95\% \CL lower mass limit on wino-like 
chargino-neutralino production exceeds 300\GeV; this can be compared with the combined lower mass limit exceeding 200\GeV 
in this $\Delta m$ range in~\cite{CMS:2024gyw}, and a similar combined lower mass limit exceeding 215\GeV in~\cite{ATLAS:2021moa}.

\begin{figure}[!ht]
  \centering
  \includegraphics[width=0.48\textwidth]{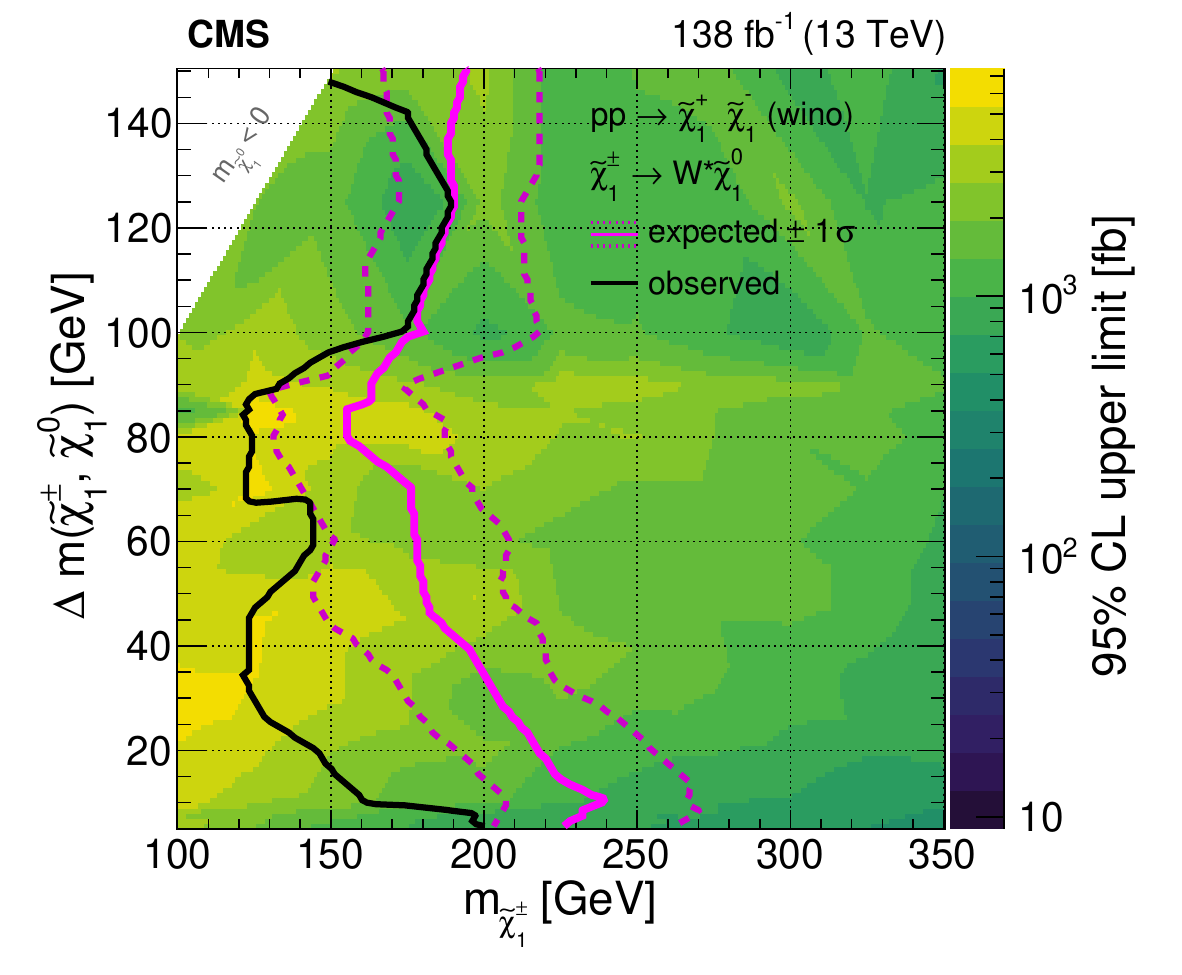}
  \includegraphics[width=0.48\textwidth]{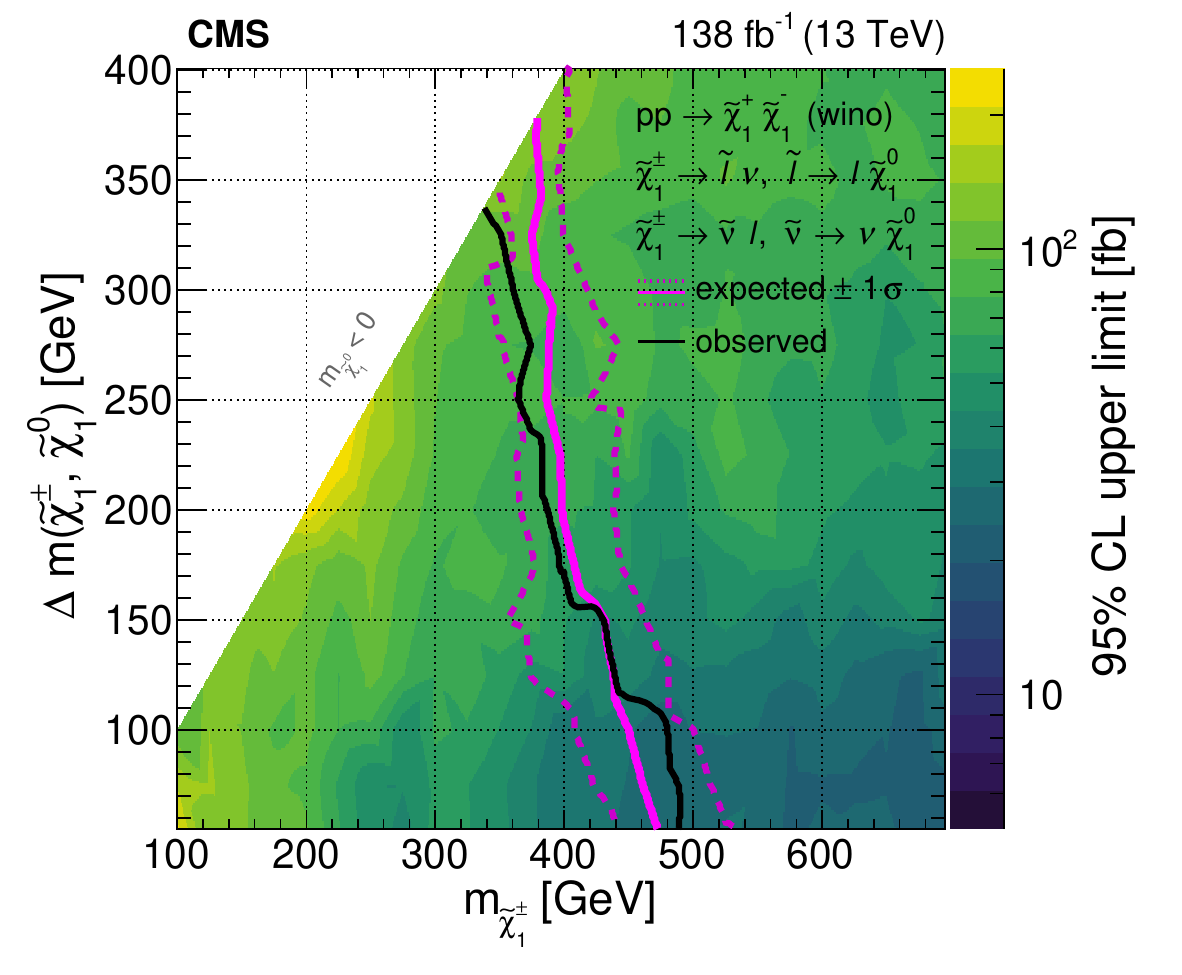}
  \caption{
 Chargino pair production. The \cmsLeft panel shows the observed upper limits at 95\%~\CL on the product of the cross section and 
 the branching fraction squared, $\sigma (\PSGcpDo \PSGcmDo) \, \mathcal{B}^{2} ( \PSGcpmDo \to \PWpm \PSGczDo ) $   
 are shown using the color scale where the $\PSGcpmDo$ mass is on the $x$-axis and 
 the mass difference between the $\PSGcpmDo$ and the LSP 
 is on the $y$-axis.
 The expected lower mass limits (magenta line) together with their ${\pm}1\sigma$ uncertainties (magenta dashed lines) and 
 the observed lower mass limits (black line) are indicated for  100\% branching fractions for 
 wino-like cross-sections. The \cmsRight panel shows the results for chargino pair production with decays as in the TChiSlepSnu model with 
 democratic decay via an intermediate sneutrino or charged slepton (\PSlLpm) with mass halfway between the 
 chargino and the lightest neutralino. These model predictions also assume wino-like cross sections.
 }
  \label{fig:tchiwwetc}
\end{figure}

Figure~\ref{fig:tchiwwetc} shows the experimental upper limits on chargino pair production 
for the decays associated with the TChiWW model and the TChiSlepSnu model. 
The TChiWW results exclude charginos with masses less than 120\GeV and mass differences exceeding 5\GeV at 95\% \CL 
for wino cross sections and $\mathcal{B} ( \PSGcpmDo\to\PWpm \PSGczDo ) = 1$. 
These TChiWW results on direct chargino pair production extend beyond the mass scales probed by the CERN LEP experiments that 
established 95\%~\CL chargino lower mass limits around 100\GeV generally also for wino-like couplings 
and mass differences exceeding 5\GeV~\cite{ALEPH:Chargino, DELPHI:2003uqw, L3:1999onh, OPAL:Chargino}. The 
generally applicable combined lower limit on the chargino mass from LEP is derived as 103.5\GeV~\cite{PDG2024}.
The TChiSlepSnu results are very competitive in the compressed mass regime reaching masses as high as 490\GeV using wino cross sections 
for this model that features favorable leptonic branching fractions. This complements other direct chargino pair 
results with the same decay assumptions such as~\cite{TChiSlepSnu} that probes to higher chargino masses but only for large mass splittings.

\begin{figure}[!ht]
  \centering
  \includegraphics[width=0.49\textwidth]{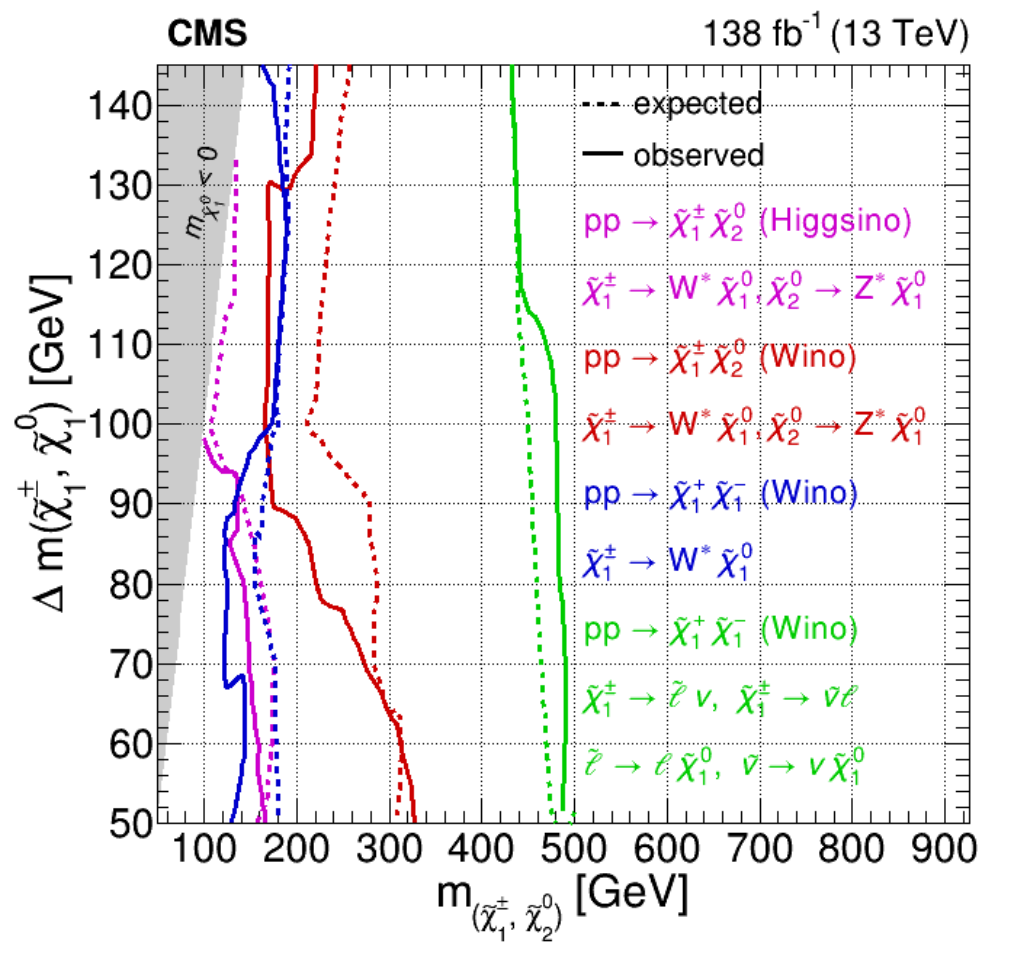}
  \includegraphics[width=0.48\textwidth]{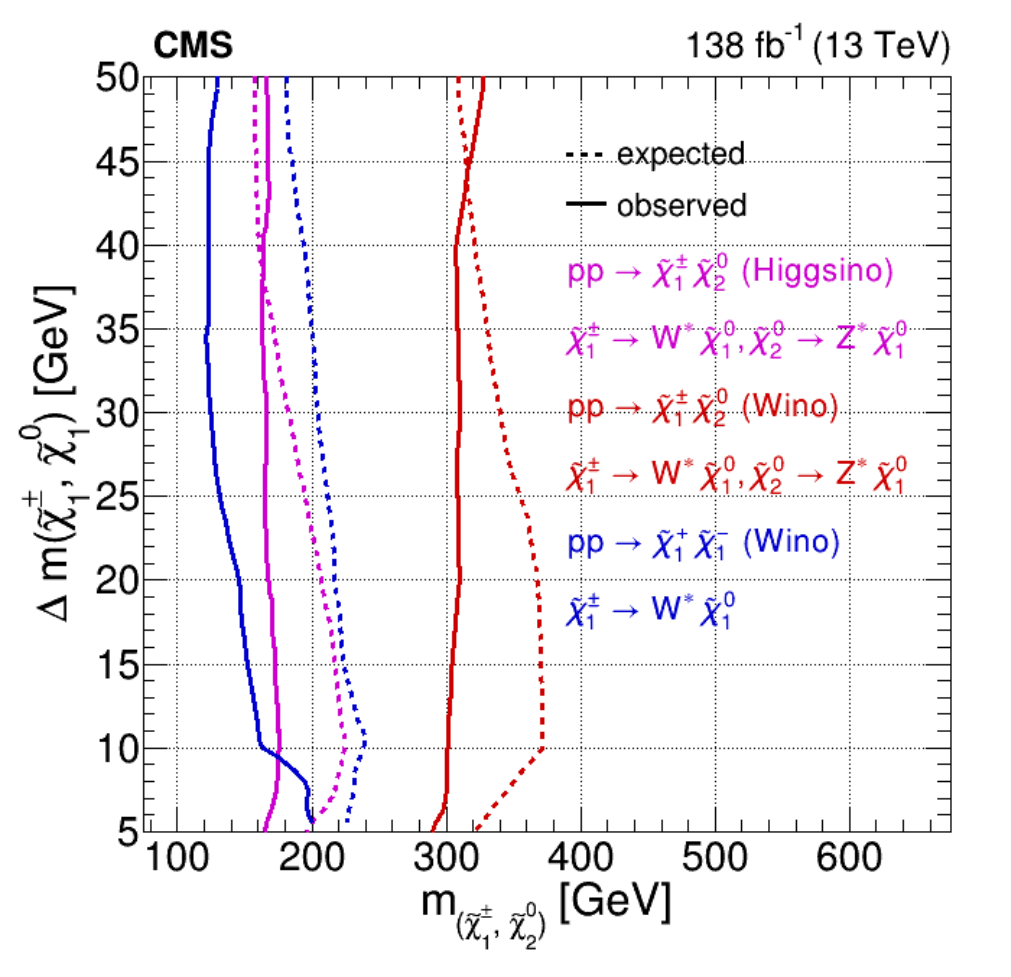}
  \caption{
 Summary of the model exclusion results on chargino-neutralino production and chargino pair production. Solid lines are 95\%~\CL observed limits 
 and dashed lines are the corresponding median expected limits. The \cmsLeft panel shows the results for mass differences exceeding 50\GeV and the \cmsRight panel 
 for mass differences below 50\GeV.
 }
  \label{fig:EWKinoSummary}
\end{figure}

The model exclusion results for chargino-neutralino production and chargino pair production 
are summarized in Fig.~\ref{fig:EWKinoSummary}.

\begin{figure}[!ht]
  \centering
  \includegraphics[width=0.49\textwidth]{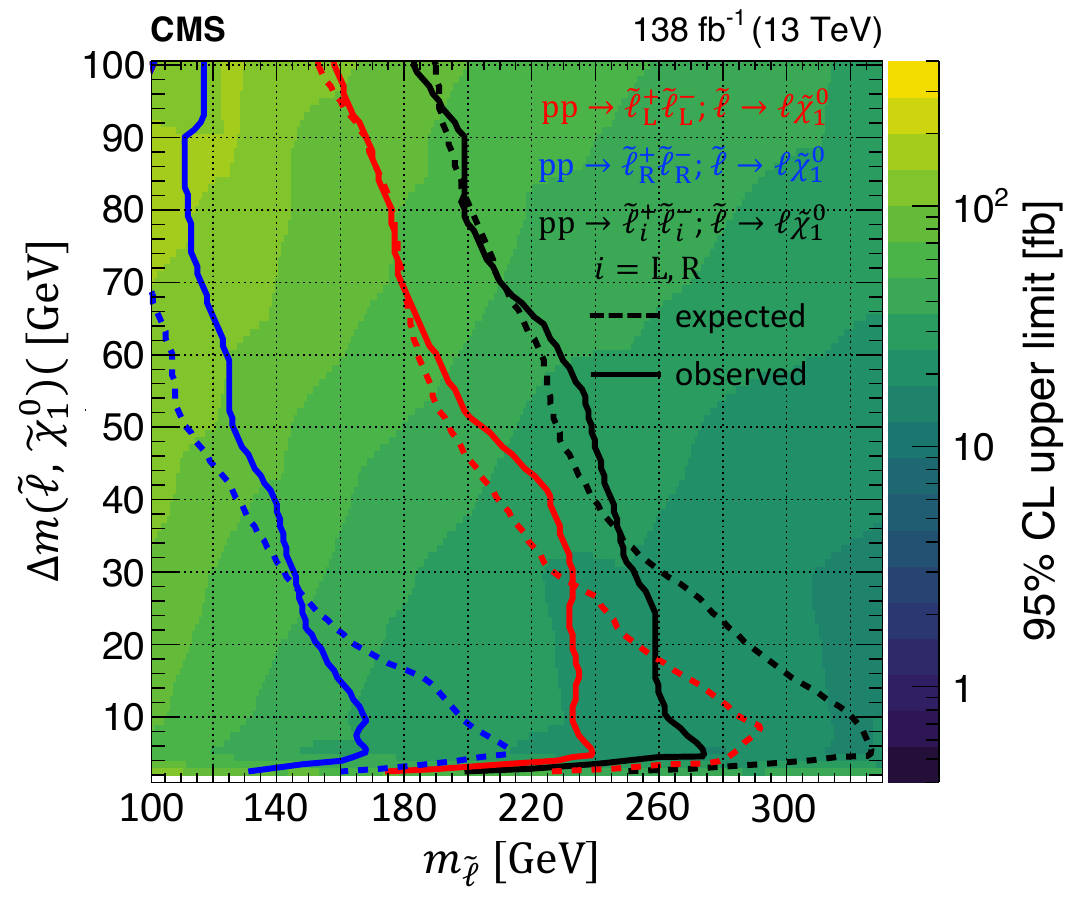}
  \caption{Slepton pair production. Observed 95\% \CL upper limits on the product of the cross section 
  and branching fraction squared for direct slepton pair production followed by decay of both sleptons to the corresponding lepton and neutralino (color scale).
  Slepton $\tilde{\Pell}_{\mathrm{L}/\mathrm{R}}$ indicates the scalar supersymmetric 
  partner of left- and right-handed electrons and muons. 
  The limit is shown as a function of the slepton mass and the mass difference 
  between the slepton and the lightest neutralino. 
  The regions to the left of the lines denote the regions excluded for a branching fraction of 100\%.
  The median expected exclusion regions for 100\% branching fraction 
  are delimited by the dashed lines.
  }
  \label{fig:slepton_xslimits}
\end{figure}

\begin{figure}[!ht]
  \centering
  \includegraphics[width=0.48\textwidth]{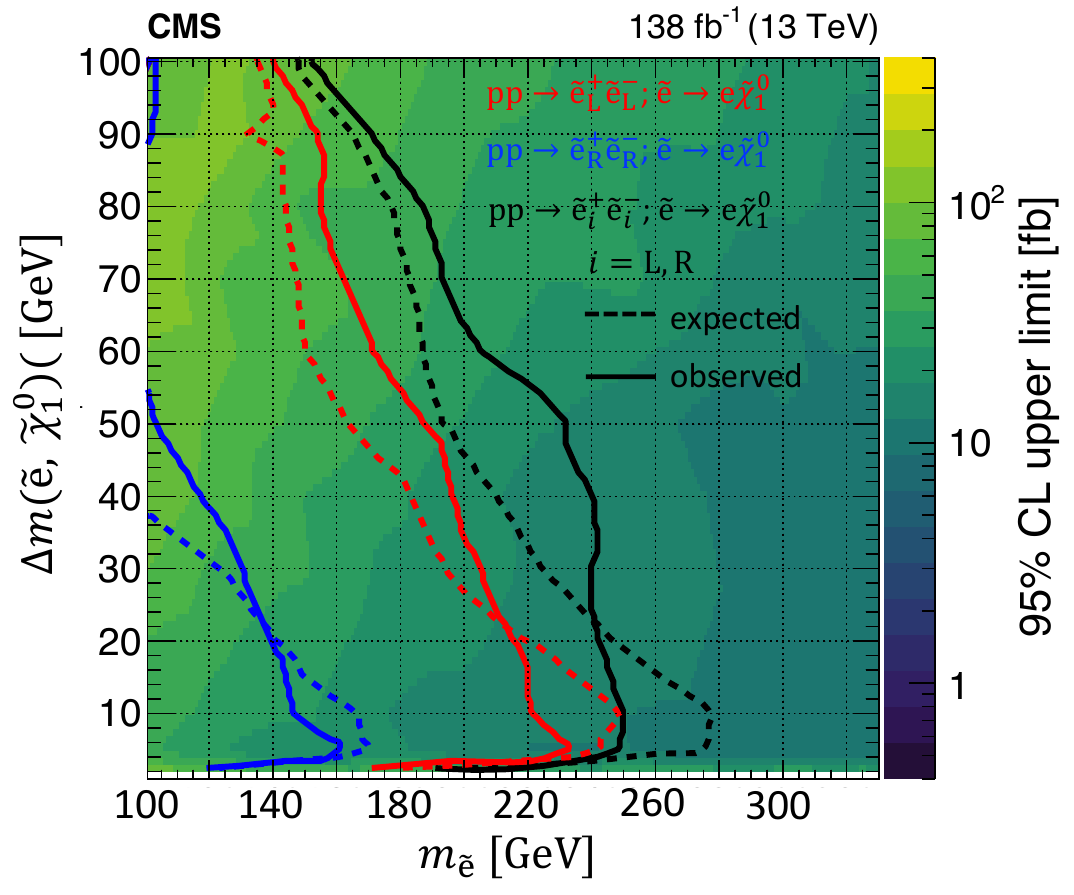}
  \includegraphics[width=0.48\textwidth]{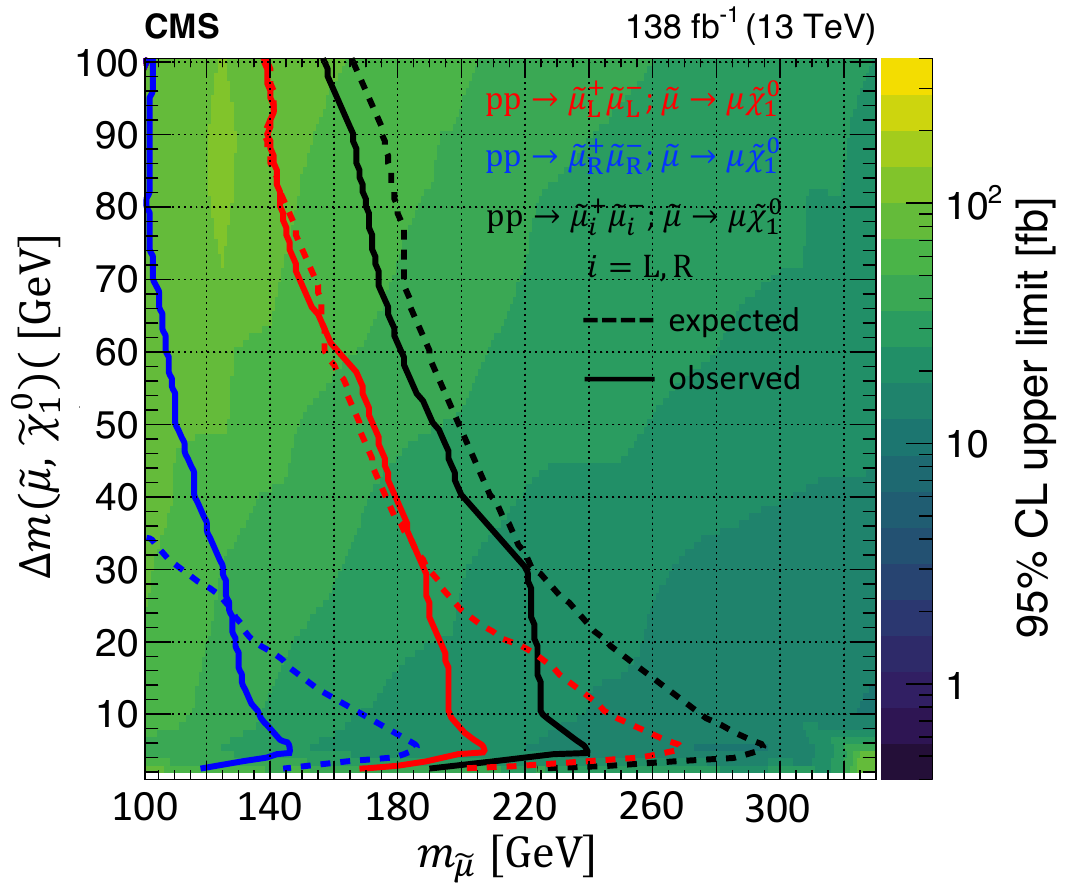}
  \caption{Slepton pair production. Observed 95\%~CL upper limits on the cross section 
  times branching fraction squared for direct selectron pair production (\cmsLeft) and smuon pair production (\cmsRight) followed by decay of both sleptons to the corresponding lepton and neutralino (color scale). 
  The limits are shown as a function of the slepton mass and the mass difference between the slepton and the lightest neutralino for the three different simplified possibilities of only RR, only LL, and 
  both RR and LL where it is assumed that the R and L masses are identical. The regions to the left of 
  the lines denote the regions excluded for a branching fraction of 100\%.
  Median expected limits for 100\% branching fraction are delimited by the dashed lines.
  }
  \label{fig:slepton_xslimits_eandmu}
\end{figure}

\begin{figure}[!ht]
  \centering
  \includegraphics[width=0.49\textwidth]{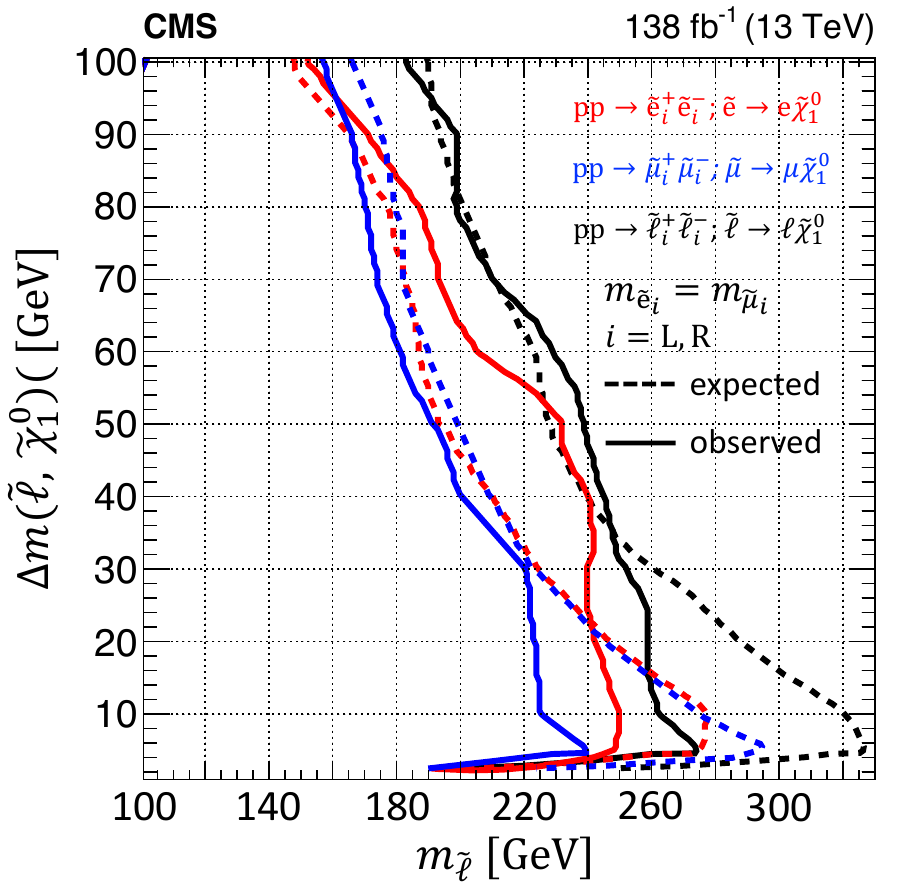}
  \caption{Slepton pair production. Observed and median expected limits for direct slepton pair production at 95\% \CL. Slepton $\tilde{\Pell}_{\mathrm{L}/\mathrm{R}}$ indicates the scalar supersymmetric partner of left- and right-handed electrons and muons. The limit is shown as a function of the slepton mass and 
  the mass difference between the slepton and the lightest neutralino. The corresponding selectron only and smuon only results of Fig.~\ref{fig:slepton_xslimits_eandmu} are shown too assuming 
  a 100\% branching fraction.}
  \label{fig:slepton_limits}
\end{figure}

\begin{figure}[!ht]
  \centering
  \includegraphics[width=0.48\textwidth]{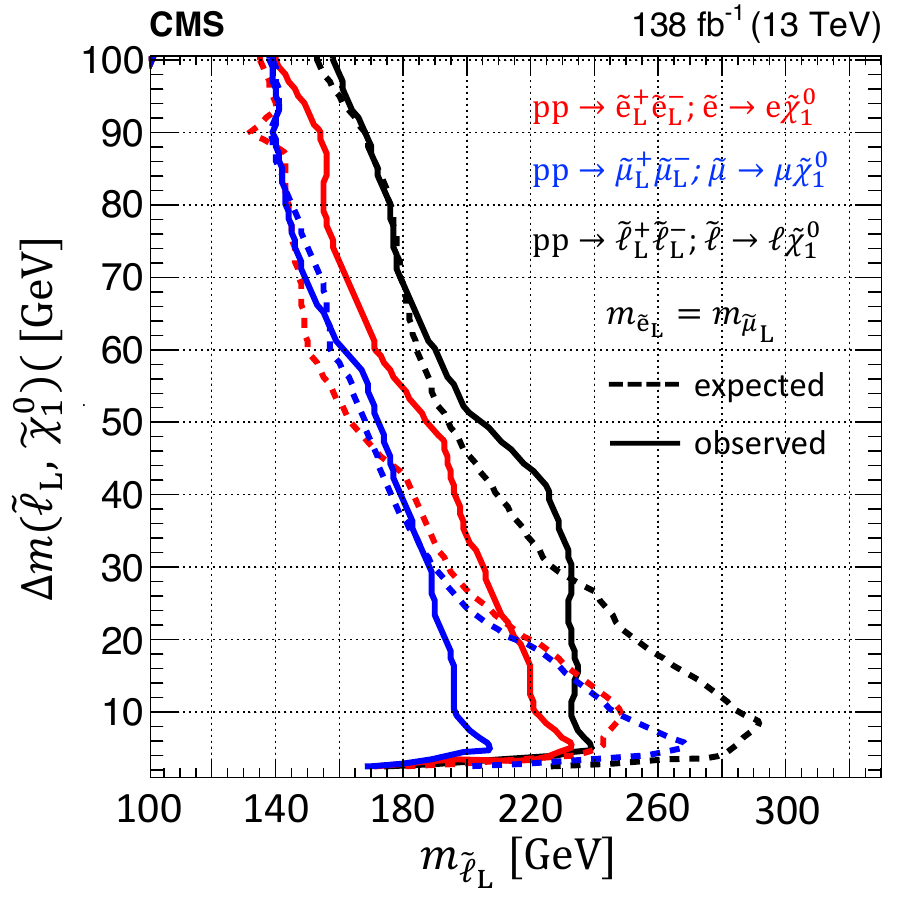}
  \includegraphics[width=0.48\textwidth]{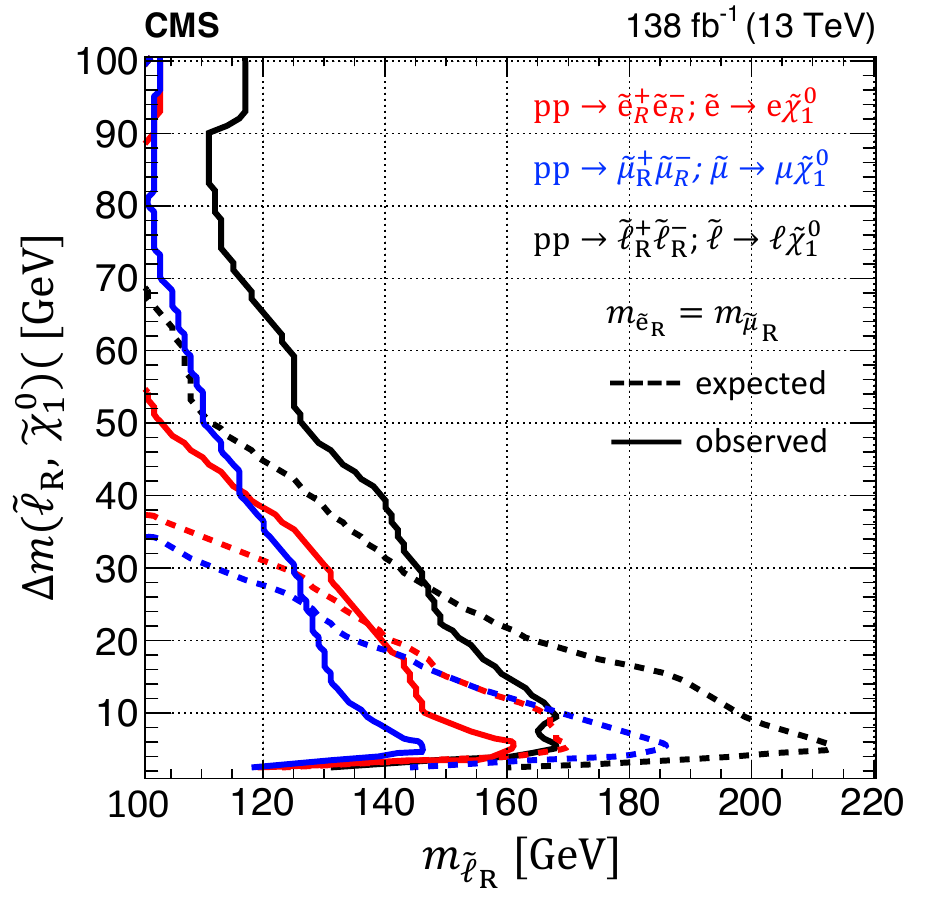}
  \caption{Slepton pair production. Observed 95\%~CL exclusion regions for direct pair production 
of the superpartners of the left-handed leptons (\cmsLeft) and direct pair production of the superpartners of the right-handed leptons (\cmsRight) 
followed by decay of both sleptons to the corresponding lepton and neutralino with 100\% branching fraction. The limits are shown as a function of the slepton mass and the  
  mass difference between the slepton and the lightest neutralino. The regions to the left of the lines denote the excluded regions.
  Median expected limits are displayed with dashed lines.
  }
  \label{fig:slepton_limits_LL-RR}
\end{figure}

Figures~\ref{fig:slepton_xslimits} and~\ref{fig:slepton_xslimits_eandmu} show the experimental 
upper limits on the product of the cross section and branching fraction 
squared for direct slepton pair production in the (mass, mass-difference) plane 
using a color scale. 
A particular focus is given to compressed mass spectra, where 
this analysis contributes substantially with 
respect to previous LHC analyses with $\sqrt{s}=13$\TeV datasets performed by ATLAS 
and CMS~\cite{Aaboud:2017leg, Sirunyan:2019zfq, Aad:2019qnd, ATLAS:2025evx}, 
and generally extends the regions probed by prior results 
from the LEP experiments~\cite{ALEPH:2001oot, DELPHI:2003uqw, L3:2003fyi, OPAL:2003nhx}. 
Figure~\ref{fig:slepton_xslimits} 
illustrates the results for selectron and smuon channels 
combined assuming degenerate selectron and smuon masses, 
while Fig.~\ref{fig:slepton_xslimits_eandmu} has the results separately 
for selectrons and smuons.
Additionally, in both figures the three different model exclusion lines for 100\% branching 
fraction show the 95\% \CL exclusion regions for pair production of 
only the superpartners of the left-handed leptons, only the superpartners of the right-handed leptons, and for both superpartners (where the two chirality partners are assumed mass degenerate).
Figures~\ref{fig:slepton_limits} and~\ref{fig:slepton_limits_LL-RR} show 
the 95\% \CL mass exclusion regions for each of the three production possibilities under the 100\% branching fraction assumption.
Each figure shows separate model exclusion lines for the three possible slepton flavor 
combinations (selectrons only, smuons only, and both light-flavor sleptons). 

These results are stronger and more comprehensive than previously reported by 
the LHC experiments for compressed masses; they include separate results 
for selectrons and smuons, and separate and combined 
results for the supersymmetric partners of the left- and right-handed charged leptons.

\section{Summary}\label{sec:summary}
A general search has been presented 
for supersymmetric particles (sparticles) in proton-proton collisions at 
a center-of-mass energy of 13\TeV with the CMS detector at the LHC using a data 
sample corresponding to an integrated luminosity of 138\fbinv.
A wide range of potential sparticle signatures are targeted 
including production of pairs of electroweakinos, sleptons, and top squarks.
The search is focused on events with a high transverse momentum system 
from initial-state-radiation jets recoiling against 
a potential sparticle system with significant missing transverse momentum.
Events are categorized based on their lepton multiplicity, jet multiplicity, 
\PQb tags, and kinematic variables sensitive to the sparticle masses and mass splittings.
The sensitivity
extends to higher parent sparticle masses than 
previously probed at the LHC for production of pairs of electroweakinos, sleptons, and top squarks 
for compressed mass spectra. The results on pair production of charginos and sleptons in the compressed mass regime 
extend well beyond the canonical 100\GeV sparticle mass scale previously explored at LEP.
The observed results demonstrate reasonable agreement with the predictions 
of the background-only model and 
model-independent event count upper limits for seven mutually exclusive event selections are reported.
Competitive 95\% confidence level (\CL) lower mass limits 
are set on sparticle pair production, especially 
in the compressed mass regime, with mass differences between the lightest and parent sparticle 
as low as 3\GeV being tested. 

Top squark mass limits for three decay models are presented in the plane of the top squark mass 
and the mass difference. Limits on the decay via a top quark extend to 780\GeV with a mass of 750\GeV excluded 
at 95\% \CL or higher for mass differences between 60 and 175\GeV; the most stringent exclusion is at a mass difference of 150\GeV.
Limits on the decay via a bottom quark and an intermediate chargino extend to 620\GeV with 
a mass of 550\GeV excluded at 95\% \CL or higher for mass differences between 35 and 140\GeV; the most stringent 
exclusion is at mass differences of between 50 and 90\GeV. Limits on the decay 
via a charm quark extend to 660\GeV with a mass of 520\GeV excluded 
at 95\% \CL or higher for mass differences between 10 and 60\GeV; the most stringent exclusion is at a mass difference of 20\GeV.

The 95\% \CL lower mass limits on chargino-neutralino production assuming heavy sleptons 
extend to 325 (175)\GeV for wino (higgsino) 
cross sections, where the most stringent mass limits are set for mass differences of 50 (10)\GeV. 
The limits with wino cross sections exceed 300\GeV for the broad range of mass differences between 8 and 65\GeV, while the 
limits with the higgsino cross section assumption exceed 163\GeV for mass differences between 3 and 50\GeV.
For chargino pair production, 95\% \CL lower mass limits are obtained for wino cross sections and decay via a \PW boson. 
These extend to 200\GeV with the most stringent mass limit set for a mass difference of 5\GeV while masses 
exceeding 120\GeV are excluded for all mass differences above 5\GeV. Related chargino pair production limits for the case 
of decays via sleptons and sneutrinos and with wino cross sections extend to 490\GeV for a mass difference of 55\GeV.

The 95\% \CL lower mass limits on pair production of charged sleptons extend to 168\GeV (slepton partner 
of right-handed lepton only), 
240\GeV (slepton partner of left-handed lepton only), and 270\GeV (both sleptons mass degenerate) for 
the most favorable mass splitting of around 5\GeV for the case of mass-degenerate first- and second-generation sleptons. 
Slepton masses exceeding 110, 175, and 200\GeV for all mass splittings ranging from 3 to 80\GeV are excluded at 95\% \CL or higher  
for the same three cases respectively. Similar results are also presented separately for selectrons and smuons 
assuming that the other slepton is not produced. For selectrons (smuons), the most stringent 95\% \CL lower mass limits 
are set at 160, 230, 250\GeV (145, 195, 240\GeV) for mass differences 
around 5\GeV for the three cases and with sensitivity to a broad range of mass differences from 3 to 100\GeV.

\begin{acknowledgments}
We congratulate our colleagues in the CERN accelerator departments for the excellent performance of the LHC and thank the technical and administrative staffs at CERN and at other CMS institutes for their contributions to the success of the CMS effort. In addition, we gratefully acknowledge the computing centers and personnel of the Worldwide LHC Computing Grid and other centers for delivering so effectively the computing infrastructure essential to our analyses. Finally, we acknowledge the enduring support for the construction and operation of the LHC, the CMS detector, and the supporting computing infrastructure provided by the following funding agencies: SC (Armenia), BMBWF and FWF (Austria); FNRS and FWO (Belgium); CNPq, CAPES, FAPERJ, FAPERGS, and FAPESP (Brazil); MES and BNSF (Bulgaria); CERN; CAS, MoST, and NSFC (China); MINCIENCIAS (Colombia); MSES and CSF (Croatia); RIF (Cyprus); SENESCYT (Ecuador); ERC PRG, TARISTU24-TK10 and MoER TK202 (Estonia); Academy of Finland, MEC, and HIP (Finland); CEA and CNRS/IN2P3 (France); SRNSF (Georgia); BMBF, DFG, and HGF (Germany); GSRI (Greece); NKFIH (Hungary); DAE and DST (India); IPM (Iran); SFI (Ireland); INFN (Italy); MSIT and NRF (Republic of Korea); MES (Latvia); LMTLT (Lithuania); MOE and UM (Malaysia); BUAP, CINVESTAV, CONACYT, LNS, SEP, and UASLP-FAI (Mexico); MOS (Montenegro); MBIE (New Zealand); PAEC (Pakistan); MES, NSC, and NAWA (Poland); FCT (Portugal);  MESTD (Serbia); MICIU/AEI and PCTI (Spain); MOSTR (Sri Lanka); Swiss Funding Agencies (Switzerland); MST (Taipei); MHESI and NSTDA (Thailand); TUBITAK and TENMAK (T\"{u}rkiye); NASU (Ukraine); STFC (United Kingdom); DOE and NSF (USA).

\hyphenation{Rachada-pisek} Individuals have received support from the Marie-Curie program and the European Research Council and Horizon 2020 Grant, contract Nos.\ 675440, 724704, 752730, 758316, 765710, 824093, 101115353, 101002207, 101001205, and COST Action CA16108 (European Union); the Leventis Foundation; the Alfred P.\ Sloan Foundation; the Alexander von Humboldt Foundation; the Science Committee, project no. 22rl-037 (Armenia); the Fonds pour la Formation \`a la Recherche dans l'Industrie et dans l'Agriculture (FRIA-Belgium); the Beijing Municipal Science \& Technology Commission, No. Z191100007219010, the Fundamental Research Funds for the Central Universities, the Ministry of Science and Technology of China under Grant No. 2023YFA1605804, and the Natural Science Foundation of China under Grant No. 12061141002 (China); the Ministry of Education, Youth and Sports (MEYS) of the Czech Republic; the Shota Rustaveli National Science Foundation, grant FR-22-985 (Georgia); the Deutsche Forschungsgemeinschaft (DFG), among others, under Germany's Excellence Strategy -- EXC 2121 ``Quantum Universe" -- 390833306, and under project number 400140256 - GRK2497; the Hellenic Foundation for Research and Innovation (HFRI), Project Number 2288 (Greece); the Hungarian Academy of Sciences, the New National Excellence Program - \'UNKP, the NKFIH research grants K 131991, K 133046, K 138136, K 143460, K 143477, K 146913, K 146914, K 147048, 2020-2.2.1-ED-2021-00181, TKP2021-NKTA-64, and 2021-4.1.2-NEMZ\_KI-2024-00036 (Hungary); the Council of Science and Industrial Research, India; ICSC -- National Research Center for High Performance Computing, Big Data and Quantum Computing, FAIR -- Future Artificial Intelligence Research, and CUP I53D23001070006 (Mission 4 Component 1), funded by the NextGenerationEU program (Italy); the Latvian Council of Science; the Ministry of Education and Science, project no. 2022/WK/14, and the National Science Center, contracts Opus 2021/41/B/ST2/01369, 2021/43/B/ST2/01552, 2023/49/B/ST2/03273, and the NAWA contract BPN/PPO/2021/1/00011 (Poland); the Funda\c{c}\~ao para a Ci\^encia e a Tecnologia, grant CEECIND/01334/2018 (Portugal); the National Priorities Research Program by Qatar National Research Fund; MICIU/AEI/10.13039/501100011033, ERDF/EU, "European Union NextGenerationEU/PRTR", and Programa Severo Ochoa del Principado de Asturias (Spain); the Chulalongkorn Academic into Its 2nd Century Project Advancement Project, and the National Science, Research and Innovation Fund via the Program Management Unit for Human Resources \& Institutional Development, Research and Innovation, grant B39G680009 (Thailand); the Kavli Foundation; the Nvidia Corporation; the SuperMicro Corporation; the Welch Foundation, contract C-1845; and the Weston Havens Foundation (USA).
\end{acknowledgments}
\bibliography{auto_generated}

\cleardoublepage \appendix\section{The CMS Collaboration \label{app:collab}}\begin{sloppypar}\hyphenpenalty=5000\widowpenalty=500\clubpenalty=5000\cmsinstitute{Yerevan Physics Institute, Yerevan, Armenia}
{\tolerance=6000
V.~Chekhovsky, A.~Hayrapetyan, V.~Makarenko\cmsorcid{0000-0002-8406-8605}, A.~Tumasyan\cmsAuthorMark{1}\cmsorcid{0009-0000-0684-6742}
\par}
\cmsinstitute{Institut f\"{u}r Hochenergiephysik, Vienna, Austria}
{\tolerance=6000
W.~Adam\cmsorcid{0000-0001-9099-4341}, J.W.~Andrejkovic, L.~Benato\cmsorcid{0000-0001-5135-7489}, T.~Bergauer\cmsorcid{0000-0002-5786-0293}, K.~Damanakis\cmsorcid{0000-0001-5389-2872}, M.~Dragicevic\cmsorcid{0000-0003-1967-6783}, C.~Giordano, P.S.~Hussain\cmsorcid{0000-0002-4825-5278}, M.~Jeitler\cmsAuthorMark{2}\cmsorcid{0000-0002-5141-9560}, N.~Krammer\cmsorcid{0000-0002-0548-0985}, A.~Li\cmsorcid{0000-0002-4547-116X}, D.~Liko\cmsorcid{0000-0002-3380-473X}, I.~Mikulec\cmsorcid{0000-0003-0385-2746}, J.~Schieck\cmsAuthorMark{2}\cmsorcid{0000-0002-1058-8093}, R.~Sch\"{o}fbeck\cmsAuthorMark{2}\cmsorcid{0000-0002-2332-8784}, D.~Schwarz\cmsorcid{0000-0002-3821-7331}, M.~Sonawane\cmsorcid{0000-0003-0510-7010}, W.~Waltenberger\cmsorcid{0000-0002-6215-7228}, C.-E.~Wulz\cmsAuthorMark{2}\cmsorcid{0000-0001-9226-5812}
% ...
\par}
\vskip\cmsinstskip
\dag:~Deceased\\
$^{1}$Also at Yerevan State University, Yerevan, Armenia\\
$^{2}$Also at TU Wien, Vienna, Austria\\
% ...
\end{sloppypar}
\end{document}